\documentclass[a4paper, 11pt]{article}
\usepackage{epsfig}

\newcommand{\be}{\begin{equation}}
\newcommand{\ee}{\end{equation}}
\newcommand{\ba}{\begin{eqnarray}}
\newcommand{\ea}{\end{eqnarray}}
\newcommand{\bi}{\begin{itemize}}
\newcommand{\ei}{\end{itemize}}
\newcommand{\tr}{{\rm Tr\,}}

\newcommand{\<}{\langle} 
\renewcommand{\>}{\rangle}  

\newcommand{\la}[1]{\label{#1}}

\begin{document}
\begin{titlepage}
\begin{flushright}
\end{flushright}
\begin{centering}
\vfill
 
{\bf \Large HIGH SPIN GLUEBALLS FROM THE LATTICE}

\vspace{1.5cm}

H.B. Meyer\footnote{meyer@thphys.ox.ac.uk},
M.J. Teper\footnote{teper@thphys.ox.ac.uk}

\vspace{0.8cm}

Theoretical Physics, University of Oxford, 
1 Keble Road,\\ Oxford, OX1 3NP, United Kingdom

\vspace*{3.0cm}
 
\end{centering}
 
\noindent 
We discuss the principles underlying higher spin glueball calculations
on the lattice. For that purpose, we develop numerical techniques
to rotate Wilson loops by arbitrary angles in lattice gauge theories 
close to the continuum. 
As a first application, we compute the glueball spectrum of the $SU(2)$
gauge theory in 2+1 dimensions for both parities and for spins ranging 
from 0 up to 4 inclusive.
We measure glueball angular wave functions directly, decomposing them
in Fourier modes and  extrapolating the Fourier coefficients to 
the continuum. This allows a reliable labelling of the continuum states
and gives insight into the way rotation symmetry is recovered.
As one of our results, we demonstrate that the D=2+1 $SU(2)$ glueball 
conventionally labelled as $J^P = 0^-$ is in fact $4^-$ and that
the lightest ``J=1'' state has, in fact, spin 3.
\vfill


\vspace*{1cm}
\vfill
\end{titlepage}
\setcounter{footnote}{0}
\section{Introduction}

There are several reasons to explore the higher spin spectrum of pure gauge
theories. One might wish to test models of glueballs.
Such models naturally divide into those where glueballs are composed of
constituent gluons, for example the bag model (see [\ref{bag}]
and references therein), and
those in which glueballs are based on closed loops of chromo-electric 
flux, for example the Isgur-Paton flux-tube model [\ref{isgur}]. 
Attempts to test these models [\ref{moretto} - \ref{model2}] against 
existing lattice spectra are severely hampered both by the lack of 
lattice results for $J \ge 3$ in D=2+1 and $J \ge 4$ in D=3+1, and
by ambiguities in assigning continuum spins to states that lie in 
representations of the lattice rotation group.
Secondly, there has been a long-standing question in QCD phenomenology 
concerning the interpretation of the (soft) Pomeron within the 
low-energy spectrum. (See [\ref{pomeron}] for recent reviews.) 
The absence of a corresponding mesonic ($q{\bar q}$)
trajectory has made glueballs the favoured candidates. Simple models
[\ref{mtjh02}] do indeed predict a Regge type behaviour of 
the glueball spectrum, namely $J \propto M^2$ at large $J$, with
a slope that is much smaller than that of the usual mesonic Regge
trajectories. A determination of the leading Regge trajectory in the 
pure gauge theory would largely resolve this issue. 
Thirdly, it is a generally interesting question to study how a 
symmetry that is broken on the lattice, in this case the rotation group, 
is restored as one reaches the continuum.

The  standard method to measure the spectrum of a gauge theory 
is to evaluate the correlation function of a gauge-invariant 
operator in Euclidean space (for a review of ``glueball technology'', see 
[\ref{mtd3}]):
\begin{equation}
\langle {\cal O}^\dagger(t) {\cal O}(t=0)\rangle =\sum_n \left|\langle 0|
 {\cal O}|n\rangle\right|^2e^{-E_nt}\label{labase}
\end{equation} 
This formula follows directly from the insertion of the complete set of 
energy eigenstates $|n\rangle$. In order to correctly  
label the latter with the quantum numbers of the
rotation group, one has to construct operators that project out the undesired
states.  In continuum Euclidean space, it is possible to construct
operators belonging to an irreducible representation of the rotation 
group; that is, with a definite spin. In
two space dimensions, where, for simplicity, we shall work from now on, 
the construction amounts to 
\begin{equation}
{\cal O}_J=\int \frac{d\phi}{2\pi} ~ e^{iJ\phi}~ {\cal O}_\phi \la{opj}
\end{equation}
where $J$ is the spin and ${\cal O}_\phi$ represents an operator 
characterising the direction $\phi$, i.e. ${\cal O}_\phi$ is obtained 
by rotating the operator ${\cal O}_{\phi=0}$ through an angle $\phi$. 
If $|j;n_j\rangle$ is a state of spin $j$ and other quantum numbers
$n_j$, we have
\begin{equation}
\langle 0| {\cal O}_J|j;n_j\rangle=\delta(J-j)
~\langle 0| {\cal O}_{\phi}|j;n_j\rangle~,
\end{equation}
where $\phi$ is now arbitrary.
The definite-spin subspaces in which the Hamiltonian can be
diagonalised separately are still infinite-dimensional, so that extracting
their lightest states requires an additional piece of strategy.
The most commonly used is the variational method 
(see [\ref{mtd3},\ref{mart_var}]
and references therein), which tries to separate pure exponentials 
in eqn(\ref{labase}).

However, on a  lattice, rotation symmetry is broken and only a 
handful of rotations leave the lattice invariant. Therefore physical 
states can only be classified into irreducible representations of the
lattice point group. This is a much less thorough classification, in
the sense that the diagonal blocks of the lattice Hamiltonian each 
contain a whole tower of smaller blocks of the continuum Hamiltonian. 
For instance, the trivial square lattice irreducible representation 
contains all multiple-of-4 spin states of the continuum, because
$\exp\{iJ\phi\}$ is unchanged if $J \to J+4n, \ n\in Z$, for 
any symmetry rotation of the lattice, 
i.e. for $\phi=n^\prime\pi/2, \ n^\prime\in Z$.

When extracting glueball masses from lattice calculations, it has been 
customary to label the obtained states with the lowest spin falling into
the lattice irreducible representation; in the example above,
it would be `spin 0'. Not only does
the procedure used till now lack the capability of extracting the higher
spin spectrum, but this labelling could very well be wrong, especially for 
excited states.  Suffice it to think of 
the hydrogen atom, where the degeneracy in $\ell$ implies, for example, 
that the $n=1, \ell=1$ is lighter than the $n=2,\ell=0$ state. In
the case of D=2+1 gauge theories, the simple flux tube model
predicts [\ref{model1},\ref{model2}] that the lightest $J^{PC}=0^{-+}$ 
state is much heavier than the lightest $4^{-+}$ state, and the mass that 
it predicts for the latter agrees with the value obtained on the 
lattice [\ref{mtd3}] for the `$0^{-+}$'. This suggests a possible
misidentification, as emphasised in [\ref{model1},\ref{model2}]. 

In the case of the 2 dimensional square lattice, the only states one can 
distinguish in the ``traditional'' fashion are those of spin 0, 1 and 2. 
This work is an attempt to go beyond that apparent limitation of lattice 
calculations: we want to check the correctness of the conventional labelling
of states as well as to extract the lowest lying states carrying spin higher 
than 2. It is
clear  that one necessary ingredient in the realisation of this program 
is a reliable way to construct operators that are rotated with respect to each
other by angles smaller than $\frac{\pi}{2}$ but that are otherwise
(nearly) identical.
Indeed one expects that as the lattice spacing becomes very much 
smaller than the dynamical length scale, this becomes possible
due to the recovery of the continuum symmetries.
In fact it has been known for a while that rotation
symmetry is restored dynamically rather early in the approach to the 
continuum. An early piece of evidence came from the ($D=3+1$) $SU(2)$ static
potential measured off the lattice axis [\ref{reb}]. Later it was shown 
[\ref{for}] that the ($D=3+1$) $SU(3)$ glueball shown dispersion relation 
$E(\vec{p})$ to a good approximation depends only on $|\vec{p}|$ already
at $\beta=5.7$. The detailed glueball spectra obtained more recently in
(2+1) [\ref{mtd3}] and (3+1) [\ref{mor}] dimensions exhibit the degeneracies
between states belonging to different lattice irreducible representations
that are expected in the continuum limit.

\paragraph{}The outline of this paper is as follows. We begin by 
discussing techniques for constructing arbitrary rotations of a 
given operator. 
We design systematic tests to evaluate how well these methods 
perform in $D=2+1$ $SU(2)$ gauge theory. (The generalisation
to $SU(N)$ is trivial.) We then discuss how to use 
these techniques to extract the high spin spectrum from lattice 
simulations. As an example, we analyse the case
of the lightest $4^-$ and $0^-$ glueballs. We find that the state
conventionally labelled as a $0^-$ is in fact a $4^-$, as was
suggested by calculations of the spectrum within the flux
tube model [\ref{model1},\ref{model2}]. We also find that the
$J=3$ ground state is lighter than the $J=1$ (again
as suggested by the flux tube model). Although all
our calculations in this paper are performed in two space
dimensions, a large part of our motivation for studying this
problem lies in $D=3+1$ and we briefly discuss the application 
of our approach  to that case.
In the conclusions we briefly comment on some other recent
attempts [\ref{rjmtrot},\ref{liurot},\ref{rjrot}] to calculate 
the masses of higher spin glueballs and summarise our results.
Our next step, to apply these ideas to determine 
the `Pomeron' Regge trajectories in  $SU(N)$ gauge theories,
will be the subject of forthcoming publications.

\section{Two methods of operator construction}

Lattice glueball operators are usually constructed out
of space-like Wilson loops.
To construct lattice operators that project onto arbitrary spins
we need linear combinations of Wilson loops rotated by arbitrary 
angles. Since we are on a cubic lattice, such a rotated loop will
usually only be an 
approximate (rotated) copy of the initial loop. The better the
approximation, the less ambiguous the spin assignment. Now, a general
Wilson loop consists of a number of sites connected by products of 
links. To obtain a good projection onto the lightest states in
any given sector, these links need to be `smeared' or `blocked'
so that they are smooth on physical rather than just ultraviolet
length scales. (See, for example, [\ref{mtd3}] and references therein.) 
To construct arbitrary (approximate) rotations of various Wilson loops,  
it is clear that we need to be able to construct `smeared' parallel
transporters between arbitrary sites in a given time-slice of the
lattice. We now describe two techniques to do this. We begin
by reviewing and elaborating upon a method [\ref{mtmatrix}] 
that has recently been used [\ref{rjrot}] in an attempt to address 
the $0^-/4^-$ ambiguity referred to in the Introduction. This method 
is however too computationally expensive to allow a realistic
continuum extrapolation with our resources, and without such an 
extrapolation one has very little control over the restoration of 
continuum rotational invariance. We therefore develop a second 
method that is much less expensive and provides a practical approach 
to the problem.

\subsection{The matrix method}
\label{subsection_matrix}

Consider the two spatial dimensions of a given time-slice of size
$L\times L$ with periodic boundary conditions. 
Each point $p$ is parametrised by integers $(m,n)$ representing its 
Cartesian coordinates in lattice units. Given a mapping 
$i \rightarrow (m(i), n(i))\equiv \varphi(i)$ 
and setting $N\equiv L^2$, we can define an $N \times N$ matrix 
$M$  by its elements
\begin{equation}
M_{ij}\equiv U(\varphi(i);\varphi(j))
\end{equation}
where $U(p;q)$ is the link matrix joining points $p$ to $q$ if they are
nearest neighbours, and vanishes otherwise. Since
$U(p,q)= (U(q,p))^\dagger$, $M$ is hermitian. For notational simplicity,
if $\varphi(i)$ is the origin and $\varphi(j)=(m,n)$, 
we shall often write $M_{ij}\equiv M[m,n]$.

It is straightforward to see that $\left(M^\ell\right)_{ij}$ 
contains all paths of length $\ell$ going from $\varphi(i)$ to 
$\varphi(j)$.  One can construct a ``superlink'' 
connecting these two points by adding up paths of all lengths, weighted by
some damping factor that ensure the convergence of the series:
\begin{equation}
K=\sum_\ell c(\ell) M^\ell
\end{equation}
In general it is very costly to calculate such a  power series
numerically. If one chooses $c(\ell) = \alpha^{\ell}$ then the
series can be resummed
\begin{equation}
K=\sum_{\ell\geq 0} \alpha^\ell M^\ell=\left(1-\alpha M\right)^{-1}. \la{kdef}
\end{equation}
and the calculation of the geometric series can be reduced to the
inversion of a matrix. For instance, one can now compute a triangular 
``fuzzy'' Wilson loop by simply multiplying together three elements 
of $K$
\begin{equation}
W = \tr K_{ij} K_{jk}  K_{ki}
\label{eqn_triangle}
\end{equation}
where $i,j,k$ are the vertices of the triangle. It is clear that
as we increase $\alpha$ the longer paths are less suppressed
i.e. increasing $\alpha$ increases the smearing of the ``superlink''.
Thus by inverting a single matrix we obtain smeared parallel 
transporters between all pairs of sites in the given time-slice. We can 
now use these to construct any Wilson loops we wish.

While the above construction is valid for an infinite volume, one
must be careful if the volume is finite. Consider, for example, the
triangular Wilson loop defined in eqn(\ref{eqn_triangle}). On a finite
spatial torus, the sum of paths contributing to $K_{ij}$ contains not
only the `direct' paths from $i$ to $j$ but also paths that go
the `long' way around the `back' of the torus between these two points.
That is to say, the Wilson loop defined in  eqn(\ref{eqn_triangle})
is not necessarily a contractible triangle; some of the 
contributions to $W$ are non-contractible closed paths that wind 
once around the torus. In the confining phase such an operator
projects onto flux loops that wind once around the torus. These
states are orthogonal to glueballs. Moreover, in the kind of
volume with which one typically works, this loop will be much lighter
than any of the excited glueballs. Such effects induce an infra-red
breaking of rotation symmetry that, unlike lattice spacing effects, 
will survive the continuum limit. Fortunately it is simple
to modify our matrix procedure so as to explicitly suppress 
such contributions and we now describe two ways of doing so.

It is convenient to label the
link matrix emanating from the site $(x,y,t)$ in the direction
$\mu$  by $U_{\mu}(x,y,t)$. Consider now the matrix $M^x$ 
which is identical to $M$ except that certain elements
acquire a minus sign:
\begin{equation}
U_{x}(x=L,y,t) \to -1 \times U_{x}(x=L,y,t) \ \ \ \ \ \forall \ y.
\label{eqn_Mx}
\end{equation}
It is clear that the corresponding matrix $K^x$ will produce
`superlinks' that are identical to those from $K$ except that
the contribution of paths that wind once (or an odd number of times)
around the $x$-torus will come in with a relative minus sign. 
Thus if we replace $K$ by $K + K^x$ in eqn(\ref{eqn_triangle}) 
the contribution of all the non-contractible paths winding 
around the  $x$-torus will cancel. In the same way we can
define a matrix  $M^y$ from  $M$ by
\begin{equation}
U_{y}(x,y=L,t) \to -1 \times U_{y}(x,y=L,t) \ \ \ \ \ \forall \ x
\label{eqn_My}
\end{equation}
and a matrix $M^{xy}$ which includes both the modifications
in equations (\ref{eqn_Mx}) and (\ref{eqn_My}). It is easy to
see that the sum of the corresponding inverse matrices,
$K + K^x + K^y + K^{xy}$, will produce superlinks that
have no contributions from non-contractible paths that
wind around the $x$-torus or the $y$-torus or simultaneously
around both tori. This is a simple and effective modification
although it would appear to suffer from the fact that it quadruples 
the length of the calculation. However it is easy to see that
one can considerably reduce this cost. We start by noting that on
a lattice with an even number of sites in both $x$ and $y$ directions, even
and odd powers of $M$ connect a given point (say, the origin) to two disjoint
sets of lattice sites. If $L_x$ and $L_y$ are odd, this is no longer the case,
but paths joining two points by going around the world have an opposite parity
weighting in powers of $\alpha$ to those connecting them directly. Therefore
in that case we can proceed as follows: use 
$K_e\equiv(1-\alpha^2M^2)^{-1}=\sum_{n\geq0}(\alpha M)^2$ to propagate by an
 even number of lattice links and $K_o\equiv \alpha M K_e$ for an odd number 
of ``jumps''. Thus paths with an odd winding number are excluded by 
construction. This way of proceeding has the additional advantage that one
can truly propagate by a distance larger than $L_{x,y}/2$, which is not the 
case with $L_{x,y}$ even.
On the other hand, the fact that with an even number of lattice sites in both
directions $K_e$ possesses a partitioned structure allows us to store any 
polynomial in the matrix $M^2$ in two matrices of size $L^2/2 \times L^2/2$.
If standard inversion algorithms are applied to compute $K_e$ (for which
the CPU time scales as $N^3$), this represents a reduction in CPU time
 by a factor four. $K_o$ is then very fast to obtain, given the sparsity
of the matrix $M$.
In summary, obtaining the superlinks free of odd winding number paths
requires either of the following computations:
if the lattice has an odd number of sites, it is sufficient
to perform one full matrix inversion, plus two multiplications by $M$; 
if $L_{x,y}$ are even, we have to use the four $Z(2)$ transformations,
as discussed above, but we can compute and store the superlinks in
matrices smaller by a factor two.

There is an interesting interpretation to the matrix construction in
an infinite volume.
If we choose the mapping such that $\varphi^{-1}(m,n) =L\cdot m+n$, then
in the frozen configuration (all link variables set to unity),
$M$ coincides with the matrix used in 
discretising partial differential equations in the finite difference scheme;
more precisely, $M-4$ is exactly the expression of a discretized
Laplacian operator on a torus. For that reason, the Klein-Gordon equation
$-\nabla^2 F+m^2 F=0$ translates into
\begin{equation}
((am)^2+4-M)_{ij}F_{j}=0,\quad \forall i=1,\dots,N
\end{equation} 
where $F$ is now a column vector containing the approximate values of the 
function $F$ on the lattice sites. If we introduce a point-like source 
$v^{(z)}$ on the RHS, that is $v^{(k)}_i=\delta_{ik}$, we obtain the 
interpretation that $[((am)^2+4-M)^{-1}]_{ij}$ is the $2d$ 
lattice propagator of a massive scalar field from point
$\varphi(i)$ to point $\varphi(j)$. For a scalar field 
minimally coupled to the gauge field at finite $\beta$, the 
ordinary derivatives are simply replace by covariant derivatives.
If the scalar field is in the fundamental representation 
of the gauge group then
our matrix $1-\alpha M$ provides a discretisation of the kinetic
term where the parameter $\alpha$ in eqn(\ref{kdef}) corresponds to
\begin{equation}
\alpha=\frac{1}{(am_0)^2+4}
\label{eqn_alphatom}
\end{equation} 
and $m_0$ is the tree level mass of the scalar particle.
Setting $\alpha$ to $\frac{1}{4}$ corresponds to $m_0=0$. The propagator
calculation amounts to introducing a scalar particle 
in the configuration, a ``test-charge'' that does not modify 
the configuration, but the closed
paths of which reveal gauge-invariant information on the background gauge
field. In other words, this method is analogous to performing
a quenched simulation of the gauge theory minimally coupled to a scalar
field (with the latter confined to a time-slice). The lighter the mass
of the scalar particle, the greater the transverse distance that
it explores as it propagates between two sites. That is to say,
we recover our earlier conclusion: as $\alpha$ increases from small 
values, the propagator is increasingly `smeared'.

We expect on general grounds that if we calculate a propagator
over some physical length scale, and for a mass $am_0$ that is fixed in
physical units, then the lattice corrections to continuum rotational
invariance will be $O(a^2)$. This provides a general theoretical
argument that our matrix method will yield `rotationally invariant'
superlinks if one chooses the parameter $\alpha$ suitably. 
Of course, since there is no
symmetry to protect against mass counter-terms, there will be both
an additive and a multiplicative renormalisation of the mass.
That is to say, choosing the mass involves a `fine-tuning'
problem that is very similar to the one that one encounters when
using Wilson fermions. 

The recovery of rotational invariance at large distances and
at weak coupling can, in general, only be seen numerically 
(see below). However, in the special case of a frozen configuration, 
it can in fact be studied analytically (see Appendix A). This leads
to the following conclusions: if $m^2+n^2=d^2$, for
$1\ll d \ll (am_0)^{-1}$ the propagation from  $(0,0)$ to $(m,n)$
results from a Brownian motion and the length of the dominating 
paths is of the order $d^2$. In this regime rotation invariance is 
recovered. For $d\gg (am_0)^{-1}$ on the other hand, the superlinks 
become more directed and rotation invariance is lost. Since one
should think of a frozen field configuration as corresponding to 
$\beta=\infty$, i.e. to the continuum limit $a = 0$, we should take 
the mass to be $am_0 = 0$ if it is to be finite in physical units.
In that case, only the range $1\ll d \ll (am_0)^{-1} = \infty$ is
relevant. If we wish to think of a very small but non-zero
value of $a$, then the regime in which a frozen configuration
is an approximation to the field is for $d \ll (am_0)^{-1}$,
if $m_0$ is on the order of a physical length scale. In either case
$d\gg (am_0)^{-1}$ does not correspond to a physically relevant 
range of distances.

The matrix method is a simple and powerful tool for obtaining
smeared `superlinks' between all pairs of sites in a given
time-slice. However even in D=2+1 SU(2) and at moderate
$\beta$ values the matrix is already large and the inversion expensive.
While a calculation with modest statistics on say a $16^3$ lattice 
may be readily performed  on a workstation, this is no longer the 
case for the $24^3$ and $32^3$ lattices that would be needed
for even a minimal attempt at a continuum extrapolation.
To circumvent this problem we develop a much faster alternative 
method in the next subsection. However, before turning to that,
we finish with some general remarks about how the matrix method
may be developed into a more realistic tool.

We have noted that the elements of our inverse matrix 
$K=\left(1-\alpha M\right)^{-1}$ 
are nothing but the propagators of a minimally
coupled scalar particle in the fundamental representation,
whose bare mass is determined by the parameter $\alpha$.
In principle we are free to consider propagators of other
particles: these should provide equally good `superlinks'.
Consider then a fermion in the fundamental representation and
suppose we discretise it as a two-dimensional staggered lattice 
fermion. The propagators are obtained by inverting a
matrix which is obtained from our matrix $M$ simply 
by multiplying the elements of $M$
by position-dependent factors of -1 [\ref{lattice}].
This lattice discretisation maintains a chiral symmetry
which protects the massless fermions from an additive
renormalisation. This removes the fine-tuning problem we
referred to earlier: a first advantage. Moreover we expect the 
long distance physics to be encoded in the lowest eigenvalues,
and corresponding eigenvectors, of our (discretised) Dirac operator.
Now we recall that the fermion propagator can be expressed 
in terms of all the eigenvectors and eigenvalues of the Dirac
operator. If we truncate this sum to include only some
suitably chosen set of these lowest eigenvalues and eigenvectors,
then this should provide us with an approximation to the propagators
that maintains the long-distance physics; for example the
restoration of full rotational invariance. That is to say,
they can be used as `superlinks' for our purposes.
As we approach the continuum limit we do not need to
increase the number of these eigenvectors, as long as the
volume is fixed in physical units, so the computational cost
scales in a way that is far better than inverting the
full Dirac operator -- a second, major, advantage. As an added 
bonus we note that we can
expect the chiral symmetry to be spontaneously broken. This
implies a non-zero density of modes near zero which generates
the chiral condensate [\ref{bankscasher}], and the choice
of what are `small' modes then becomes unambiguous.
This is of course only an outline of a strategy; its
practical application is something we do not attempt here.

\subsection{The path constructor method}

We turn now to a simpler, more direct and, above all, faster alternative
method for constructing superlinks between any two sites.

In order to define a path from site $A$ to site $B$, we first
introduce a ``d-link'' in the diagonal direction of the lattice:
\begin{equation}
U_{\mu\nu}(x)={\cal U}\left(U_\mu(x)U_\nu(x+a\hat{\mu})+
	U_\nu(x)U_\mu(x+a\hat{\nu})\right)
\end{equation}
where $\cal{U}$ represents a unitarisation procedure (in $SU(2)$, the 
operation amounts to dividing the matrix by its determinant).
From a point $x$, there are now 8 directions available.
It is easy to write an algorithm that finds the path following 
the straight line from $A$ to $B$ as closely as possible.
 Indeed, at each step, it is 
sufficient to try all directions by adding the corresponding (d-)link
to the current state of the path and select the result that has maximal
 projection on the
$\overrightarrow{AB}$ vector. As this can lead to a path that is not invariant
under a $\pi$-rotation, one can average with the opposite path 
 obtained with the same algorithm by starting from $B$
and inverting it at the end.
In practise, before starting to calculate the path , we may, as usual,
 smear the ordinary links to reduce short-wavelength fluctuations and
achieve better overlap onto physical states. 

\subsection{A test for the operator construction methods}

We will now calculate appropriate Wilson loops using the two kinds 
of superlinks introduced above,  so as to test the extent to which
rotation symmetry is violated. The result is obviously determined
by the dynamics of the lattice gauge theory and the operator being measured.
Therefore, to obtain rotation invariance to a good accuracy, 
 the two following conditions must be satisfied: 
\begin{equation}
aL\gg \sigma^{-1/2}\qquad {\rm and } \qquad \sigma^{-1/2}\gg a
\label{eqn_conditions} 
\end{equation}
It is our task, when constructing rotated copies of operators, to ensure that
these conditions are also sufficient. 

We will refer to the 
method using the propagator as ``method I'', to the path constructor method
as ``method II''. In both cases,
in order to have a gauge-invariant operator, we must form a closed path.
Since a great number of paths contribute to the superlinks in method I, 
an operator of the type 
\begin{equation}
{\cal O}=\mathrm{Tr}~{K(y,x)\cdot K(x,y)  }
\end{equation}
is a perfectly acceptable operator characterising the direction determined
by the points $x,y$. We call it a ``segment'' operator. (It has an
interesting physical interpretation which we shall elaborate upon below.)
We can choose pairs of points $x,y$ that have approximately 
the same length -- up to the percent level -- and which
are rotated by an approximately constant angle. These ``approximations''
mean that there is an intrinsic limitation to the rotation invariance 
that we can expect to observe at a ``fixed'' length $|x-y|$. 
One can finesse this problem by plotting the values of 
${\cal O}(x,y)$ for all points  $x,y$ and seeing to what extent 
they fall on a single smooth curve. For our purposes an alternative
two-part strategy is more illuminating. In the first step, working 
at a ``fixed'' value of $|x-y|$, we ask if the 
violations in rotation symmetry are of the same order of magnitude
as the differences in the lengths. In the second step, as
a more  direct test of rotational invariance, 
we first normalise the operators  ${\cal O}(\phi)$ to a common
value and then calculate the  correlation between rescaled segment 
operators at different angles. 

We perform these tests  on a $16^3$ lattice at $\beta=6$, where we know
[\ref{mtd3}] that the string tension is
$a\surd\sigma \simeq 0.254$ so that the conditions
in eqn(\ref{eqn_conditions}) would appear to be satisfied. 
We choose $\alpha=0.24$ and there is no preliminary smearing of the links. 
Using the tree-level relation in eqn(\ref{eqn_alphatom}) 
and the above value of $a\surd\sigma$ 
we see that this value of $\alpha$ corresponds to a mass for the
scalar particle of $m_0\simeq 1.6\surd\sigma$, i.e. a physical rather 
than an ultra-violet scale. The segment we use is of length $7a$, 
and is rotated by multiples of
approximately $\frac{\pi}{16}$. To illustrate the necessity of the 
torelon-suppression procedure, we shall present our results with and 
without implementation of the latter.

We present in Fig.~\ref{meaI} the results of the first step of
our test, as applied to method I. The points
$x,y$ that we use lie on an (approximate) semicircle and are joined by 
solid lines for clarity. We do not label the $x$ and $y$ axes, but the
on-axis distance from the origin of the (semi)circle is 7 in lattice
units and this sets the (separate) scales for the $x$ and $y$ axes. 
Using this information, the Euclidean length $R(\phi)$ of the segment  
in each direction $\phi$ can be read directly off this polar plot;
the x-axis of the plot corresponds to the lattice x-axis. 
For each point $x,y$  we plot the average value of the segment operator,
as a point along the same  direction, with the distance to the
origin representing its value. For clarity these points have been
joined up by dashed lines. Both sets of points have
been rescaled so that they can be plotted on the same graph,
and there are separate plots with and without torelon removal.

\paragraph{}With method II,
the superlink from  a point $A$ to a point $B$  is a
unitary matrix (or a sum of two such matrices).
Therefore we cannot use segment operators since
they would be trivial. Instead we use long rectangular Wilson loops, 
typically $7\times1$; they each characterise a specific direction $\phi$.
We present the results in Fig.~\ref{meaII} in a similar polar plot
to that used in Fig.~\ref{meaI}: in each direction $\phi$, the 
Euclidean length $R(\phi)$ of the segment is given, as well as the average
value of the rectangle operator pointing in that direction. Note that
in this case there are no torelon contributions that need to be subtracted.

\begin{figure}[tb]

\centerline{\begin{minipage}[c]{8cm}
    \psfig{file=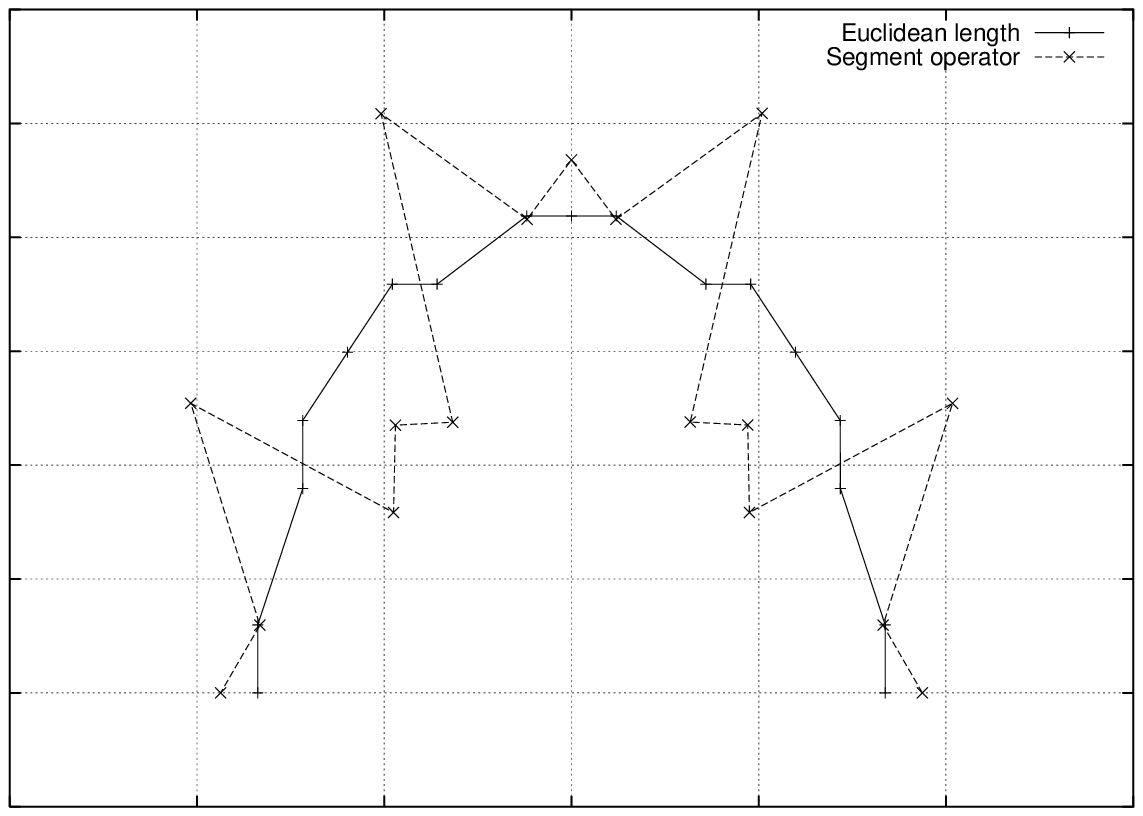,angle=0,width=8cm, height=4cm}
    \end{minipage}
    \begin{minipage}[c]{8cm}
    \psfig{file=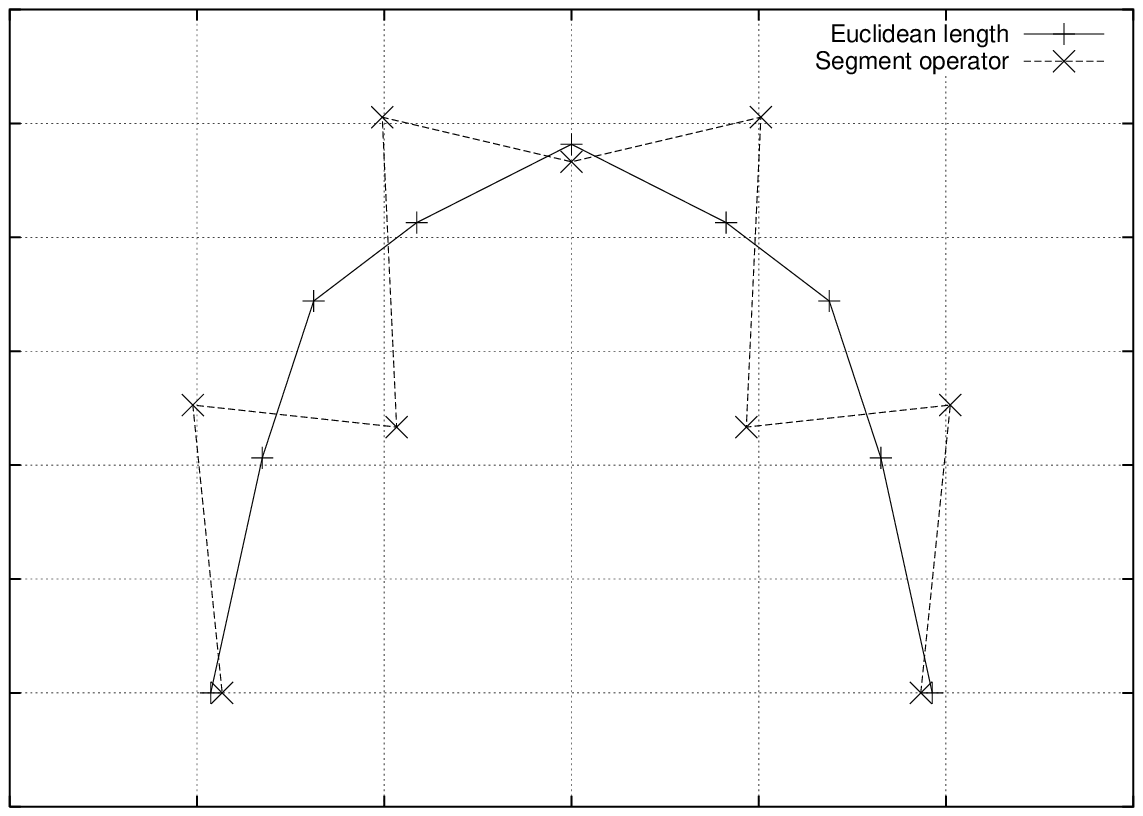,angle=0,width=8cm, height=4cm}
    \end{minipage}}

\caption[a]{In polar coordinates: the Euclidean length and the average
 value of segment operators in different directions $\phi$; left: 
method I without torelon-suppression, right: method I with 
torelon-suppression}

\la{meaI}
\end{figure}
\begin{figure}[bt]

\centerline{\begin{minipage}[c]{8cm}
    \psfig{file=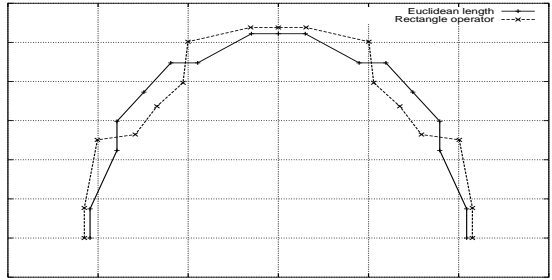,angle=0,width=8cm, height=4cm}
    \end{minipage}}
\vspace*{0.5cm}

\caption[a]{ In polar coordinates: the Euclidean length and the average
 value of rectangle operators in different directions $\phi$;  method II}

\la{meaII}
\end{figure}

%
\paragraph{Remarks\\} 
Method I: The average operators have a significantly larger
vacuum expectation value (vev) in the $\phi=\frac{\pi}{8}$ direction, 
and a smaller vev in the $\frac{\pi}{4}$ direction. Notice that the data
is symmetric around the $\frac{\pi}{4}$ direction. 
The observed distortions are not due to  winding paths, since the data 
with the torelon suppression implemented shows the same pattern.

Method II:  Although there are still variations of the operators'
vev's along the semicircle, they are of the same order as the geometric
distortions. It must be said that the right angles of the rectangles also
get distorted, so that in  general, the rectangle becomes a parallelogram
at an arbitrary angle $\phi$.

The above comparison provides a first hint as to the rotational
properties of these operators, although it is not as direct as
it might be because the Wilson loops used for method I and method II 
are somewhat different. In 
the case of the segment operator used with method I, there is a 
simple and useful physical interpretation. As we saw earlier, the
superlink $K(x,y)$ is the propagator of a scalar particle
in the fundamental representation of the gauge group in
two Euclidean dimensions. Its tree-level mass  $m_0$ is given 
by eqn(\ref{eqn_alphatom}). Thus the segment operator in
${\cal O}(x,y)$ is the quenched propagator of
`mesons' composed of a scalar and its antiparticle. Such a
point to point propagator includes contributions from 
all allowed energies and from excited as well as ground state 
masses. Since our value of $\alpha$ corresponds to $2am_0 \simeq 0.8$,
the propagator will vary rapidly with distance. In addition, the
short distance part of the propagator will certainly vary strongly
with the angle $\phi$. Note that if we were to use in the
case of method I
the same rectangular loops as we used for method II, then the physical
interpretation would remain the same, except that the `meson'
wavefunctional would now be smeared, extending over roughly
one lattice spacing, which one would expect to favour the
contribution to the propagator of the lighter intermediate states,
leading to a weaker dependence on distance.

\paragraph{}In any case it is clear from the above that if we want
to construct trial wave-functionals with definite rotational
properties, then we should  renormalise the individual operators  
${\cal O}(\phi)$ in such a way that they have exactly the same vev. 
Doing so we shall now investigate how far we can restore 
rotational invariance by looking at the correlation between 
rescaled segment operators at different angles.

\paragraph{Correlation of rotated operators\\}
Using the values of the  segment operators calculated above,
we calculate the correlation function
\begin{equation}
\langle \bar{\cal O}(\phi-\frac{\Delta \phi}{2})\bar{\cal O}
(\phi+\frac{\Delta \phi}{2}) \rangle
\label{corphi}
\end{equation}
for a fixed $\Delta \phi$ (the bar indicates that the operators are now 
rescaled so that $\<\bar{\cal O}(\phi)\>=1$). This quantity is plotted 
in the direction $\phi$ in the  polar plots in Fig.~\ref{cor_standI}
and Fig.~\ref{cor_standII} for methods I and II respectively.
Ideally it should be independent of $\phi$.

\begin{figure}[tb]

\centerline{\begin{minipage}[c]{6.5cm}
    \psfig{file=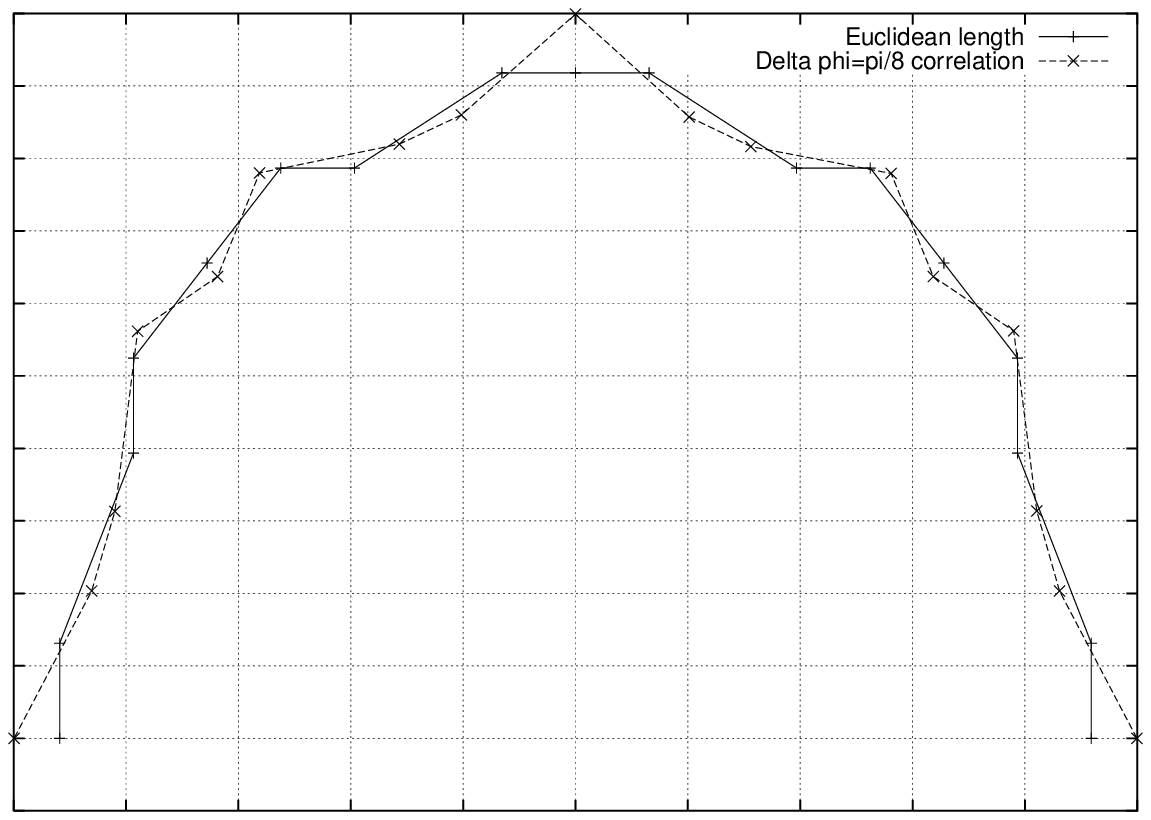,angle=0,width=6.5cm, height=3.25cm}
    \end{minipage}
    \begin{minipage}[c]{6.5cm}
    \psfig{file=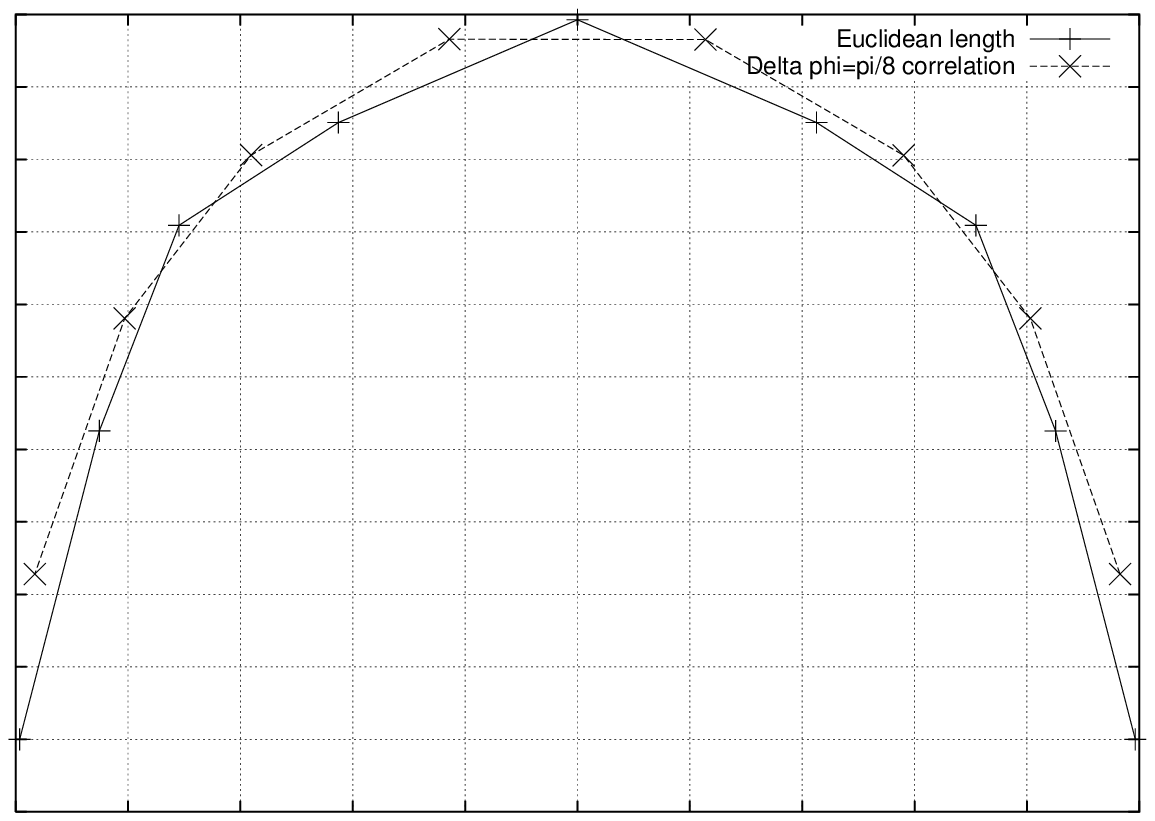,angle=0,width=6.5cm, height=3.25cm}
    \end{minipage}}
 \centerline{
 \begin{minipage}[c]{6.5cm}
    \psfig{file=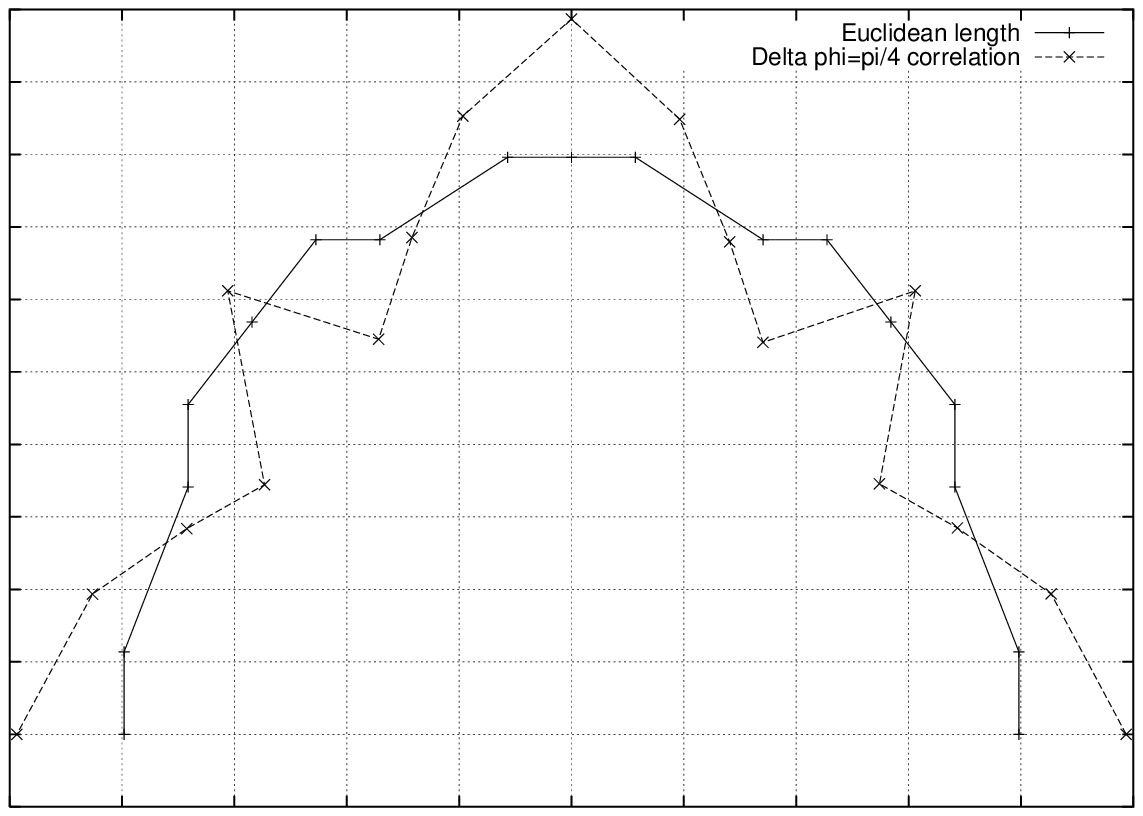,angle=0,width=6.5cm, height=3.25cm}
    \end{minipage}
    \begin{minipage}[c]{6.5cm}
    \psfig{file=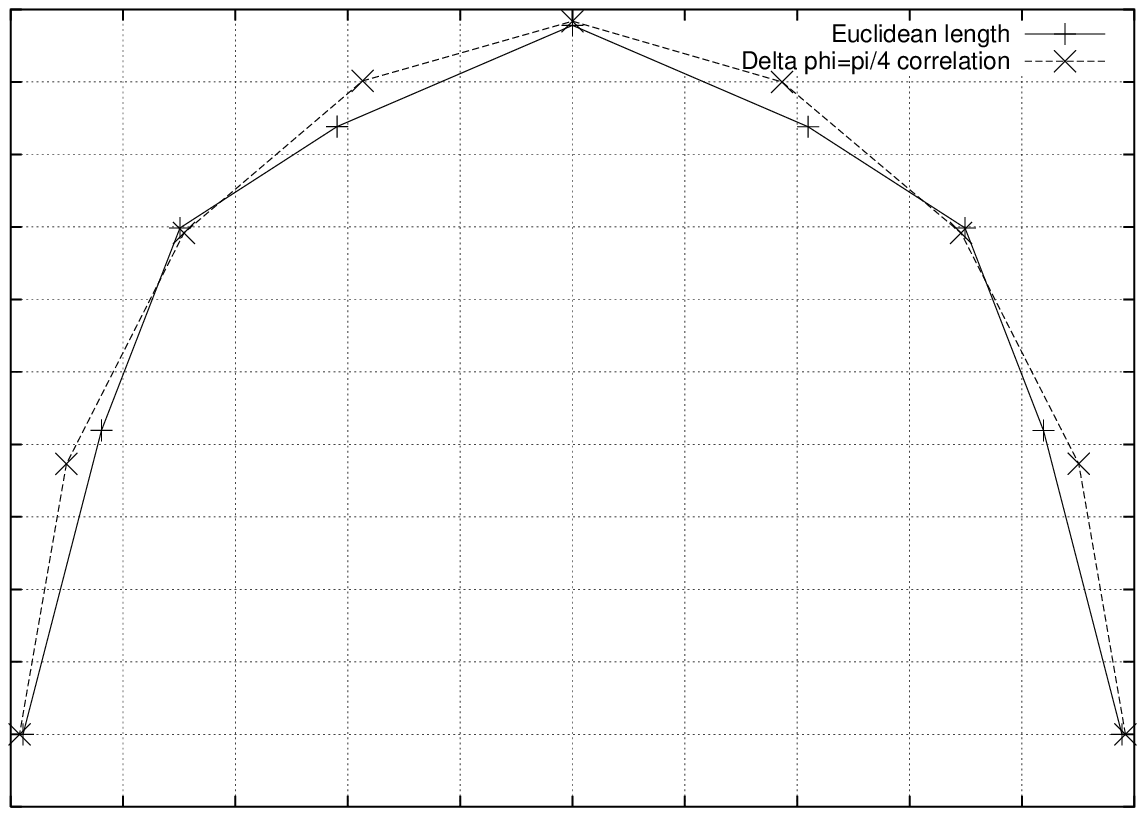,angle=0,width=6.5cm, height=3.25cm}
    \end{minipage}}
 \centerline{
 \begin{minipage}[c]{6.5cm}
    \psfig{file=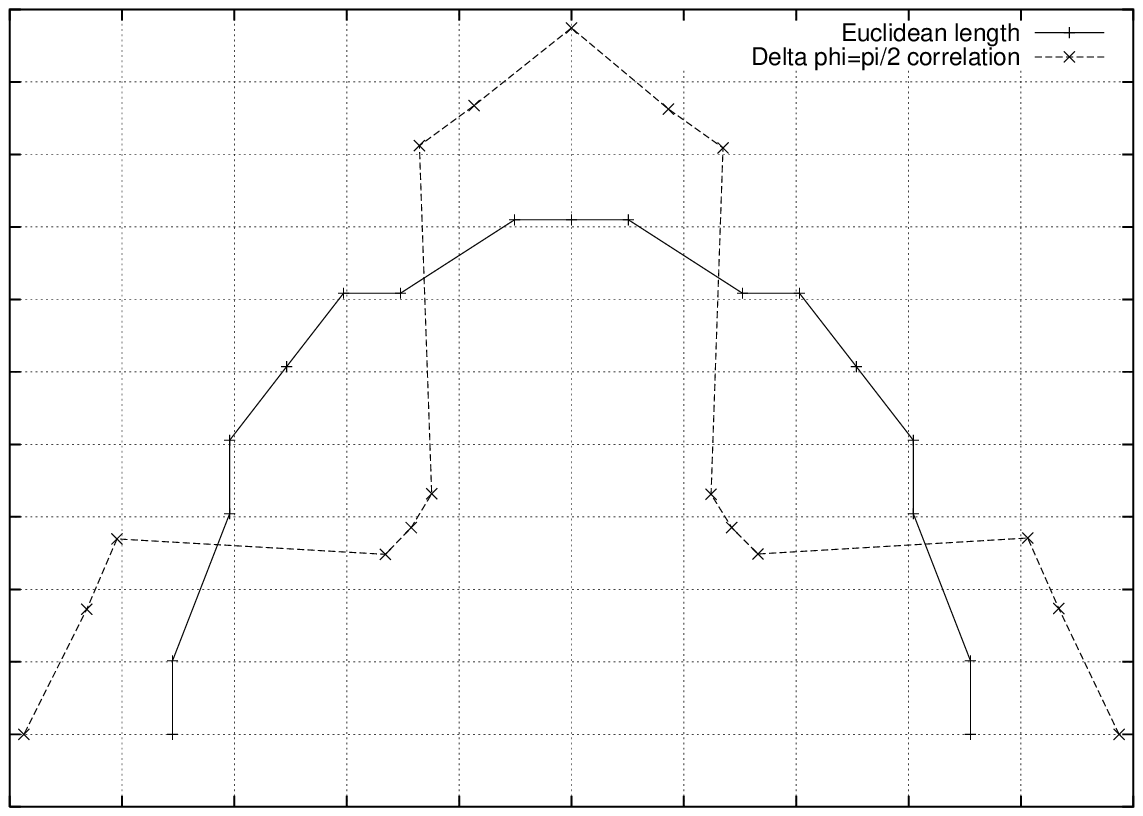,angle=0,width=6.5cm, height=3.25cm}
    \end{minipage}
    \begin{minipage}[c]{6.5cm}
    \psfig{file=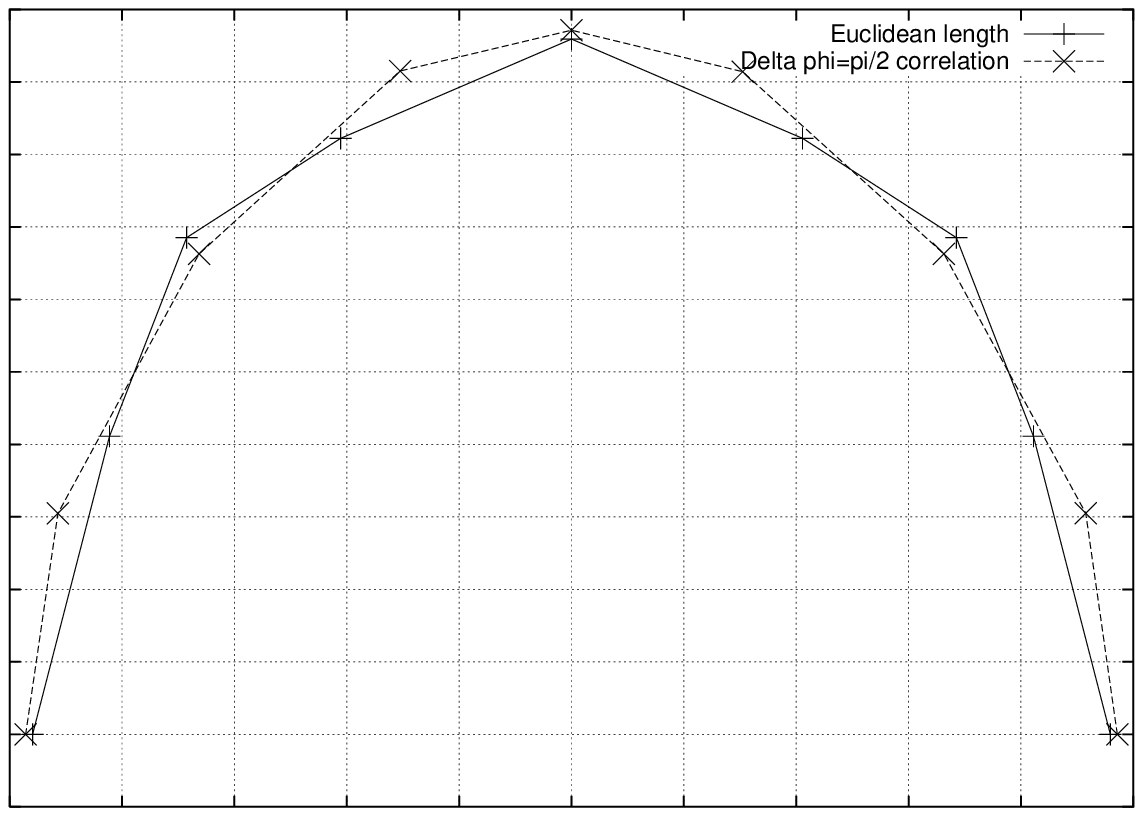,angle=0,width=6.5cm, height=3.25cm}
    \end{minipage}}
\vspace*{0.5cm}

\caption[a]{Method I: the Euclidean length and the correlation
 function (\ref{corphi}) in different directions $\phi$, 
for $\Delta\phi=\frac{\pi}{8},~\frac{\pi}{4},~\frac{\pi}{2}$ (from top to
bottom); left: without torelon-suppression, right:  with 
torelon-suppression.}
\la{cor_standI}
\end{figure}
\begin{figure}[bt]

  \centerline{
    \begin{minipage}[c]{6.5cm}
    \psfig{file=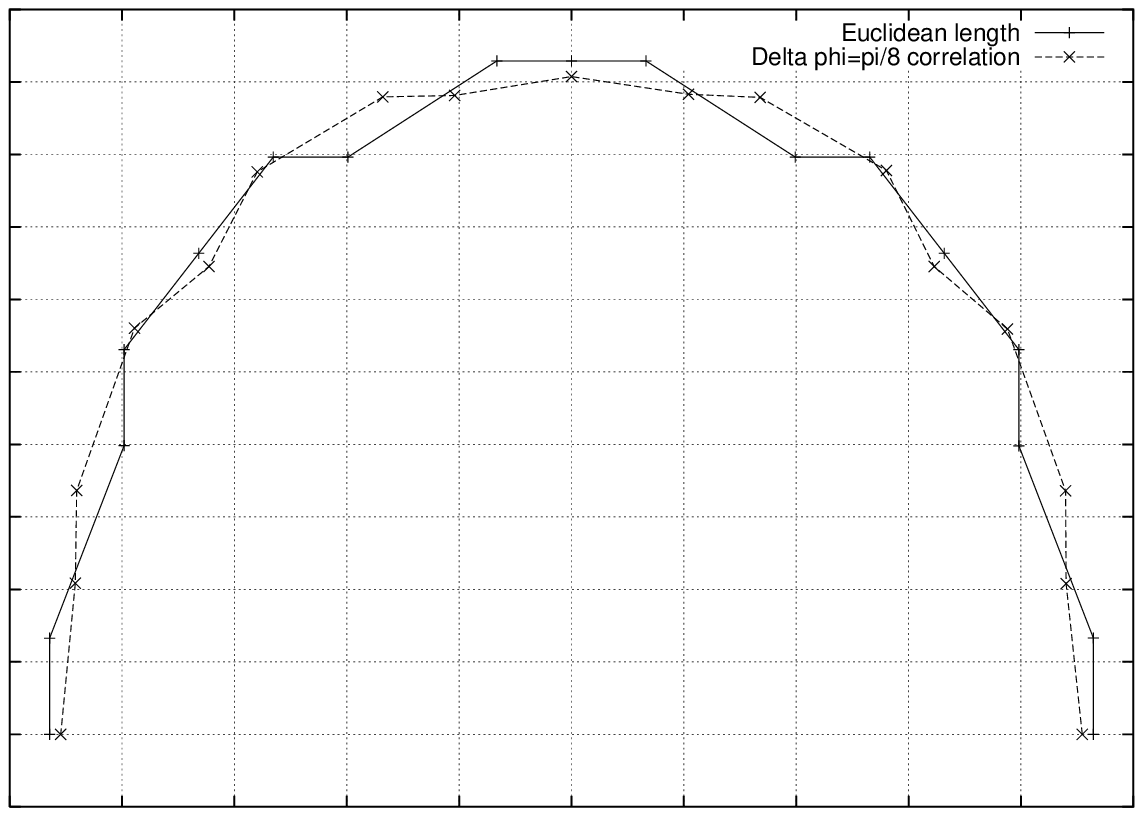,angle=0,width=6.5cm, height=3.25cm}
    \end{minipage}}
 \centerline{
    \begin{minipage}[c]{6.5cm}
    \psfig{file=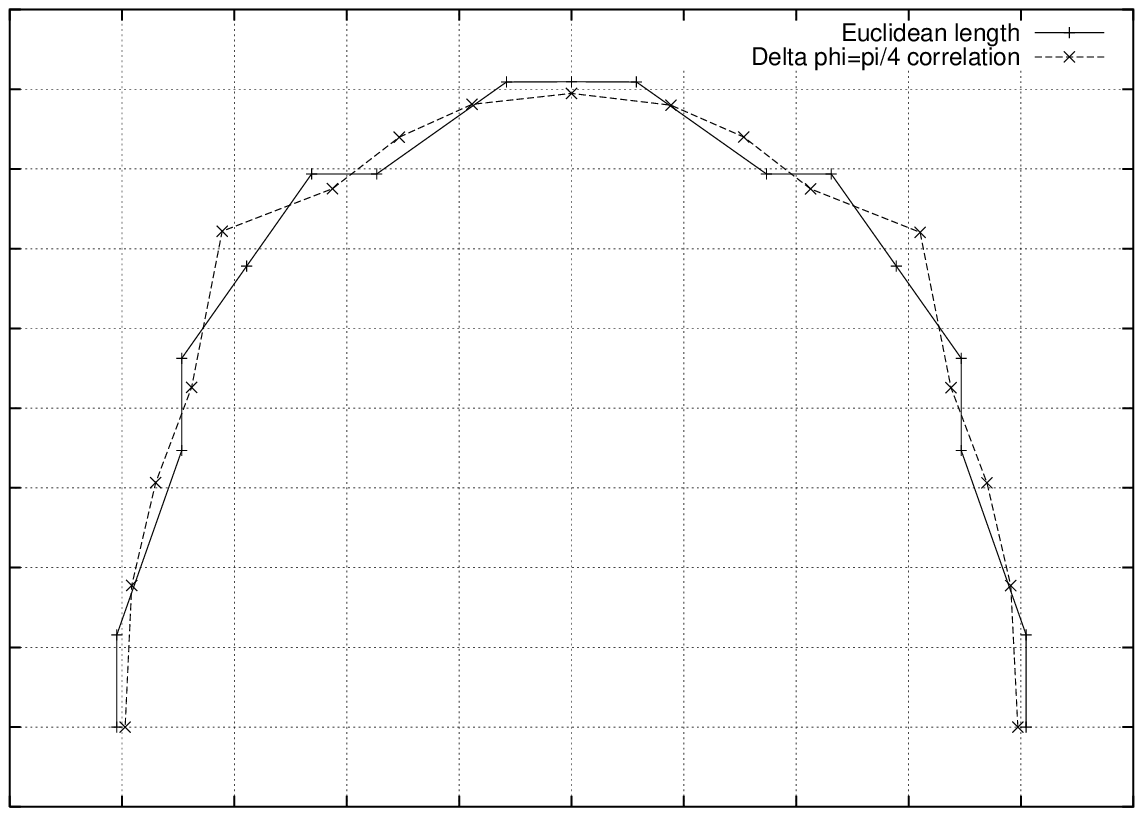,angle=0,width=6.5cm, height=3.25cm}
    \end{minipage}}
 \centerline{
    \begin{minipage}[c]{6.5cm}
    \psfig{file=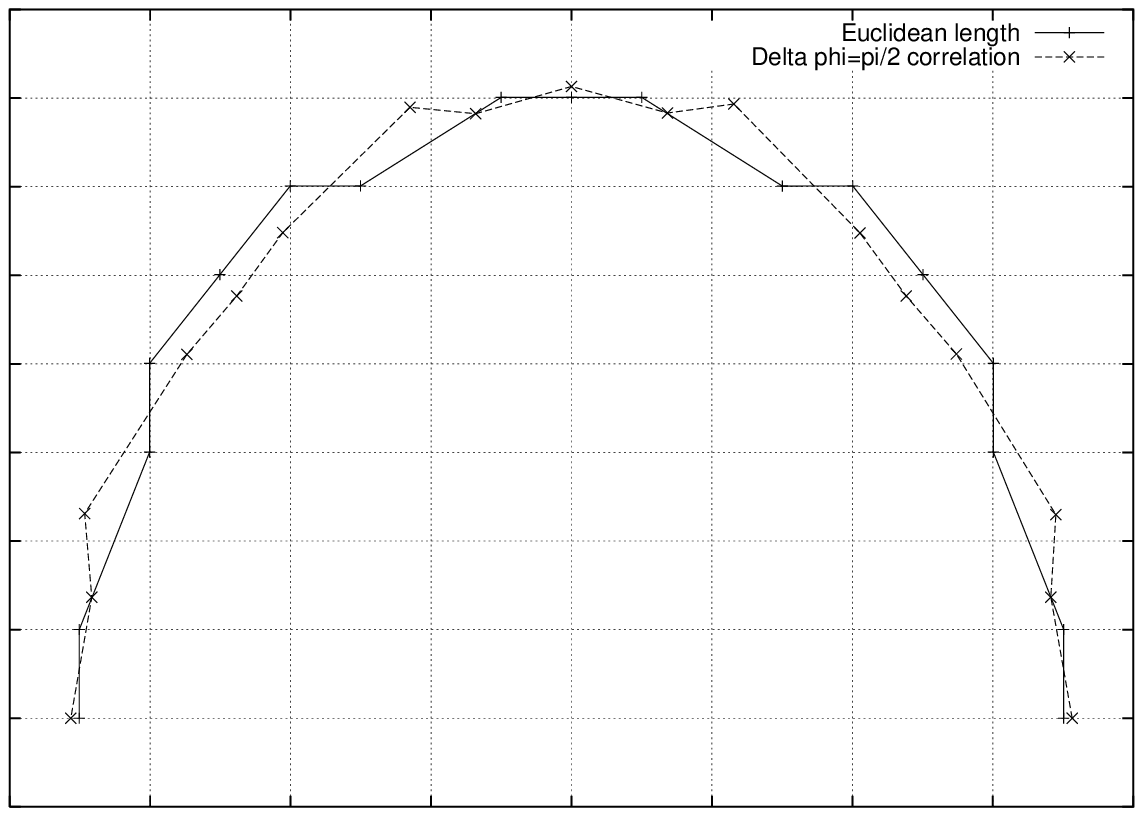,angle=0,width=6.5cm, height=3.25cm}
    \end{minipage}}
\vspace*{0.5cm}

\caption[a]{Method II: the Euclidean length and the correlation
 function (\ref{corphi}) in different directions $\phi$, 
for $\Delta\phi=\frac{\pi}{8},~\frac{\pi}{4},~\frac{\pi}{2}$ (from top to
bottom)}
\la{cor_standII}
\end{figure}

%
\paragraph{Remarks} 
Method I: without the torelon suppression, we see that for a small
 angle between the operators 
(first graph, $\Delta\phi=\frac{\pi}{8}$), the correlation function has
only small variations, of the order of the errors and 
geometric length and angle variations. However for a larger angle (second 
graph,  $\Delta\phi=\frac{\pi}{4}$), the variations are significantly larger.
   The worst case is the difference between the
 $\Delta\phi=\frac{\pi}{2}$ correlation of two segment operators along 
the lattice main directions or along the diagonals; here there is no
geometric error on $\Delta\phi$ and hardly any ($\sim 1\%$) on the length:
yet the two correlation functions differ by a factor 2.5.
This can be easily understood in terms of paths winding around the torus. 
Once the latter are suppressed (right column in Fig.~\ref{cor_standI}), 
the rotation invariance is restored to a good approximation.

Method II:
 We notice that the correlation function is independent of $\phi$, at the 
level of a few percent. In particular, the variations are practically
the same at the three values $\Delta\phi=\frac{\pi}{8},~\frac{\pi}{4},~
\frac{\pi}{2}$ and are of the same order as the geometric distortions.

\paragraph{Conclusion\\}
The fact that we observe correlations between Wilson loops that are 
approximately independent of their orientation with respect to the lattice
axes constitutes evidence for a dynamical restoration of rotation invariance.
From the average values and correlation functions of rectangle operators
constructed with a very simple algorithm, 
we conclude that both methods are suitable
to calculate operators rotated by angles smaller than $\frac{\pi}{2}$. 
In view of the practical argument that method II is very much faster
than method I, we shall use the former in practical calculations. Indeed,
constructing the d-links takes less time than a smearing iteration, whereas
the calculation of the propagator scales roughly as the sixth power of the 
lattice size.

\section{High spin states on the lattice?}

We now suppose that we dispose of a reliable way to construct operators  
rotated by angles of the type $\frac{2\pi}{n}$.
How can we use this tool to resolve the spectrum of the theory in the 
continuum limit?

\subsection{Lattice vs. continuum symmetry group}

The symmetry group of the square lattice contains 2 
 rotations by $\frac{\pi}{2}$, one rotation
by $\pi$ and two types of symmetry axes  
($x$ and $y$ axes, and $y=\pm x$ axes). For convenience, the character table
of  this group is given in appendix C. There are four one-dimensional 
representations, and one two-dimensional representation. 

The continuum rotation group only has 1-dimensional irreducible 
representations (irrep's), due to
the commutativity of rotations in the plane. However, because parity does not
commute with rotations (see appendix B), we ought to consider 2-dimensional 
representations, which are irreducible under the full symmetry group 
(rotations + parity). 

Each of the continuum two-dimensional representations can be
decomposed onto the irreducible representations of the square. For instance,
the spin $4^\pm$ representation $D_4$ decomposes onto two one-dimensional 
irreducible representations of the lattice group: 
\begin{equation}
D_4=A_1\oplus A_2
\end{equation} 
This tells us in what lattice irrep's to look in order to extract information
on the $D_4$ states.

The Hamiltonian eigenstates belonging to $A_1$ 
can be written as linear combinations
\begin{equation}
|\psi_1\>=\sum_{n\geq 0} c_n(\rho) |(j=4n)^+\>|_{lat} \quad {\rm with } \quad 
\sum_n |c_n|^2=1 
\label{se}
\end{equation} 
and correspondingly  for $A_2$
\begin{equation}
|\psi_2\>=\sum_{n\geq 0} c_n'(\rho)  |(j=4n)^-\>|_{lat}~.
\end{equation} 
Here $|j^\pm\>$ are eigenvectors of both angular momentum and parity, 
the reference axis of which we have chosen to coincide with a lattice axis, 
and $|j^\pm\>|_{lat}$ denotes their restriction to the lattice sites.
In other words, $\<\phi|j^+\>\propto\cos{j\phi}$ and 
$\<\phi|j^-\>\propto\sin{j\phi}$~(\footnote{
The $|0^-\>$ state is somewhat special in that it cannot be represented as 
the wavefunction of an object containing a symmetry axis. We shall exploit
this property later on.}).

We now set $\rho$ to a fixed physical length scale $\bar{\rho}\sim 
1/\sqrt{\sigma}$.
As we evolve from a small lattice spacing $a\ll \xi$
to coarser and coarser lattices, we imagine the following scenario in terms
of the coefficients $c_n\equiv c_n(\bar{\rho})$:

\bi
\item close to the continuum, for any particular Hamiltonian eigenstate
$\psi_1\in A_1$, 
the $c_n$ are close to $\delta_{mn}$ for some $m$. If for instance $m=1$, 
these states  ``remember'' that their wave function changes sign under 
approximate $\frac{\pi}{4}$ rotations that are available on the lattice 
at length scales much greater than $a$. Moreover, the state in $A_2$ with 
$c_n'\simeq \delta_{nm}$ is almost degenerate with  $\psi_1$.

\item as we move away from the continuum, the sharp 
dominance of one particular $c_n$ in the series becomes smoother, and we can 
think of the angular wave functions as having a ``fundamental mode'' $m$, plus
some fluctuations due to ``higher modes''. It is as if we started with a sound
of pure frequency, and the effect of the lattice is to add
 contributions from higher harmonics, giving the sound a richer timbre.
 The degeneracies between the states in $A_1$ and $A_2$ 
are broken more and more badly. This is due to the nonequivalence of the
two classes of parity transformations available on the lattice.

\item in general, more and more terms contribute to the series 
in eqn(\ref{se}). 
Thus it seems that the angular dependence of a general state 
in $A_1$ or $A_2$ becomes very intricate. 
 However as $a\rightarrow \xi$, higher modes in the expansion  must 
become irrelevant, because there are no lattice points to support
 their fluctuations on the length scale of the theory.
 We know from the strong coupling expansion 
that the lowest lying states have a simple behaviour as $\beta\rightarrow 0$:
the wave function of the fundamental state is simply a plaquette.
\ei

In fact, we have ignored a possible complication. We have assumed that
 no phase or roughening transition occurs, and that 
we can define smooth trajectories of the states in an E vs. $a$ plane. 
However, in general we must expect crossings of states to occur. 
For any given range of energies, $E\le E_0$,
there will exist a lattice spacing $a_0$ such that for $a<a_0$, 
there are no more crossings until the continuum is reached\footnote{There is 
a possible exception to that: we know 
that states come as parity doublets, which means that pairs of trajectories
must converge and could possibly cross many times in doing so. This is not
a problem for the present discussion.}. At $a_0$, 
the states  represent the continuum spectrum faithfully, with only
small numerical deviations on their energies. We now follow the trajectory of
one particular state as the lattice spacing is increased. 
Suppose we  meet another trajectory at $a=a_1$. At that particular 
lattice spacing, there will seem to be an ``accidental'' degeneracy. 
Nearly-degenerate states will mix with the mixing driven by the matrix
element of the lattice Hamiltonian between the `unperturbed' 
eigenstates, i.e. $\langle 1 | H(a) | 2\rangle$. Near the continuum
limit the unperturbed states will be close to continuum spin
eigenstates, $H(a)$ will be close to the continuum Hamiltonian and
so the mixing parameter $\langle 1 | H(a) | 2\rangle$ will be close to
zero. Nonetheless sufficiently close to the crossing, the states
will mix completely and so will the angular Fourier components of the
state. That is to say, the Fourier components need not have a simple
 behaviour with $a$ as $a\to 0$, and care must be taken to
identify any near-degeneracies in following the Fourier components
toward the continuum. 

\subsection{Two strategies}
\label{subsec_twostrat}

With these ideas on the evolution of the rotation properties of the 
physical states as functions of $a$, 
at least two (related) strategies are available in order to extract the
high spin continuum spectrum numerically:
\begin{enumerate}
\item[{\bf I}.] if we can afford to work close to the continuum, we can 
construct operators with an approximate continuum wave function (cwf)
$e^{i J \phi}$, using the operator construction technique presented earlier.
This kind of operator belongs to one of the irreducible representations of 
the lattice symmetry group. But because the perturbation of such an operator 
off  the cwf is different from that of the Hamiltonian eigenstates, 
the expected behaviour of  the local effective mass in the correlation 
function is the following: we should see an almost-flat plateau 
(corresponding 
to the excited state of the lattice irrep that will evolve into a high spin
state in the continuum limit), followed by a breakdown into another, flat and
stable plateau (corresponding to the fundamental state in the given lattice
irrep).
\item[{\bf II}.] First we construct a set of lattice irrep operators, 
$\{W^1_i\}_{i=1}^{N}$. Next we construct (approximate) rotated copies of these. 
We thus have a large basis of operators, $W^r_i$, $r$ labelling the rotation.
We  diagonalise the  correlation matrix (using the variational 
method~[\ref{mart_var}]) of  $\{W^1_i\}_{i=1}^{N}$  in order to 
extract the energy eigenstates  in this lattice irrep. These states $\psi^1_i$
are encoded by their components in the original basis $W^1_i$:
\begin{equation}
 \psi^1_i = \sum_j v_{ij} W^1_j 
\end{equation} 
 Now we need to determine the angular wave function of these glueball states. 
We do so by building the linear combinations
\begin{equation}
\psi^r_i  = \sum_j v_{ij} W^r_j,\quad \forall r 
\end{equation} 
and looking at the correlation function
\begin{equation}
G_t(r,r') = \< \psi^{r_i\dagger}(0)  \psi^{r'}_i(t) \>~. 
\label{eqn_cwfprobe}
\end{equation} 
In practise the time-separation $t$ is chosen so that the local effective
mass of the $ \psi^1_i$ 2-pt function has reached a plateau.
If we are reasonably close to the continuum, we should observe
an approximate cwf behaviour of this 
correlation function, i.e. $G_t(r,r')\propto e^{iJ (\phi_r-\phi_{r'})}$ for 
some $J$, with small fluctuations of other modes (cf. eqn(\ref{se})). 
(As remarked above, care has to be taken near any level crossings.)
Note that in principle this method can be generalised so that the operators
used in eqn(\ref{eqn_cwfprobe}) are based on different loops.
\end{enumerate}
We note that the data needed for both analyses is the same, so that 
they can easily be used in parallel. 
The second method has the advantage that there is no need to restrict
ourselves to $\frac{2\pi}{n}$-type angles in order to project out  states
corresponding to unwanted spins.  On the other hand, if high spin states
are very heavy, a large number $N$ of trial operators will be needed in order
that the variational method can resolve them. A simple case of 
this method, that does not employ the variational method and allows one
to determine the mass and quantum numbers of the lowest-lying 
state in a given lattice irrep, consists in measuring the correlation matrix 
of one operator $W$ with its rotated copies $\{W^r\}$ at sufficiently large
 Euclidean-time separation  so that the local 
effective mass has reached a plateau.

\subsection{A caveat}
\label{subsection_caveat}

Before we present our actual results, we first illustrate how one can
easily be misled by a naive application of `strategy I'. 
In the continuum, the simplest operator coupling to
$4^+$, but not to $0^+$ states, is the linear combination of two square
Wilson loops rotated by 45 degrees with a relative minus sign. At $\beta=6$ 
on a $16^3$ lattice for instance, we might choose a set of four 
square-shaped operators:
\begin{center}
\begin{tabular}{|c|c|}
\hline
&Squares:\\ 
\hline
1& $(4,0)\pm(3,3)$ \\
2& $(5,0)\pm(4,4)$\\
3& $(6,0)\pm(4,4)$\\
4& $(7,0)\pm(5,5)$\\
\hline
\end{tabular}
\end{center}
The notation here is that each pair of points represent the
coordinates of the end of a side of the square that begins
at the origin. This specifies the operator uniquely since we
sum all translations of the square in forming zero momentum
operators. The $\pm$ sign determines whether the linear combination
of the two squares is a trial $J=0$ or $J=4$ operator.
We calculate these operators using the $M$ matrix method
with $\alpha=0.24$. If we diagonalise the four $4^+$ operators 
according to the variational procedure [\ref{mart_var}], 
we obtain the following components of the two lightest operators (primed)
in the normalised initial basis (unprimed):
\begin{center}
\begin{tabular}{|c||c|c|c|c||c|}
\hline
components & 1 & 2 & 3 & 4 & $am$ (1 lat. sp.) \\
\hline
$1^\prime$ & -0.97& 7.2&3.7& -0.89& 1.754(58)\\
\hline
$2^\prime$ & -1.6& 18.2& -13.8& 0.32 & 1.882(46) \\
\hline
\end{tabular}
\end{center}
Must we conclude that $am=1.754$ is an upper bound on 
the lightest $4^+$ glueball mass at $\beta=6$? In fact, these masses are
completely compatible with the masses of the $0^{+*}$ and $0^{+**}$ obtained
in [\ref{mtd3}]. Indeed, the equal-time overlaps of the operators $1^\prime$
and $2^\prime$ onto our
best `$0^+$', `$0^{+*}$' and  `$0^{+**}$' operators, obtained from the
variational method, are found to be (-0.13,0.98,-0.04) and (0.096,0.16,-0.12)
where the operators have been
normalised in such a way that their equal-time 2-pt function is 1.
This clearly shows that the correlator of the candidate ground state
`$4^+$' operator yields nothing but the $0^{+*}$ mass.
A possible explanation of how it can be so is the following.
The very smeared squares  constructed here with superlinks ``look'' 
almost like rings of different sizes. Thus the relative minus
signs in the linear combinations of the orthogonalised operators 
would suggest a radial wave function for the glueball that contains 
nodes and what we have actually constructed are operators suited
to measure excited  $0^+$ states.

\paragraph{}Thus the naive application of the continuum notion of 'spin'
can lead to a wrong labelling of the states extracted from the lattice.
The general point is that since the
rotated loops are only approximate copies of each other, it is 
quite possible that the cancellations induced by the oscillating
coefficients induce not only a piece of the wavefunction that
has the desired angular oscillations, but also a piece
where the cancellations, and resulting oscillations, are in
the radial rather than angular direction. The latter can
project onto an excited $0^+$. (Since one expects the ground
state radial wavefunction to be smooth, a significant
overlap onto the ground state  $0^+$ would be unexpected.)
Now, since the lightest $0^+$ is
very much lighter than the lightest $4^+$, even such an
excited $0^+$ may be lighter than the lightest $4^+$
-- as turns out to be the case here -- and may undermine a 
variational calculation. As $a\to 0$ and the rotated
loops can be chosen to be better copies of each other,
the radial cancellations become more extreme, the $0^+$
states being projected upon become more highly excited and
more massive, and once they become more massive than the
lightest $4^+$ states of interest the problem disappears for
all practical purposes. The general lesson  is that one can
never be completely confident that one has isolated
a high spin state at a fixed value of $a$. It is only
through varying $a$ toward the continuum limit that
such confidence can be achieved.

\section{Applications of Strategy I}

As remarked in the Introduction, a robust prediction [\ref{model1},\ref{model2}]
of the Isgur-Paton flux tube model [\ref{isgur}] is that the lightest
$0^-$ state should be much heavier than the mass that has been
obtained in lattice calculations 
[\ref{mtd3}] of the lightest SU(2) glueball in the $A_2$ lattice 
representation. Moreover the latter mass is close to the mass 
of the $4^-$ glueball as predicted by the flux tube model
[\ref{model1},\ref{model2}]. Since
the $A_2$ lattice representation contains the continuum
$0^-,~4^-,\dots$ states this has led to the conjecture
[\ref{model1},\ref{model2}] that the lightest $A_2$ state is in
fact $4^-$ rather than  $0^-$.
All this has motivated some lattice calculations 
[\ref{rjmtrot},\ref{rjrot}] which suggest
that this is in fact so. 

In this Section we shall use the first of the two strategies outlined 
in Section \ref{subsec_twostrat} 
to address this question in some detail.
We shall begin with a simple approach applied directly
to states in the $A_2$ representation (the conclusions of which
will be confirmed in a quantitatively controllable way
when we apply  'strategy II'). The reader will note
that the approach we use here is similar to the one we  presented earlier
as a caveat of how a naive approach can fail. However in 
that earlier example it was applied to the $A_1$ representation where
the $4^+$ state that we were trying to isolate  is much heavier
than the lightest, and first excited $0^+$ states. Here by contrast
the candidate $4^-$ state turns out to be much lighter than the $0^-$,
so that it is the $A_2$ ground state that is being identified as $4^-$.
Now, as pointed out in  Section \ref{subsection_caveat}, we do not
expect a significant projection of a trial $4^-$ operator 
onto the $0^-$ ground state. Since we expect the $A_2$ ground state
to be either the $0^-$ or the  $4^-$ ground state, it must in fact be 
the latter. Thus in this rather special case, which will
not be characteristic of other high spin calculations, our caveat 
loses its force.

We then return to the more difficult question of how one isolates
the $4^+$ from the $0^+$ in the $A_1$ representation. We provide 
a procedure that appears to work well. Because of parity doubling
for $J\not= 0$ this provides another way to calculate
the $4^-$ mass. (And indeed the $4^+$ and $4^-$ masses we
obtain are entirely compatible.) After checking that
finite volume corrections for the higher spin states are 
under control, we perform calculations for several
larger $\beta$ values and  extrapolate our mass ratios to
the continuum limit, as one needs to do if one is to be
confident that one has indeed separated the different spin states.

\paragraph{Simulation details.-} 
All our calculations use the standard Wilson action.
The update is a 1:4 mixture of heat-bath ([\ref{hb}], [\ref{kenpen}]) and
over-relaxation [\ref{ovr}] sweeps. Because we measure a large number 
of operators, we can do 6-8 of these compound sweeps between
measurements without significantly increasing the total cost of
the calculation. We use an increasing number of smearing [\ref{smear}] 
iterations (and also increase the 
smearing weights) as we approach the continuum.
For the lattice data that we eventually extrapolate to the continuum (Tables
\ref{sIdata} and \ref{sIIdata}),
we perform  ${\cal O}(10^5)$ sweeps and collect the data 
typically in 30 bins. Errors are estimated with a standard jackknife analysis.

\subsection{The $0^-~/~4^-$ puzzle\label{04m}}

To distinguish the  $0^-$ from the $4^-$ glueball we construct 
trial $0^-$ and $4^-$ wavefunctionals using suitable 
linear combinations of (approximate) rotated copies of asymmetric 
operators (as described earlier). We first do this for a set
of three operators and their twelve rotations. We simultaneously
calculate operators in the $A_2$ lattice representation that are 
of the same kind as were used in [\ref{mtd3}], where it was 
assumed that the lightest state was the $0^-$. We find that the
overlap of these $A_2$ operators is large onto our trial
$4^-$ operators and small onto the $0^-$ ones. In addition
the mass of the former is much lighter than the mass of the
latter. This strongly suggests that the lightest $A_2$ state
becomes the $4^-$ glueball in the continuum limit.

Subsequently, we use two further independent sets of operators, 
each rotated eight times, so as to obtain an extra check on our 
calculation. The masses obtained at one lattice spacing are found 
to agree within errors. 

The calculations in this subsection are performed at $\beta=6$ on a 
$16^3$ lattice. The links are smeared before the paths joining lattice 
sites are constructed.

\paragraph{Twelve-fold rotated operators.-}
The three operators on the left of Fig. (\ref{sh1}) are rotated 
through 12 angles while maintaining, approximately, their shapes.
Those on the right -- which are only rotated by 90 degrees -- 
are of a type that has been used previously [\ref{mtd3}]
to measure the lightest state in the $A_2$ representation.
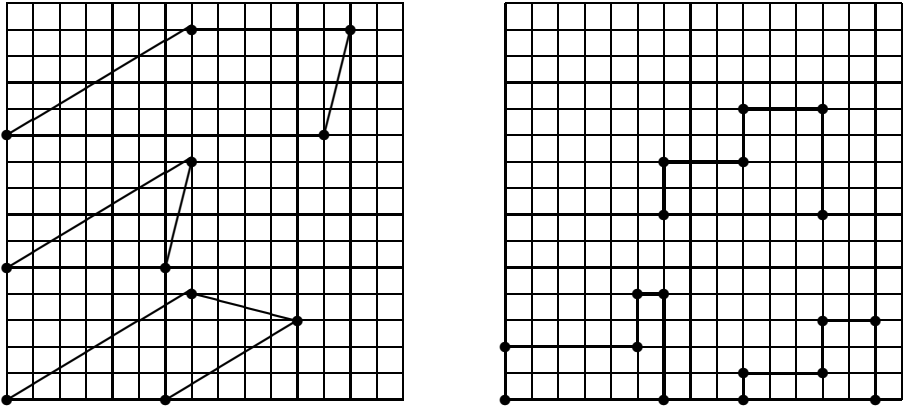
\begin{figure}[ht]
\centerline{
\begin{minipage}[c]{6.5cm}
\begin{picture}(100,160)(10,20)
\multiput(0,0)(0,10){16}{\line(1,0){150}}
\multiput(0,0)(10,0){16}{\line(0,1){150}}
{\thicklines
\put(0,0){\line(1,0){60}}
\put(60,0){\line(5,3){50}}
\put(110,30){\line(-4,1){40}}
\put(0,0){\line(5,3){70}}}
\put(70,40){\circle*{4}}
\put(0,0){\circle*{4}}
\put(60,0){\circle*{4}}
\put(110,30){\circle*{4}}
{\thicklines
\put(0,50){\line(1,0){60}}
\put(60,50){\line(1,4){10}}
\put(0,50){\line(5,3){70}}}
\put(0,50){\circle*{4}}
\put(60,50){\circle*{4}}
\put(70,90){\circle*{4}}
{\thicklines
\put(0,100){\line(1,0){120}}
\put(120,100){\line(1,4){10}}
\put(130,140){\line(-1,0){60}}
\put(0,100){\line(5,3){70}}}
\put(0,100){\circle*{4}}
\put(120,100){\circle*{4}}
\put(130,140){\circle*{4}}
\put(70,140){\circle*{4}}
\end{picture}
\end{minipage}
\begin{minipage}[c]{6.5cm}
\begin{picture}(100,160)(10,20)
\multiput(0,0)(0,10){16}{\line(1,0){150}}
\multiput(0,0)(10,0){16}{\line(0,1){150}}
\thicklines{\put(0,0){\line(1,0){60}}
\put(60,0){\line(0,1){40}}
\put(60,40){\line(-1,0){10}}
\put(50,40){\line(0,-1){20}}
\put(50,20){\line(-1,0){50}}
\put(0,20){\line(0,-1){20}}}
\put(0,0){\circle*{4}}
\put(60,0){\circle*{4}}
\put(60,40){\circle*{4}}
\put(50,40){\circle*{4}}
\put(50,20){\circle*{4}}
\put(0,20){\circle*{4}}
\thicklines{
\put(60,70){\line(1,0){60}}
\put(120,70){\line(0,1){40}}
\put(120,110){\line(-1,0){30}}
\put(90,110){\line(0,-1){20}}
\put(90,90){\line(-1,0){30}}
\put(60,90){\line(0,-1){20}}}
\put(60,70){\circle*{4}}
\put(120,70){\circle*{4}}
\put(120,110){\circle*{4}}
\put(90,110){\circle*{4}}
\put(90,90){\circle*{4}}
\put(60,90){\circle*{4}}
\thicklines{
\put(90,0){\line(1,0){50}}
\put(140,0){\line(0,1){30}}
\put(140,30){\line(-1,0){20}}
\put(120,30){\line(0,-1){20}}
\put(120,10){\line(-1,0){30}}
\put(90,10){\line(0,-1){10}}}
\put(90,0){\circle*{4}}
\put(140,0){\circle*{4}}
\put(140,30){\circle*{4}}
\put(120,30){\circle*{4}}
\put(120,10){\circle*{4}}
\put(90,10){\circle*{4}}
\end{picture}
\end{minipage}
}
\vspace*{1cm}
\caption{On the left, operators used to construct a $4^-$ wave function;
(I) trapeze  (II) triangle (III) pentagon.
On the right, a ``conventional'' set of operators (1) bottom left (2) top (3)
bottom right}
\la{sh1}
\end{figure}
With the former three operators and their rotations, we form 
linear combinations that
correspond to trial $4^-$ and $0^-$ operators, while with the latter
three operators we construct the $A_2$ representation of the square group.
We now look at the overlaps between these two sets of operators.
The overlap here is defined as 
\be
{\rm overlap}\equiv\frac{\langle {\cal O}_1 {\cal O}_2\rangle
  }{\left(\langle{\cal O}_2
 {\cal O}_2\rangle \langle {\cal O}_1{\cal O}_1 \rangle \right)^
{\frac{1}{2}}}   
\la{overlap}
\ee
and the values we obtain are:
\begin{center}
\begin{tabular}{|c|c|c|c|}
\hline
overlap & 1($A_2$)  & 2($A_2$) & 3($A_2$) \\
\hline
I($4^-$) & 0.89 &  0.88 &  0.73\\
II($4^-$) & 0.95 & 0.96 &  0.97 \\
III($4^-$) &  0.38 &  0.38 & 0.24\\
\hline
I($0^-$) &  0.24 &  0.20 &  0.15 \\
II($0^-$) & - & - & -\\
III($0^-$) &  0.16 &  0.13 &  $<0.01$\\
\hline
\end{tabular}
\end{center}
The horizontal labelling refers to the three $A_2$ operators based 
on the three loops on the right of Fig.\ref{sh1}, while the
vertical labels are the three `$4^-$' and `$0^-$' operators
obtained using the left-hand loops. (Errors are smaller than $1\%$ 
here; shape II produced a very noisy $0^-$ operator.)
Clearly, the approximate $4^-$ operators overlap much more onto
the operators of the $A_2$ representation.
This, and the fact that we observe a much smaller
mass for our trial $4^-$ operator than for the $0^-$, is a strong 
indication that the labelling in [\ref{mtd3}] of the lightest  
$A_2$ glueball as $0^-$ was mistaken, and that it is in fact a $4^-$.
Performing a variational analysis of the correlation matrix of the 
first set of operators, we obtain effective masses at one lattice
spacing of $am(4^-)=2.556(68)$ and $am(0^-)=  3.34(31)$. The much 
larger  error on the latter is a reflexion of a much poorer signal.
At this stage it would be useful, as a check, to use the usual 
variational calculation to identify the lightest few $4^-$ and $0^-$ 
states, so as to verify explicitly their small mutual overlaps.
However the very heavy $0^-$ mass discourages us from attempting a
variational identification of excited $0^-$ states. 

\paragraph{Eightfold rotated operators.-}
As a check we now repeat the above analysis using the quite different 
set of operators drawn in Fig. (\ref{sh2}). 
There are two parallelograms, and three completely asymmetric shapes. 
We rotate them by 45 degree angles, so each is
rotated 8 times, along with the parity images. Proceeding as before
to construct approximate $4^-$ and $0^-$ wave functions, we extract
masses at one lattice spacing of $am(4^-)=2.519(52)$ and 
$am(0^-)=3.016(42)$. We see that these estimates agree within
errors with our earlier estimates.
Thus we have obtained a consistency check, namely that 
the mass estimates and labelling are independent of the particular set of
operators we are using.

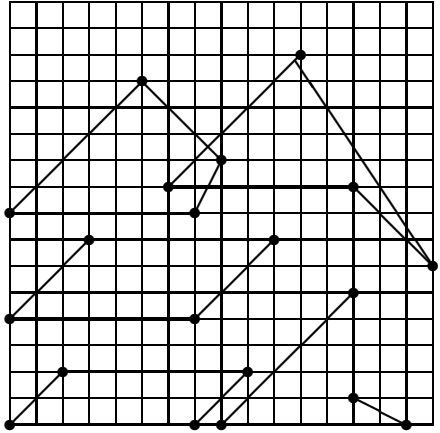
\begin{figure}[ht]
\centerline{
\begin{minipage}[c]{6.5cm}
\begin{picture}(100,160)(10,20)
\multiput(0,0)(0,10){17}{\line(1,0){160}}
\multiput(0,0)(10,0){17}{\line(0,1){160}}
{\thicklines
\put(0,0){\line(1,0){70}}
\put(70,0){\line(1,1){20}}
\put(90,20){\line(-1,0){70}}
\put(0,0){\line(1,1){20}}}
\put(0,0){\circle*{4}}
\put(70,0){\circle*{4}}
\put(90,20){\circle*{4}}
\put(20,20){\circle*{4}}
{\thicklines
\put(0,40){\line(1,0){70}}
\put(70,40){\line(1,1){30}}
\put(100,70){\line(-1,0){70}}
\put(0,40){\line(1,1){30}}}
\put(0,40){\circle*{4}}
\put(70,40){\circle*{4}}
\put(100,70){\circle*{4}}
\put(30,70){\circle*{4}}
{\thicklines
\put(80,0){\line(1,0){70}}
\put(150,0){\line(-2,1){20}}
\put(130,10){\line(0,1){40}}
\put(130,50){\line(-1,-1){50}}}
\put(80,0){\circle*{4}}
\put(150,0){\circle*{4}}
\put(130,10){\circle*{4}}
\put(130,50){\circle*{4}}
{\thicklines
\put(0,80){\line(1,0){70}}
\put(70,80){\line(1,2){10}}
\put(80,100){\line(-1,1){30}}
\put(50,130){\line(-1,-1){50}}}
\put(0,80){\circle*{4}}
\put(70,80){\circle*{4}}
\put(80,100){\circle*{4}}
\put(50,130){\circle*{4}}
{\thicklines
\put(60,90){\line(1,0){70}}
\put(130,90){\line(1,-1){30}}
\put(160,60){\line(-2,3){52}}
\put(110,140){\line(-1,-1){50}}}
\put(60,90){\circle*{4}}
\put(130,90){\circle*{4}}
\put(160,60){\circle*{4}}
\put(110,140){\circle*{4}}
\end{picture}
\end{minipage}}
\vspace*{1cm}
\caption{The operators used to construct  $4^-$ and $0^-$ wave functions}
\la{sh2}
\end{figure}

\paragraph{Square operators.-}
Yet another evaluation of the lightest $A_2$ state
 is provided by the  operators shown in Fig. (\ref{sh3}). There are two 
squares and one rectangle. This time the mass at one lattice spacing is
$am(4^-)=2.610(65)$; again consistent with our earlier values. 

\begin{figure}[ht]
\centerline{
\begin{minipage}[c]{6.5cm}
\begin{picture}(100,160)(10,20)
\multiput(0,0)(0,10){16}{\line(1,0){150}}
\multiput(0,0)(10,0){16}{\line(0,1){150}}
{\thicklines
\put(20,0){\line(5,2){100}}
\put(120,40){\line(-2,5){20}}
\put(100,90){\line(-5,-2){100}}
\put(70,20){\line(-2,5){20}}
\put(0,50){\line(2,-5){20}}}
{\thicklines
\put(110,100){\line(3,1){30}}
\put(140, 110){\line(-1,3){10}}
\put(130, 140){\line(-3,-1){30}}
\put(100, 130){\line(1,-3){10}}}
\put(110,100){\circle*{4}}
\put(140,110){\circle*{4}}
\put(130,140){\circle*{4}}
\put(100,130){\circle*{4}}
\put(20,0){\circle*{4}}
\put(120,40){\circle*{4}}
\put(100,90){\circle*{4}}
\put(70,20){\circle*{4}}
\put(0,50){\circle*{4}}
\put(50,70){\circle*{4}}
\end{picture}
\end{minipage}}
\vspace*{1cm}
\caption{The square operators used for the $A_2$ representation}
\la{sh3}
\end{figure}
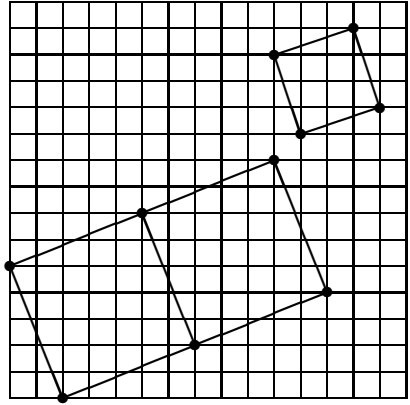

\subsection{The even-spin spectrum at $\beta=6$}

We now return to the problem of distinguishing $4^+$ and $0^+$  
states in the $A_1$ representation. Since the  $0^+$ is much
lighter than the $4^+$ there is the danger that what we will 
claim to be a  $4^+$ will in fact be an excited  $0^+$; as we
warned in Section \ref{subsection_caveat}. Thus one needs to 
perform enough checks to avoid such a fate. This leads to
a rather involved procedure which we will describe in
the context of a calculation at $\beta=6$. After checking
that finite volume corrections on our higher spin masses
are insignificant, we perform higher $\beta$ calculations
so as to be able to perform a continuum extrapolation.

To construct our trial states of spin 0 and 4 we use the four
operators in Fig. (\ref{sh4}) together with their twelve rotations.
(Their coordinates can be found in Table \ref{coords}).

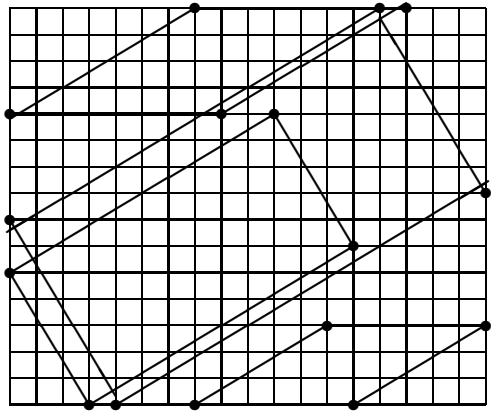
\begin{figure}[ht]
\centerline{
\begin{minipage}[c]{6.5cm}
\begin{picture}(100,160)(10,20)
\multiput(0,0)(0,10){16}{\line(1,0){180}}
\multiput(0,0)(10,0){19}{\line(0,1){150}}
{\thicklines
\put(30,0){\line(5,3){100}}
\put(130,60){\line(-3,5){30}}
\put(100,110){\line(-5,-3){100}}
\put(0,50){\line(3,-5){30}}}
\put(30,0){\circle*{4}}
\put(130,60){\circle*{4}}
\put(100,110){\circle*{4}}
\put(0,50){\circle*{4}}
{\thicklines
\put(70,0){\line(1,0){60}}
\put(130,0){\line(5,3){50}}
\put(180,30){\line(-1,0){60}}
\put(120,30){\line(-5,-3){50}}}
\put(70,0){\circle*{4}}
\put(130,0){\circle*{4}}
\put(180,30){\circle*{4}}
\put(120,30){\circle*{4}}
{\thicklines
\put(40,0){\line(5,3){141}}
\put(180,80){\line(-3,5){40}}
\put(140,150){\line(-5,-3){141}}
\put(0,70){\line(3,-5){40}}}
\put(40,0){\circle*{4}}
\put(140,150){\circle*{4}}
\put(180,80){\circle*{4}}
\put(0,70){\circle*{4}}
{\thicklines
\put(0,110){\line(1,0){80}}
\put(80,110){\line(5,3){70}}
\put(150,150){\line(-1,0){80}}
\put(70,150){\line(-5,-3){70}}}
\put(0,110){\circle*{4}}
\put(80,110){\circle*{4}}
\put(150,150){\circle*{4}}
\put(70,150){\circle*{4}}
\end{picture}
\end{minipage}}
\vspace*{1cm}
\caption{The operators used for the determination of the even spin spectrum; 
(1) small rectangle (2) large rectangle  (3) small parallelogram (4) large 
 parallelogram}
\la{sh4}
\end{figure}

\subsubsection{A recipe for data analysis\label{recipe}}

We construct approximate rotations of the four shapes given above and 
we suppose that we have obtained the correlation of each operator
in any orientation with any operator in any orientation. How 
can we separate with confidence the $4^+$ from the $0^+$ signal?
Our experience has shown the following procedure to be reliable
at $\beta=6$. While the appropriate choice of various selection
parameters is empirical, in the continuum limit the procedure
should be unambiguously valid if these parameters are made to
approach obvious limiting values, and if the operators used
are scaled appropriately and are chosen to be increasingly
better copies of each other. 
\begin{description}

\item[Preselection of the operators.-] Look at the correlation matrix of each
individual 'shape' with its rotated copies. 
This $n\times n$ matrix should (ideally) be a symmetric Toeplitz one, 
with the additional cyclic property ($M_{i,j}=M_{n+2-i,j}$, $i=j+1,\dots,n$,
reflecting the fact that the correlation between angle $0$ and 
$\frac{2\pi}{n}$ is the same as between $-\frac{2\pi}{n}$ and $0$. 
 For instance, in the present case the correlation matrices of
 the operators read (op. 1 is top left, 2 top right, 3 bottom left and 4
bottom right):
\begin{center}
\begin{tabular}{|c@{$\quad$}c@{$\quad$}c@{$\quad$}c@{$\quad$}c@{$\quad$}c|c@{$\quad$}c@{$\quad$}c@{$\quad$}c@{$\quad$}c@{$\quad$}c|}
   \hline
	 1.00 &&&&& &1.00 &&&&&\\
 
    0.52 &   1.00&&&& &  0.22&    1.00&&&&\\
 
    0.24  &  0.52  &  1.00&&& & 0.26 &   0.22  &  1.00&&&\\

    0.18 &   0.22 &   0.62  &  1.00&& &0.03 &   0.01  &  0.16 &   1.00&&\\
    0.22  &  0.13 &   0.22 &   0.52  &  1.00& &0.01 &   0.00  &  0.01  &  0.21 &   1.00&\\
 
    0.62 &   0.22  &  0.18 &   0.24 &   0.52  &  1.00 & 0.16 &   0.01 &   0.03 &   0.26  &  0.22 &   1.00\\
\hline
 1.00&&&&& &  1.00&&&&&\\
 
    0.84  &  1.00 &&&& &  0.65 &   1.00&&&&\\
 
    0.73  &  0.86 &   1.00&&& & 0.47  &  0.69 &   1.00&&&\\
 
    0.60  &  0.65 &   0.87  &  1.00&& & 0.37 &   0.42  &  0.69 &   1.00&&\\
 
    0.65  &  0.60 &   0.73  &  0.84  &  1.00& &0.42  &  0.37  &  0.48  &  0.65  &  1.00&\\
 
    0.87  &  0.73  &  0.76  &  0.73  &  0.86  &  1.00 & 0.69  &  0.48  &  0.49  &  0.48 &   0.69  &  1.00\\
\hline
 \end{tabular}
\end{center}
Notice that the second is worse than the others; we therefore discard it and 
only operators 1, 3 and 4 remain in the subsequent steps.
\item[Selection of the operators.-] We now concentrate on the linear 
combinations corresponding to  spin $J_1$
and spin $J_2$ operators (with $J_1=0$ and $J_2=4$ being the case
of interest here). Each of the spin $J_1$ operators has a certain overlap
with each of the spin $J_2$ operators. This is due partly 
to the imperfect rotations, and partly because the lattice
Hamiltonian eigenstates do not diagonalise the spin operator.
First look at the diagonal quantities, that is the overlap  
of the $J_1$ and $J_2$ operators constructed with the same shape. 
Eliminate those which have more than $\sim 20\%$ overlap. 
(As $a\to 0$ the cut-off should gradually be reduced to zero as well.) 
Once bad operators are eliminated, this whole set
of overlaps should contain none larger than $20\%$. If this cannot be 
achieved, it either means that we are too far from the continuum --
in the sense that the wave functions of the
physical states are very different from the continuum ones--, or that our
rotated operators are not sufficiently faithful copies of the initial
ones.  In the present case, we find the following overlaps for our three
candidate $J_2$ operators onto the corresponding $J_1$ operators: 
(0.04,0.0,0.12), (0.13,0.20,0.07) and (0.09,0.05,0.10) respectively
for 1, 3 and 4, and so we retain these operators for the  subsequent 
calculation.
\item[Diagonalisation of the $J_1$  operators.-] We now
diagonalise the remaining $n_{sel}$ $J_1$ operators using the
 variational procedure. To decide how many of the orthogonal states one should
keep, the following criteria can be applied.
From the comparison of  the components
of each  linear combinations to $\chi\equiv(\mathrm{det}O)^{1/n_{sel}}$,
where $O$ is the 
transformation matrix leading to the orthogonal operators with unity 
equal-time correlator, only keep those lightest states whose components 
are not significantly larger than this determinant\footnote{
If the determinant itself is large, 
try removing shapes that could be too similar to one another
and diagonalise again.}. In practise, linear combinations with large
components are found to have a very poor signal. 
 In the present case, the coordinates in units of $\chi$ read 
(0.090, 0.24, 0.37), (0.37, -1.1, 1.1) and  (-0.93, -1.19, 1.7); we keep all
three states. 
\item[An intermediate check.-]
Look at the overlaps $\langle {\cal O}_{J_2} {\cal O}^D_{J_1}\rangle$. 
Here the ${\cal O}^D_{J_1}$ are the operators obtained from the variational
procedure, and correspond to our best estimates of the  relevant 
(i.e. the lightest) $J_1$ glueball wavefunctionals. The operators
${\cal O}_{J_2}$ are the original un-diagonalised (but normalised) 
$J=1$ loop combinations. We require that the total overlaps,
which can now be calculated as  
\be
\left(\sum_i \left(\langle{\cal O}_{J_2} 
{\cal O}^{(D)i}_{J_1}\rangle\right)^2\right)^{1/2},
\ee
are less than $10-15\%$.  Again we should take this cut-off $\to 0$ as  
$a\to 0$. The overlaps in the present case are 
(0.07,0.21,0.26), (0.14,-0.21,-0.38) and (0.088, 0.10, 0.039).
Therefore we only keep the last operator as a candidate $J_2$
operator.

\item[Diagonalisation of the $J_2$ operators.-]
Diagonalise the $J_2$ operators. In the present case the operation is trivial
since we are left with only one operator.

\item[Final check.-] Now consider the same overlaps as above
but with both $J_1$ and $J_2$ operators being diagonalised ones. 
The total overlaps between these final $J_1$ and $J_2$ operators 
is required to be still less than $\sim 20\%$ ($\to 0$ as $a\to 0$).
Here we obtain a total overlap of $14\%$.

\end{description}

\subsubsection{Results}
We give our results in terms of effective masses at 1 and 2 lattice spacings
(see Table \ref{sIdata}). The ``quality'' of an operator is defined as
$|\langle \Omega | {\cal O} | n\rangle |^2$, where $n$ is the state being 
measured and ${\cal O}$ is our  operator. In calculating this quantity
we assume that the bold-faced mass values in the Tables represent the
corresponding mass plateaux. The ``overlap'' calculates the overlap,
as defined in equation (\ref{overlap}), between the state of interest
and each of the other spin eigenstates lying in the 
same square irreducible representation (e.g. spin 4 with the
various spin 0 states obtained through the variational procedure)
and adds these in quadrature. Thus it provides a measure of the
overlap of the wavefunctional onto the basis of states of the `wrong'
spin. 

We note that our operator construction method leads to excellent overlaps onto
the physical states, while the overlaps of operators with different quantum
numbers typically remain well under the $10\%$ level. This quality requirement
 is obviously dependent on the spectrum itself: if there is a large gap 
between the spin $0$ and the spin $4$, the spin 4 operator will have to be 
of exceptionally high purity, since the heavier state contribution to the 
correlator, even with a much larger projection onto the operator, 
 becomes negligible with respect to the lighter state at large 
time-separations.

\subsection{Finite volume corrections}

In physical units the size of our standard spatial volume is 
$aL\simeq 4/\sqrt{\sigma}$. This is the same size that was used in 
[\ref{mtd3}], where the choice was motivated by an explicit 
finite volume study, showing this to be more-or-less the 
smallest volume on which the lightest states in the $A_1$ and $A_3$ 
lattice irreducible representations  did not suffer significant finite 
volume corrections. Since we would expect that the size
of a glueball grows with its spin $J$, at least for large enough $J$,
it is important to check that our volumes are indeed large enough to 
accommodate a glueball with $J=4$ .

We have therefore carried out an explicit check of finite volume
effects at $\beta=9$. Here $1/\sqrt{\sigma}\simeq 6a$, so 
our standard lattice size corresponds to $L=24$. To check for
finite volume corrections we also perform calculations on
$L=16$ and $L=32$ lattices. The results are shown in  Fig.\ref{vdep} 
and listed in Table \ref{vdep_tab}. We see that there is no sign of 
any substantial correction to the $J=4$ glueball mass between $L=16$, 
$L=24$ and $L=32$, confirming that our mass calculations in this paper
are not afflicted by significant finite volume corrections.

While this reassures us that our `standard' volume size can 
be used safely at other values of $\beta$, it might 
seem paradoxical that we see no volume corrections on the
smaller $L=16$ lattice where the corrections to the $J=2$ glueball mass 
are quite visible (in Fig. \ref{vdep}). This is in fact less odd than 
it appears. The reason is that the leading finite volume corrections to 
the $J=2$ mass arise from the presence of states composed of
pairs of flux tubes that wind around the spatial torus [\ref{mtd3}].
The mass of such a winding state is approximately $\propto L$ and as
we decrease $L$ it will at some point mix with the `true'
glueball state and then become lighter than it. (The
same occurs at smaller values of $L$ in the $J=0$ sector). Thus
the appearance of finite volume effects in the $J=2$  sector
should not be taken as a signal that the volume has become
too small to accommodate that glueball and has no immediate
implications for finite volume corrections for states of
other spins.

\begin{figure}[htb]

\centerline{~~\begin{minipage}[c]{10cm}
 \psfig{file=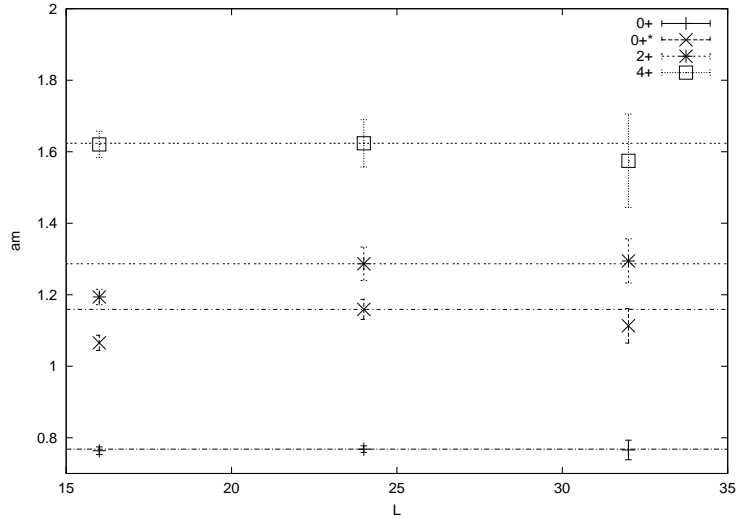,angle=0,width=10cm}
 \end{minipage}}

\caption[a]{Volume dependence of some of the glueball masses at $\beta=9$; 
the horizontal lines going through the data points at $L=24$ are drawn to
guide the eye.}
\la{vdep}
\end{figure}

\subsection{Smaller lattice spacings and the continuum limit}

Following the same procedure as we used for our $\beta=6$ calculation
in section \ref{recipe}, we perform further  mass calculations
at $\beta=9,~12~$ and 14.5 on lattices of the same physical size. 
The operators used are described in Table \ref{coords} and the results 
of our calculations are presented in Table \ref{sIdata}. 
We now express the glueball masses in units of the string tension,
$am/a\sqrt{\sigma} \equiv m/\sqrt{\sigma}$, using values of
the string tension, $a^2\sigma$, obtained at the same values of
$\beta$. The leading lattice correction should be $O(a^2)$ and
so if we plot our values of $m/\sqrt{\sigma}$ against $a^2\sigma$,
as in Fig. \ref{extrapol}, then we can extrapolate linearly to $a=0$ 
for sufficiently small $a$. Such extrapolations are shown in
Fig. \ref{extrapol} and the resulting continuum values of the
glueball masses are given in Table \ref{contI}.

\begin{figure}[htb]

\centerline{~~\begin{minipage}[c]{10cm}
 \psfig{file=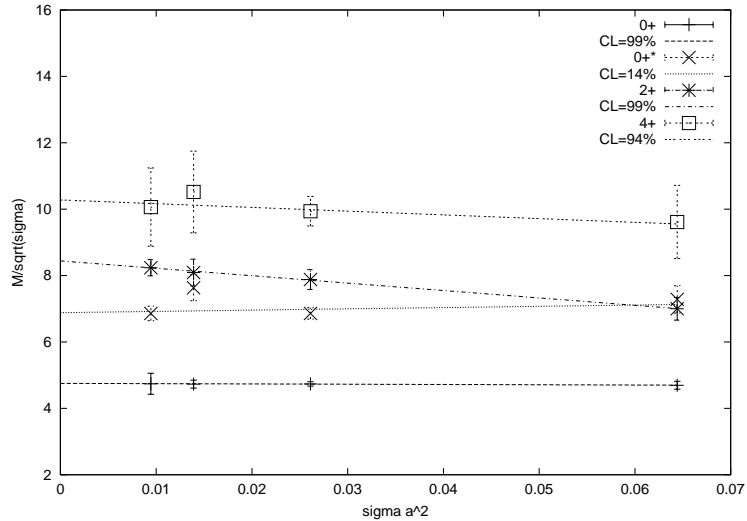,angle=0,width=10cm}
 \end{minipage}}

\caption[a]{Strategy I: continuum extrapolation of the $J^P=(2n)^+$ masses}
\la{extrapol}
\end{figure}

We note that we obtain the expected parity-degeneracies in the continuum 
limit for the $J=2$ and $J=4$ spins that we consider. We also observe
that the corrections to the $4^+$ mass are reasonably well
fitted by a simple $a^2$ term for $\beta \geq 9$, just as they are
for the  $0^+$ and $2^+$.

\section{Applications of Strategy II}

In this section we apply the second strategy introduced
in Section~\ref{subsec_twostrat} in which we probe glueball
wave-functions with
variously rotated operators so as to directly extract the
coefficients of the Fourier modes contributing to the angular 
variation of those wavefunctions.

As an illustration we first apply the method to a case where we 
believe we know the answer, i.e. the lightest states in the $A_1$ 
and $A_3$ representations. We confirm that these states are
indeed $J=0$ and $J=2$ respectively. We then return to the
$0^-~/~4^-$ puzzle. We establish that the lightest state in the 
$A_2$ representation is indeed spin 4 and that this is much lighter
than the $0^-$ ground state. The evidence is more convincing
than before not only because of the greater transparency of this
approach, but also because we repeat the calculation closer
to the continuum limit. We then go on to investigate
the angular behaviour of the lightest states falling in the 
two-dimensional $E$ representation which contains all of the 
continuum odd-spin states. Our conclusion will be that
the lightest states have quantum numbers  $3^\pm$, rather than 
$1^\pm$. Finally, we reanalyse the spectrum of 
states in the $A_1$ representation, and perform a continuum 
extrapolation, obtaining results that are consistent with
those that we obtained earlier using our first method.

\subsection{Wave functions of the lightest $A_1$ and $A_3$ states}

To analyse the angular content of the lightest states lying in the 
$A_1$ and $A_3$ representations we use the four operators in
Fig. (\ref{sh4}) together with their twelve rotations.
 We first use the exact lattice symmetries
to form operators in each of these lattice representations, and 
then we use the variational method to determine what are the linear
combinations of these operators that provide the best approximations 
to the ground state glueball wavefunctionals. We then construct
the same linear combinations of (approximately) the same operators
rotated by different angles. This provides us with rotated
versions of the ground state wavefunctionals. From the correlation 
at (typically) two lattice spacings between the original and rotated 
copies of our ground state wavefunctional, we can extract
the angular variation, as displayed in Fig. (\ref{a1a3}).

\begin{figure}[htb]

\centerline{~~\begin{minipage}[c]{6.5cm}
    \psfig{file=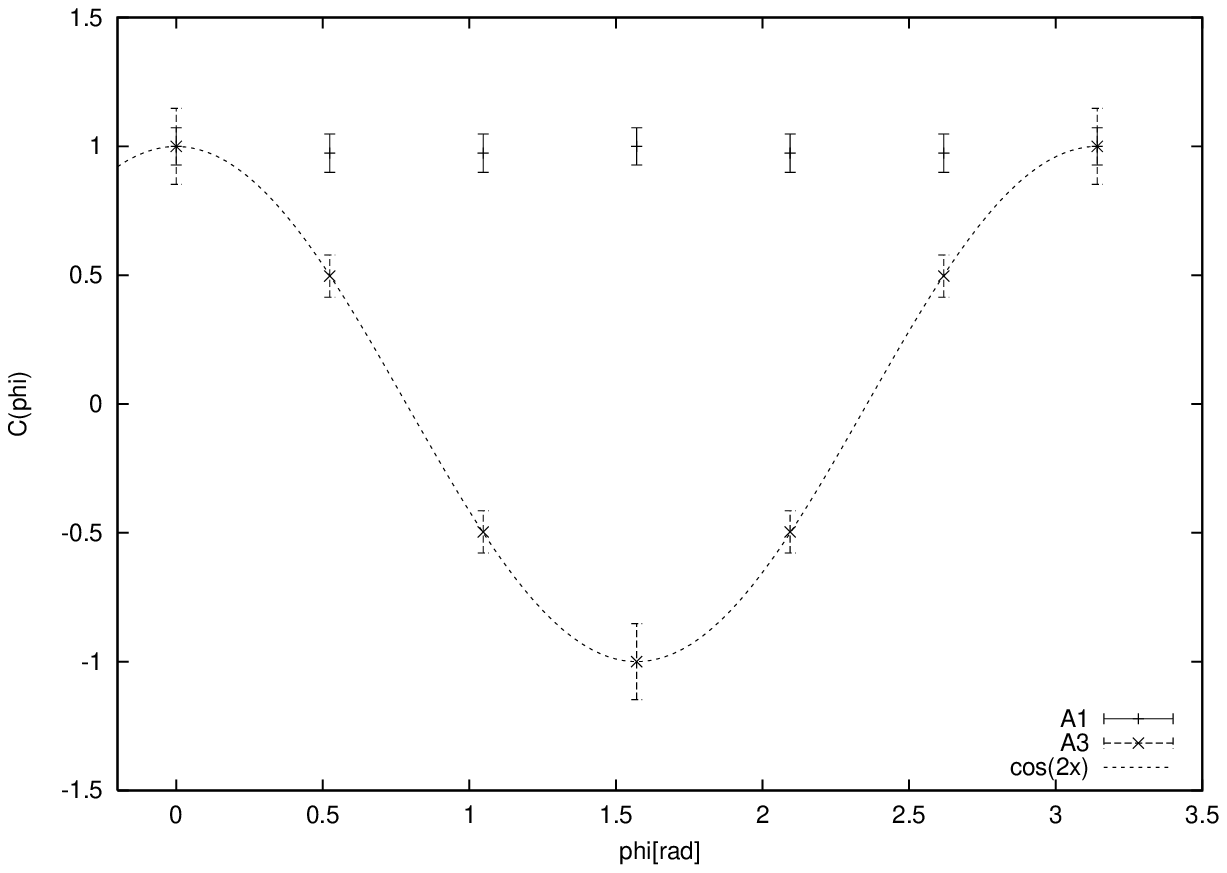,angle=0,width=6.5cm}
    \end{minipage}
  ~~\begin{minipage}[c]{6.5cm}
    \psfig{file=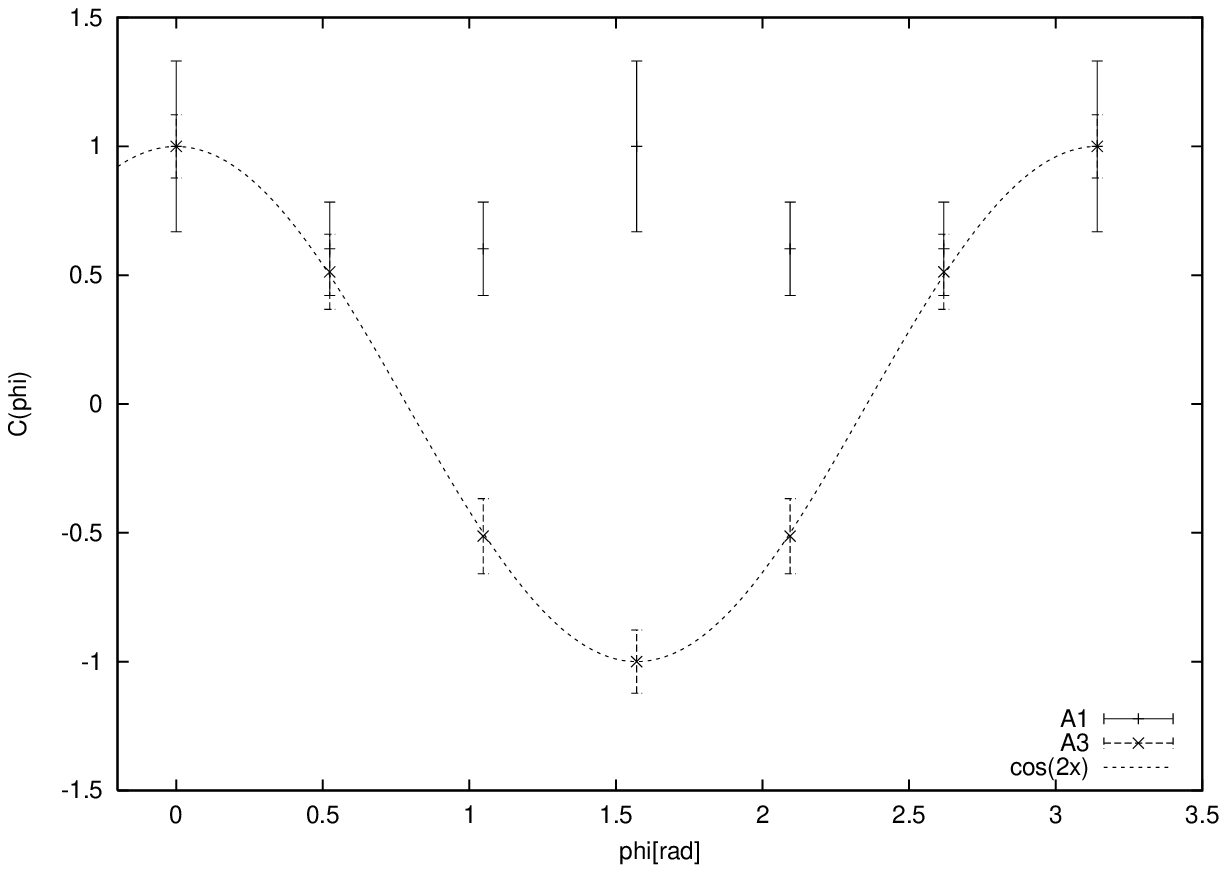,angle=0,width=6.5cm}
    \end{minipage}}

\centerline{~~\begin{minipage}[c]{6.5cm}
    \psfig{file=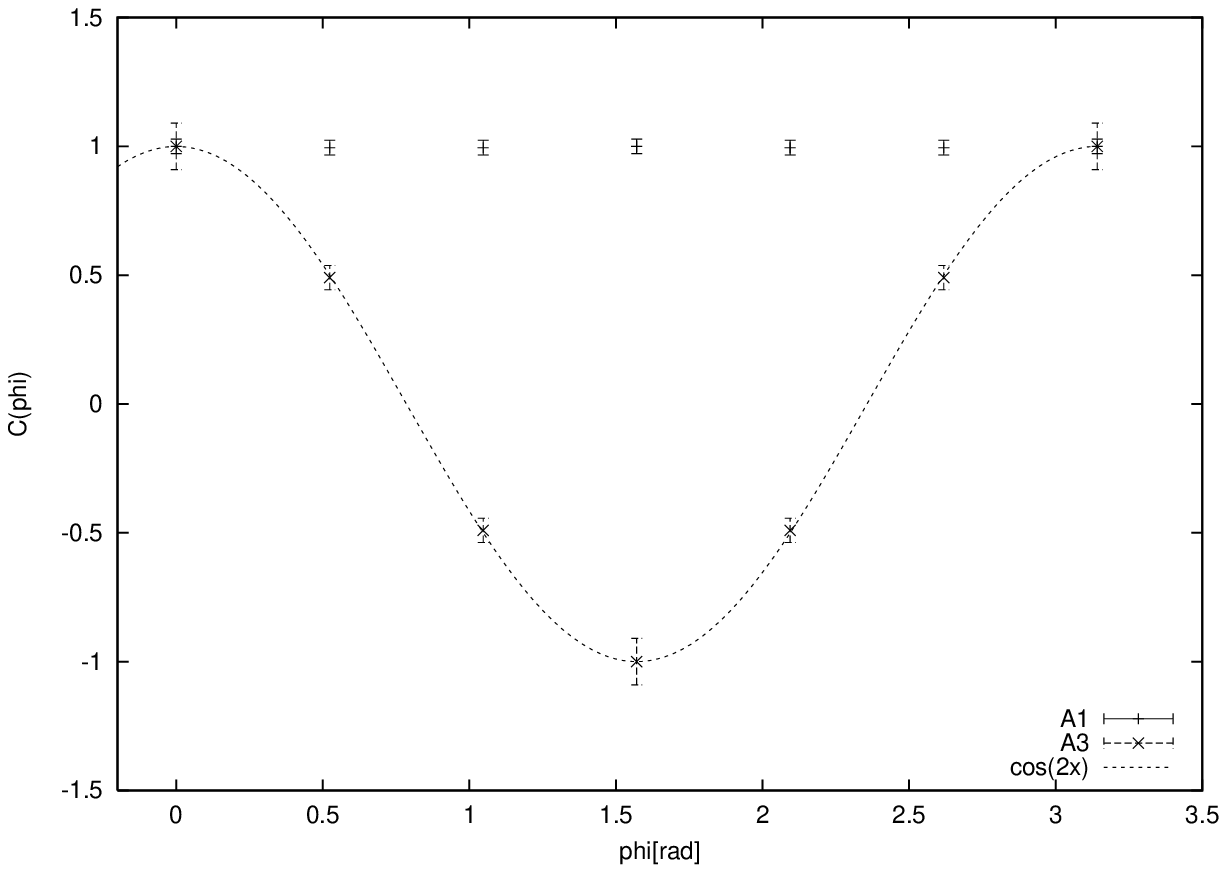,angle=0,width=6.5cm}
    \end{minipage}
  ~~\begin{minipage}[c]{6.5cm}
    \psfig{file=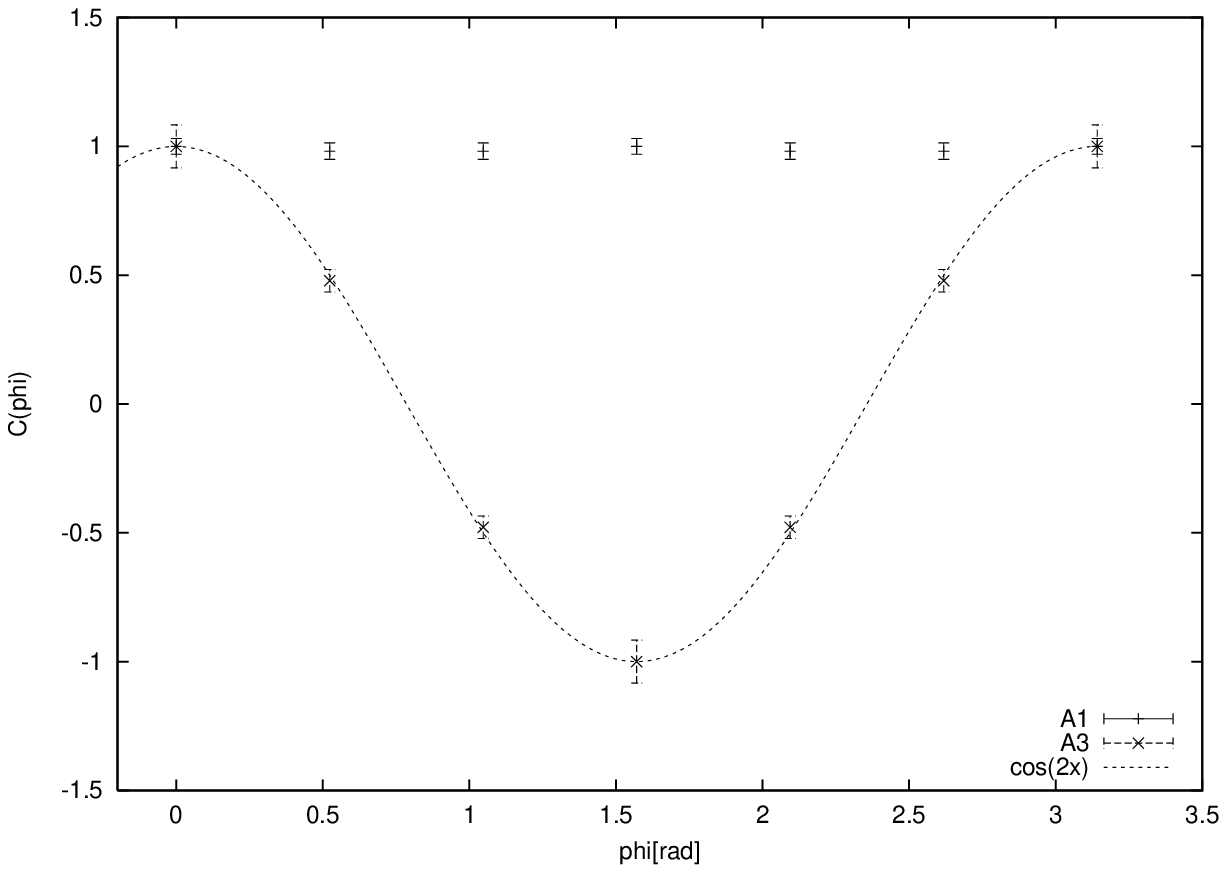,angle=0,width=6.5cm}
    \end{minipage}}
\vspace*{0.5cm}

\caption[a]{The wave function of the lightest state in the $A_1$ and $A_3$ 
lattice representations, as measured with our four operators at $\beta=6$.
The plots can be compared to the correlation matrices given 
in section (\ref{recipe}).}
\la{a1a3}
\end{figure}
We clearly observe the characteristic features of $0^+$ and $2^+$ wave 
functions. (The $A_1$ representation of the second operator varies
more with the angle $\phi$ than the others. We had already noted that 
its correlation matrix was far from being Toeplitz
and the selection criteria in Section \ref{recipe} had led 
us to remove it from the analysis.) This provides a simple
illustration of the method in a non-controversial context.

\subsection{The $0^-~/~4^-$ puzzle revisited}

We now proceed to analyse the angular variation of the wave function
of the ground state $A_2$ glueball on our $16^3$ lattice
at $\beta =6$. We begin with the first set of operators of 
Section \ref{04m}, as displayed in Fig. \ref{sh1}.  We obtain 
the following mass (effective, at 1 lattice spacing) 
and mode decomposition (at 2 lattice spacings):
\begin{center} 
\begin{tabular}{|c|c|c|c|}
\hline
$\mathbf{\beta=6}$ & $am$  & $c_0$ & $c_4$ \\
\hline
$4^-$   & 2.575(71) & -0.0525(60)  &  0.9986(53) \\ 
 \hline
\end{tabular}
\end{center}
The fit, with two parameters, is to three independent points 
and possesses a goodness of fit given by $\chi^2=2.28$ 
(see Fig. (\ref{a2})). With the second  set of operators of 
Section \ref{04m} (which are rotated by $\frac{\pi}{4}$
angles, Fig. \ref{sh2}), we obtain the following mass 
and wave function (at one lattice spacing):
\begin{center} 
\begin{tabular}{|c|c|c|c|}
\hline
$\mathbf{\beta=6}$ & $am$  & $c_0$ & $c_4$ \\
\hline
$4^-$   & 2.455(60) & -0.224(71)  &    0.975(71) \\
 \hline
\end{tabular}
\end{center}
The coefficients clearly show that the lightest $A_2$ state wave
function is completely dominated by the spin 4 Fourier component. 
Thus, we can conclude that the $4^-$ ground state 
is lighter than the $0^-$. Of course this
statement holds at $\beta = 6$ and one needs to check that
it is robust against lattice spacing corrections. This we now
do by performing a calculation at  $\beta=12$ on a $32^3$ lattice. 

In principle we could proceed as before: constructing the 
$A_2$ square representation and looking at the 2-lattice-spacing 
correlations with an operator oriented in different directions. 
If we saw a $\sin{(4x)}$ behaviour  then we could conclude
that the lightest state of the $A_2$ representation does indeed 
become a $4^-$ state in the continuum. On the other hand, if
the result was independent of the direction, then this would
mean that the lightest state of the $A_2$ representation is
a $0^-$ state at this $\beta$; and probably also in the continuum 
limit. However there is an interesting subtlety associated with the 
$0^-$ state and the unfamiliar parity operation which we can 
exploit so as to proceed somewhat differently. 
In $3+1$ dimensions, where parity commutes
with all rotations, it is impossible to construct a $P=-1$ operator
with Wilson loops that are $P=+1$. In general this restriction does 
not exist in $2+1$ dimensions, and we have made 
use of this fact when using operators that have a symmetry axis
in order to obtain the masses of $P=-1$ states.
However, it does apply to the spin 0 sector, where parity commutes with
rotations: in other words, a linear combination of symmetric operators
corresponding to the quantum numbers of the $A_2$ representation does not 
couple to the $0^-$ component of the lattice  states. The 
reason is that the image of a symmetric operator under an axis-symmetry
can also be obtained by a rotation, so that the relative minus sign cancels
the contribution of any $0^-$ component. Thus we can 
measure the projection of the wave function onto the space orthogonal to the 
$0^-$ subspace. If the overlap onto the state whose mass we extract with
this operator is not dramatically decreasing  as we approach the continuum 
limit, we can safely conclude that the state has quantum numbers $4^-$.

The kind of operators we used are triangular, as drawn in Fig.\ref{sh5}.
The corresponding wave function shown in Fig.\ref{a2}, 
confirms the last paragraph's conclusions: 
the data points fall perfectly on a $\sin{4x}$ type curve. This result
has been obtained with all the operators we employed.
From the fact that our overlaps onto the state at each of 
$\beta=9,12,14.5$  are  better than $90\%$, we confidently
conclude that the state carries quantum numbers $4^-$ in the continuum 
limit. As expected from parity doubling, it is found to be degenerate with
the lightest $4^+$ glueball, with a mass of 
$am(4^-,\beta=12)= 1.365(56)$.
%
\begin{figure}[htb]
\centerline{
\begin{minipage}[c]{6.5cm}
\begin{picture}(100,80)(10,20)
\multiput(0,0)(0,10){8}{\line(1,0){150}}
\multiput(0,0)(10,0){16}{\line(0,1){70}}
\thicklines{
\put(0,0){\line(1,0){130}}
\put(130,0){\line(-1,5){10}}
\put(0,0){\line(5,2){120}}
}
\put(130,0){\circle*{4}}
\put(0,0){\circle*{4}}
\put(120,50){\circle*{4}}
\end{picture}
\end{minipage}}
\vspace{1cm}
\caption[a]{The triangular operator rotated by $\frac{\pi}{8}$ angles used
to determine  the lightest $A_2$ state wave function at $\beta=12$}
\la{sh5}
\end{figure}
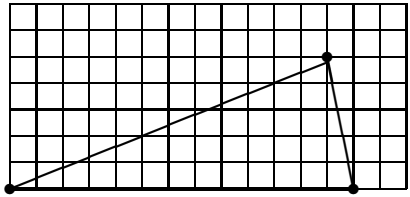

\begin{figure}[hbt]

\centerline{\begin{minipage}[c]{5cm}
    \psfig{file=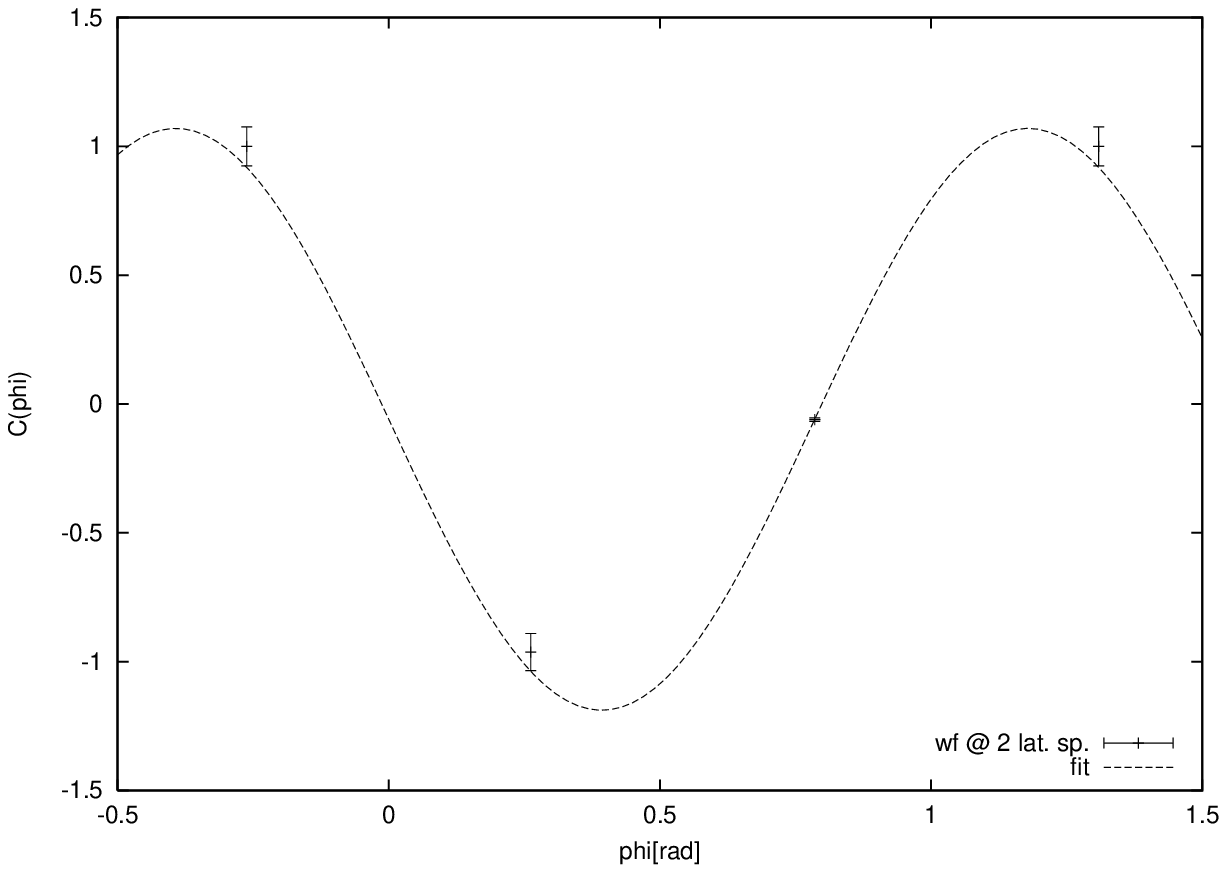,angle=0,width=5cm, height=5cm}
    \end{minipage}~~\begin{minipage}[c]{5cm}	
    \psfig{file=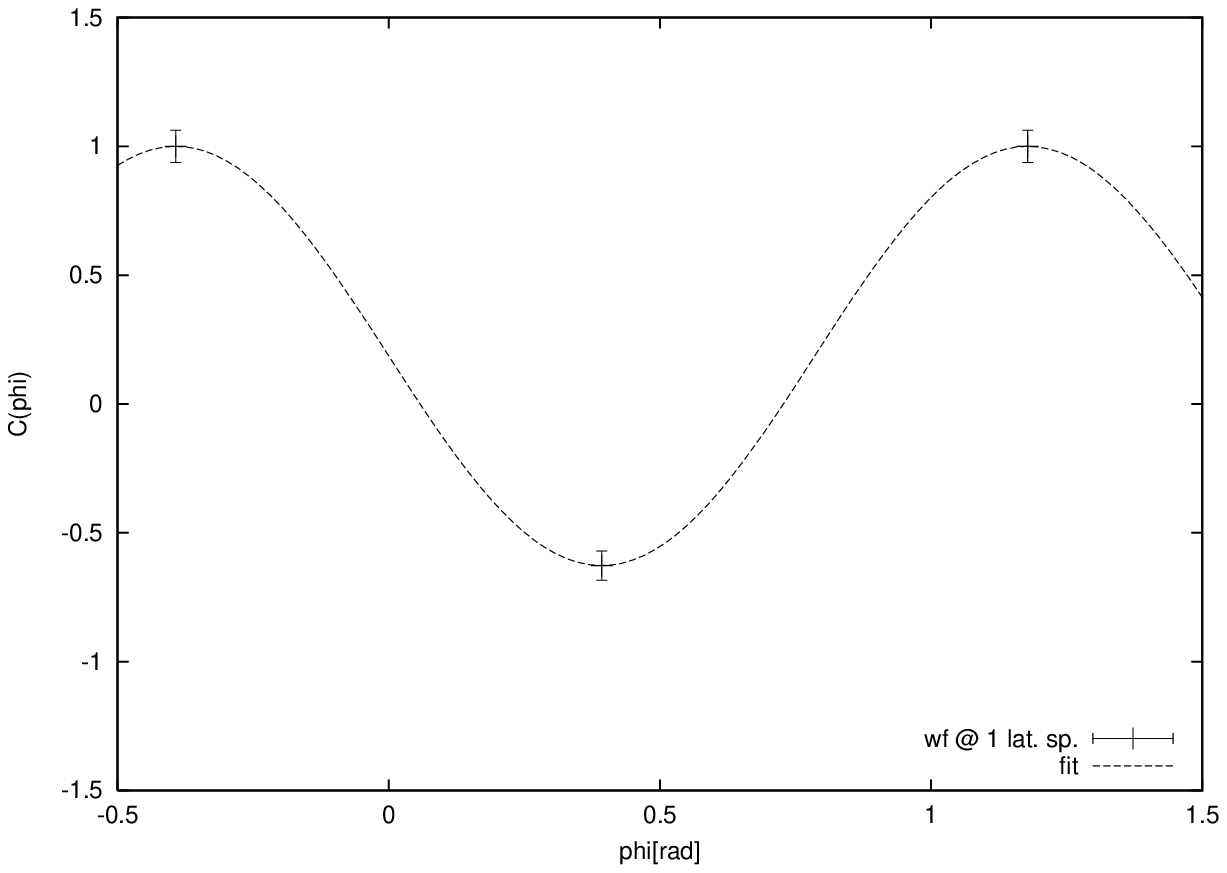,angle=0,width=5cm, height=5cm}
    \end{minipage}~~\begin{minipage}[c]{5cm}
    \psfig{file=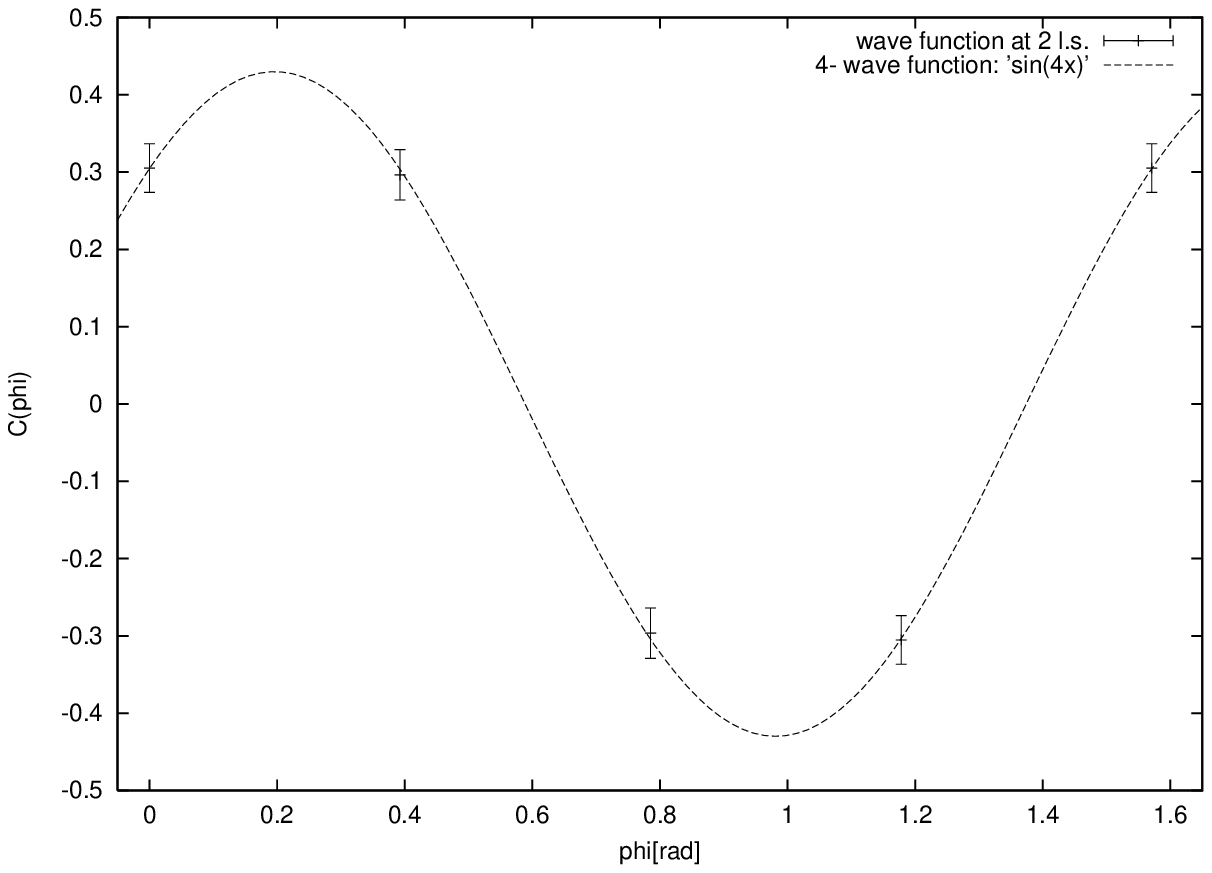,angle=0,width=5cm, height=5cm}
    \end{minipage}}

\caption[a]{The wave function of the lightest state in the $A_2$
lattice representation; left, at $\beta=6$ and measured with 12-fold operators
; centre, at $\beta=6$ with 8-fold operators; right, at $\beta=12$ with
16-fold operators. The vertical axis has arbitrary scale.}
\la{a2}
\end{figure}

\subsection{Wave functions of the lightest $E$ states}

Proceeding as above, we extract the  angular wave function of the 
lightest states lying in the $E$ representation at $\beta=12$ 
(Fig. (\ref{E12})). These are obtained from correlations 
separated by 2 lattice spacings, so as to allow the excited modes
to decay relative to the contribution from the lightest states.
This representation contains the continuum spin 1 and 3 states, 
both in the positive and negative parity channels. We clearly observe
the characteristic behaviour of a spin 3 wave function. 
More quantitatively, the $3^+$ wave function is well fitted by 
$0.1880(83)\cos{3\phi}$  ($\chi^2=0.291/3$) and the $3^-$ by
($0.957(38)\sin{3\phi}$, $\chi^2=0.285/3$).  From this we conclude that
the spin 3 glueball is lighter than the spin 1.

\begin{figure}[htb]

\centerline{~~\begin{minipage}[c]{6.5cm}
    \psfig{file=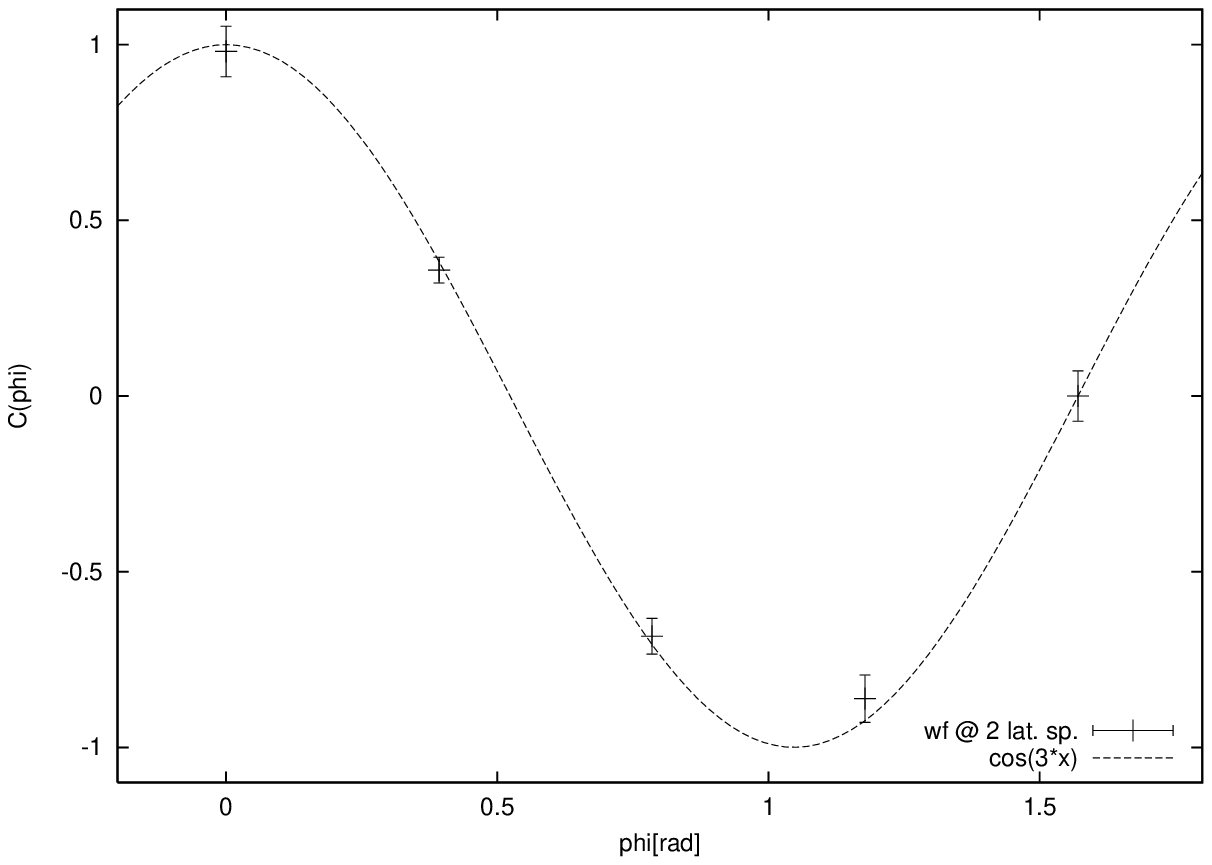,angle=0,width=6.5cm}
    \end{minipage}
  ~~\begin{minipage}[c]{6.5cm}
    \psfig{file=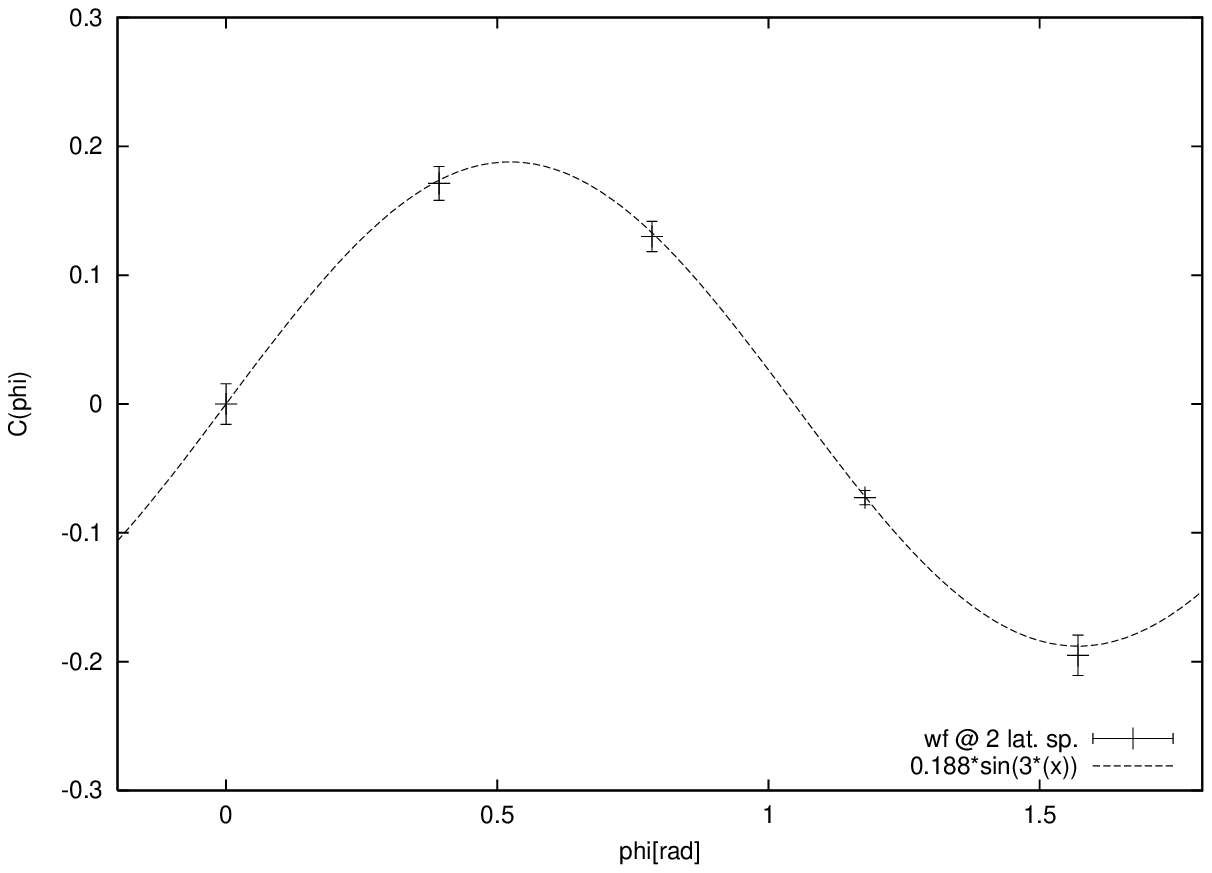,angle=0,width=6.5cm}
    \end{minipage}}

\caption[a]{The wave functions of the two lightest states 
in the two-dimensional lattice representation $E$, 
as measured at $\beta=12$.}
\la{E12}
\end{figure}

\subsection{The masses and Fourier coefficients in the continuum limit}

Having identified at various values of $\beta$ the states
corresponding to different continuum spins, we can extrapolate
them to the continuum limit in the usual way. This is
shown in Fig.\ref{extrapol2}. The masses agree with the
calculations performed earlier using our ``strategy I''. 
In identifying the lattice states as belonging to particular
continuum spins, we assume that the appropriate Fourier
coefficient  $|c_n|^2$ will extrapolate to unity in the
continuum limit. A check that this is so is provided in
Fig.\ref{coeffs} where we show an extrapolation of the $4^+$  
component of the $A_1^{***}$ state, as well as the $0^+$ component 
of the $A_1^{**}$ state. Although the error bars are relatively large, 
the data allows use to draw  definite conclusions about the 
quantum number of these states in the continuum.

\begin{figure}[htb]

\centerline{~~\begin{minipage}[c]{10cm}
 \psfig{file=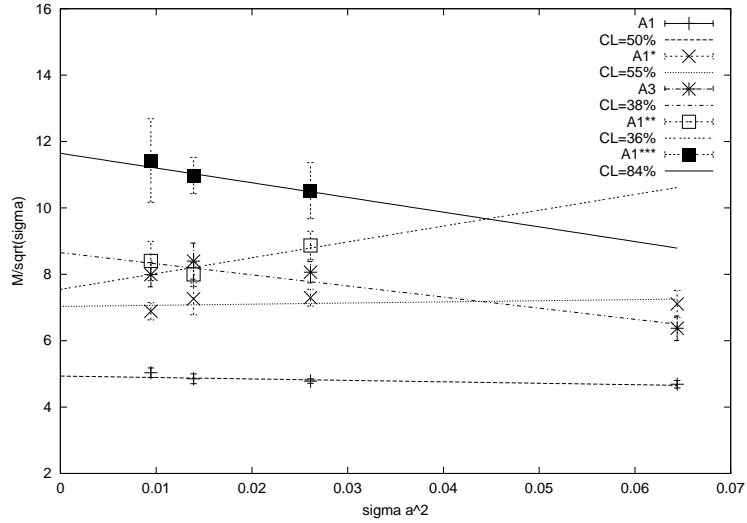,angle=0,width=10cm}
 \end{minipage}}

\caption[a]{Continuum extrapolation of the $J^P=(2n)^+$ masses, as obtained
with strategy II}
\la{extrapol2}
\end{figure}

\begin{figure}[htb]

\centerline{~~\begin{minipage}[c]{10cm}
 \psfig{file=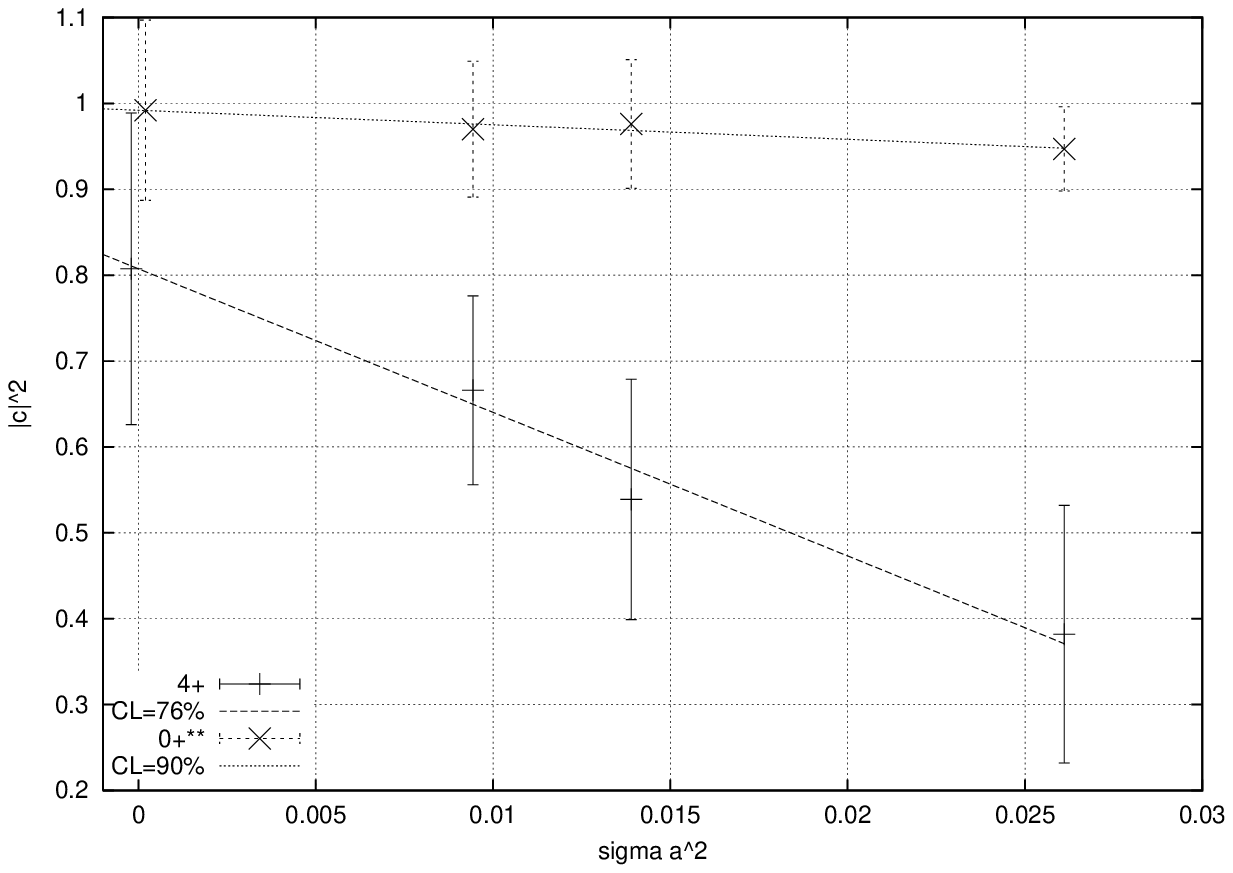,angle=0,width=10cm}
 \end{minipage}}

\caption[a]{Continuum extrapolation of coefficient $c_0$
 for the $A_1^{**}$ state and coefficient $c_4$ for the  $A_1^{***}$ state. 
The continuum values are 0.996(53) and 0.81(18) respectively. Thus the former
state evolves to a $0^+$, the latter a $4^+$ state in the continuum limit.}
\la{coeffs}
\end{figure}

\section{Generalisation to (3+1) dimensions}
We note that the ideas on high spin wave functions 
described for simplicity in the plane carry over 
directly to three dimensions. The wave functions are now spherical harmonics; 
the ${D^{(j)m}}_{m'}(\alpha, \beta, \gamma)$  matrices  tell us
how they transform under rotations. Here $(\alpha, \beta, \gamma)$ are Euler
angles. Recall that the point of this parametrisation is that
\be
R(\alpha, \beta, \gamma)\equiv
~R_{\hat z''}(\alpha)~R_{\hat y'}(\beta)~R_{\hat z}(\gamma)=
R_{\hat z}(\gamma)~R_{\hat y}(\beta)~R_{\hat z}(\alpha)
\ee
where indexes indicate the axis of rotation and the argument the angle.
A prime denotes the image of an axis under the preceding rotation in the 
definition of $R(\alpha, \beta, \gamma)$. The nice feature of the second
expression for $R$ is that it contains only rotations around the Cartesian
axes, making it very easy to compute. Now, by definition, ${D^{(j)m}}_{m'}$
is a representation of the rotation group:
\be
U(\alpha, \beta, \gamma)|j,m\>=\sum_{m'}{D^{(j)m}}_{m'}(\alpha, \beta, \gamma)
|j,m'\>
\ee
so that the 3-dimensional equivalent of (\ref{opj}) reads
\be
{\cal O}^{(m')}_{j,m}\propto\int d\gamma d(\cos{\beta})d\alpha~ 
{D^{(j)m}}_{m'}(\alpha, \beta, \gamma) {\cal O}(\alpha, \beta, \gamma)\la{3dg}
\ee
where ${\cal O}(\alpha, \beta, \gamma)$ represents an operator rotated 
in the direction $(\alpha, \beta, \gamma)$. Thus, if we want to measure the
mass of a state $|j,m\>$, there is a choice for $m'\in\{-j,\dots,j\}$.
Although a general expression exists for $D$, in order 
to construct a $(j,m)$ operator it is sufficient to know
 the following property:
\be
{D^{(j)m'}}_{m}(\alpha, \beta, \gamma)=e^{-i\alpha m'}~d^{(j)}(\beta)^{m'}_m
e^{-i\gamma m}
\ee
with
\be
{d^{(j)}(\beta)^{m}}_0=\left(\frac{(j-m)!}{(j+m)!}\right)^\frac{1}{2}
P_j^m(\cos{\beta})
\ee
Here $P_j^m(z)$ is a Legendre function. Thus a simpler, yet general enough
version of (\ref{3dg}) is 
\be
{\cal O}_{j,m}\equiv {\cal O}^{0}_{j,m} =
\int d\gamma d(\cos{\beta}) d\alpha   ~ 
P^m_j(\cos{\beta})~e^{im\gamma}~ {\cal O}(\alpha, \beta, \gamma)\la{3dop}
\ee
The higher complexity of these wave functions will make it more difficult to 
project out states with a discretised approximation to (\ref{3dop}) 
of the kind used in our strategy I. For that reason we expect the second 
strategy, where the wave function information is extracted from the 
lattice rather than being assumed, to become our main method. The $2j+1$
degeneracy of states (in the continuum limit) should provide us with a 
powerful consistency check.

\section{Previous work}

There have been two recent calculations of the $4^+$ glueball mass 
in D=2+1 SU(2) gauge theories [\ref{rjmtrot},\ref{rjrot}], 
and a calculation of the $4^{++}$ glueball mass in the
D=3+1 SU(3) gauge theory [\ref{liurot}].  We shall now briefly
comment upon these earlier calculations.

The calculation in [\ref{rjmtrot}] starts with rectangular
Wilson loops constructed out of simple, blocked link
matrices and constructs other Wilson loops that are
approximate $\pi/4$ rotations of these. Linear combinations of
these are taken to produce, altogether, twelve trial $0^+$ and
$4^+$ operators. The overlaps at $t=0$ of corresponding
$J=0$ and $J=4$ operators are calculated and operators with
substantial overlaps are discarded. This leaves nine pairs
of operators with smaller overlaps. Effective masses at $t=a$ are 
obtained from the diagonal correlators of these nine pairs
of operators, and the minimum 
effective mass in each set provides an estimate of the
lightest glueball of the appropriate spin. The calculation
is at $\beta=9$ on a $L=24$ lattice and the estimate of the 
lightest $J=4$ glueball mass, $am(J=4) = 1.607(27)$, is 
consistent with the estimates in the present paper at
that value of $\beta$. However the method is very similar
to the naive precursor of our Strategy I, which served as a
caveat in Section \ref{subsection_caveat}. No attempt is
made to determine if the `J=4' state is not in fact an excited 
$J=0$ state and no approach to the continuum is attempted.
That is to say, the control over the relevant systematic errors
is almost entirely lacking.

The calculation in [\ref{rjrot}] is similar to that in 
[\ref{rjmtrot}] except that it uses the matrix method described
in Section \ref{subsection_matrix} to construct the `sides'
of suitably chosen Wilson loops. These involve (approximate) copies 
rotated by angles that are multiples of $\pi/3$. Such
Wilson loops are allowed to come in four different sizes 
and calculations are performed with the parameter $\alpha$
in eqn(\ref{kdef}) taking values  $\alpha =0.1,0.15,0.2,0.25$.
(There is no smearing of the link matrices that enter as the
elements of the original matrix.) Taking suitable linear combinations 
one thus obtains sixteen trial  $0^+$ and $4^+$ operators. The 
overlaps at $t=0$ of corresponding $J=0$ and $J=4$ operators are 
calculated and the operator for which the overlap is minimal
(and indeed negligible) is used to extract a $J=4$ mass estimate
from the diagonal correlator at $t=a$. The calculation is on a $L=16$ 
lattice at $\beta=6$. The operator corresponds to the largest loop
and $\alpha=0.2$. Using eqn(\ref{eqn_alphatom}) such a value of 
$\alpha$ translates into a scalar propagator mass of about
$am_0 \simeq 1.0 \simeq 4 a\surd\sigma$ which is rather large
and would suggest that the effective smearing is weak.
The $t=a$ mass estimate for the $4^+$ glueball
is $am(J=4) = 2.13(4)$ which is significantly lower than our
$\beta = 6$ mass estimate.  Also puzzling is the fact that
some of the $t=a$  $2^+$ glueball effective masses are significantly 
lower than the usually accepted mass estimate for this supposedly
uncontroversial state. In any case,
this calculation suffers from a poor control over most of the
systematic errors that we listed earlier in discussing the calculation
of [\ref{rjmtrot}]. 

The calculation in [\ref{liurot}] is in D=3+1 so we cannot, for
the moment, compare their results with ours.
Their approach is to expand small Wilson loops in  terms 
of continuum fields and their covariant derivatives
so as to identify a linear combination that is $J=4$. 
Such a linear combination is then used as a  $4^+$ trial
operator. The calculations are performed  for very coarse 
spatial lattice spacings, presumably because only there can
small unsmeared Wilson loops act as useful glueball wavefunctionals.
However for such coarse spatial lattice
spacings one might worry about the relevance of the
continuum field expansion that motivates the spin assignment.
On the positive side, an improved action is used
which one might hope would reduce any lattice corrections. A
continuum extrapolation is performed, albeit over a range of
spatial lattice spacings that are large. It would certainly
be worthwhile to pursue this interesting method further
incorporating some of the checks described in our paper.

\section{Conclusions}

To calculate the mass of a glueball of spin $J$ in a lattice
calculation, one must identify the lattice energy eigenstate
that tends to that state in the continuum limit. The limited
rotational invariance on a lattice introduces ambiguities
which means that at a fixed value of $a$ one cannot be
confident in one's spin assignment. Only by performing
a continuum extrapolation while monitoring the angular
content of the glueball wavefunctional, can one be confident
in the mass one extracts for a high spin glueball.

In practise one needs to identify likely candidates for
such lattice eigenstates and we introduced two related strategies
to do so. The first tries to construct 
wavefunctionals with the required rotational symmetry,
which of course can only be approximate at finite $a$.
The second probes the angular variation of eigenstates obtained 
through a conventional lattice calculation, using probes
that have the required rotational symmetry, which again
 can only be approximate at finite $a$. In either case
one needs to be able to easily calculate smeared Wilson loops
with arbitrary shapes, and we developed methods for
doing so.

To test our methods we applied them to the relatively simple
problem of determining whether the $4^-$ is lighter than
the $0^-$ in  D=2+1 gauge theories. The physical interest of 
this possibility lies in the fact that it is predicted by flux 
loop models of glueballs and it is plausible that such models are 
relevant to glueball structure in D=3+1. Our calculations
confirmed unambiguously that the ground state $4^-$ glueball
is indeed much lighter than the ground state $0^-$ glueball,
so that the usual identification of the ground state of the 
$A_2$ representation as being $0^-$ is mistaken.
Similarly we showed that the $4^+$ state degenerate with this  
$4^-$ is usually misidentified as the $0^{+***}$.
(It is of course possible that with a larger basis of
operators we would resolve additional excitations of the $0^+$
that are lighter than the  $4^+$.)

In applying our methods we found that the second strategy
was in practise the more transparent and reliable. Although
it is more affected by the mixing that occurs
if two levels that  correspond to different (continuum) spins
cross as we reduce $a$, such crossings should be easy to spot 
and to take into account.

Our calculations make us confident that calculating higher
spin glueball masses is a practical task. 
Currently we are extending our work to a full continuum extrapolation 
covering the range $\beta=9$ to 18 and extending to higher spins 
than four [\ref{evspin}]. This will enable us to see if
the glueball mass spectrum defines a  `Pomeron' trajectory
in two space dimensions. At the moment our calculations are for
SU(2) but it is plausible that the  simplicity of a linear
Pomeron trajectory will only become manifest in the SU($N\to\infty)$
theory and we hope to explore that limit as well.
Of course the question of real physical interest is whether 
glueballs define the Pomeron(s) in D=3+1 with the SU(3) gauge group of
the strong interactions. As we pointed out, the general ideas, applied 
here for simplicity in the plane, carry over directly to three dimensions, 
where the wave functions are spherical harmonics and  the $2j+1$
degeneracy of states  provides a useful consistency check in the 
determination of the continuum spectrum. We are currently
performing calculations of higher spin glueballs in the D=3+1
SU(3) gauge theory at intermediate lattice spacings.

\paragraph{Acknowledgements.-} One of us (HM) thanks the Berrow Trust 
for financial support.

\vfil\eject

\paragraph{References}
\begin{enumerate}
\item \label{bag} T. H. Hansson, in {\it Non-perturbative Methods}
(Ed. S. Narison, World Scientific 1985)
\item \label{isgur} N.~Isgur and J.~Paton, Phys. Rev. D31 (1985) 2910.
\item \label{moretto} T.~Moretto, M.~Teper, hep-lat/9312035.
\item \label{model1} R.~Johnson, M.~Teper, Nucl. Phys. Proc. Suppl. 63
(1998) 197 (hep-lat/9709083).
\item \label{gkjp} G.~Karl and J.~Paton, Phys. Rev. D61 (2000) 074002
(hep-ph/9910413).
\item \label{model2} R.~Johnson, M.~Teper, Phys. Rev. D66 (2002) 036006 
(hep-ph/0012287).
\item \label{pomeron} P. Landshoff, hep-ph/0108156. \\
A. Kaidalov, hep-ph/0103011.
\item \label{mtjh02} M.~Teper, Invited Talk at the John Hopkins
Workshop 2002.
\item \label{mtd3} M.~Teper, Phys. Rev. D59 (1999) 014512  (hep-lat/9804008).
\item \label{mart_var} M.~L\"uscher, U.~Wolff, Nucl. Phys. B339 (1990)
222.
\item \label{mtmatrix} M.~Teper, Nucl. Phys. Proc. Suppl. 4
(1988) 41 and unpublished.
\item \label{reb} Lang, C.~Rebbi, Phys. Lett. B115(1982) 137
\item \label{for} Ph. de Forcrand, G.~Schierholz, H.~Schneider, M.~Teper, 
Z. Phys. C31(1986) 87
\item \label{mor} C.~Morningstar, M.~Peardon, Phys. Rev. D60:034509, 1999 
\item \label{rjmtrot} R.~Johnson, M.~Teper, Nucl. Phys. Proc. Suppl. 73
(1999) 267 (hep-lat/9808012).
\item \label{liurot} Da Qing Liu, Ji Min Wu, Mod.Phys.Lett. A17 (2002)
(hep-lat/0105019).
\item \label{rjrot}  R.~Johnson, D.Phil. thesis, Oxford University,
2001 and hep-lat/0206005.
\item \label{lattice} 
J. Smit, {\it Introduction to Quantum Fields on a Lattice},
Cambridge 2002. \\
H. Rothe, {\it Lattice Gauge Theories} 2nd Edition,
World Scientific 1997. \\
I. Montvay and G. Munster, {\it Quantum Fields on a Lattice},
Cambridge 1994. \\
M. Creutz, {\it Quarks, gluons and lattices},
Cambridge 1983. 
\item \label{bankscasher} T. Banks and A. Casher, 
Nucl. Phys. B169 (1980) 103.
\item \label{block} M. Teper, Phys. Lett. B183 (1987) 345.
\item \label{smear} M. Albenese et al (APE Collaboration),
Phys. Lett. B192 (1987) 163.
\item N.~Cabibbo and E.~Marinari, Phys.\ Lett.\ B {119} (1982) 387.\label{hb}
\item Kennedy, Pendleton, Phys.~Lett., 156B (1985) 393\label{kenpen}.
\item S.L.~Adler, Phys.\ Rev.\ D {23} (1981) 2901.\label{ovr} 
\item \la{evspin} H.~Meyer, M.J.~Teper, in preparation.
\end{enumerate}
\newpage
\appendix
\section{Properties of the $M$ matrix}
\paragraph{Basic properties}
\begin{eqnarray}
\mathrm{Tr} M^{2n+1}&=&\sum_i \lambda_i^{2n+1}=0 \quad\forall~ n \nonumber\\
\mathrm{Tr} M^2&=&\sum_i \lambda_i^2=4n^2 \quad\Rightarrow\quad 
\sqrt{\langle \lambda^2 \rangle}=2  \nonumber\\
\mathrm{Tr} M^{4}&=&\sum_i \lambda_i^{4}=4\sum_i \mathrm{Tr}P_i+
32 n^2  \quad\Rightarrow\quad \langle \lambda^4 \rangle =4 \langle P\rangle
+32 \nonumber
\end{eqnarray}
\paragraph{The frozen configuration\\}
The spectrum of the matrix $M$ in the frozen configuration (all links set to
1) is the following:
\begin{equation}
\lambda_{ab}=2\left(\cos{\left(\frac{2\pi a}{L}\right)}+\cos{\left(\frac{2\pi
 b}{L}\right)}\right)=4\cos{\left(\frac{2\pi(a+b)}{L}\right)}
\cos{\left(\frac{2\pi(a-b)}{L}\right)},\qquad a,b=1,2,\dots,L
\end{equation}
with corresponding eigenvectors
\begin{equation}
v_{ab}=\left(1,\lambda_a,\lambda_a^2,\dots,\lambda_a^{L-1},\lambda_b,\lambda_b
\lambda_a,\dots,\lambda_b \lambda_a^{L-1}, \dots, \lambda_b^{L-1},
\lambda_b^{L-1}\lambda_a,\dots,\lambda_b^{L-1}\lambda_a^{L-1}\right)
\end{equation}
where
\be
\lambda_a=\exp{\left(\frac{2\pi i a }{L}\right)}.
\ee
The  $(\ell=2k+m+n)^{th}$ power of $M$ reads:
\begin{equation}
M^{2k+m+n}[m,n]= \left(\begin{array}{c} 2k+m+n\\ k+n\end{array}
\right) \left(\begin{array}{c} 2k+m+n\\ k\end{array}\right),\quad k\geq 0
\end{equation}
and of course $M^{m+n+2k+1}[m,n]=0,~\forall k$. 
Thus the expression  of the superlink from point $x(0,0)$ to $y(m,n)$ reads:
\begin{equation}
K[m,n]=\sum_{k\geq 0} \left(\begin{array}{c} 2k+m+n\\ k+n\end{array}
\right) \left(\begin{array}{c} 2k+m+n\\ k\end{array}\right) \alpha^{ 2k+m+n}
\label{kfroz}
\end{equation}
Asymptotically, the binomial coefficients tend to
\begin{equation}
\left(\begin{array}{c} 2k\\ k\end{array}\right)\sim \frac{4^k}{\sqrt{k}},
\quad k\rightarrow \infty,
\end{equation}
 \emph{so that $K[m,n]$ becomes singular at
 $\alpha=\frac{1}{4}$ for any $(m,n)$}. This is as expected from the spectrum
of $M$, the maximum of which is 4, and corresponds to our test-charge 
becoming massless. We will be mainly interested in the 
case where $\alpha$ is close to its critical value. However one must realise
that this is the opposite limit of that in which the quenched approximation
would become valid, namely $m \rightarrow \infty$. Therefore $K$
 is far away from the propagator of the physical (unquenched) theory
 mentioned above.

The decay rate of the terms in (\ref{kfroz}) the series is easily found: 
since the terms behave like $(16\alpha^2)^k$, the correlation length  is
\begin{equation}
\frac{\xi_{\ell}}{a}=\frac{-1}{\log{(4\alpha)}} \label{corlen}
\end{equation}
As $\alpha\rightarrow\frac{1}{4}$, the correlation length diverges 
logarithmically in $\alpha$. It might seem surprising that we have already
seen another distance scale related to $\alpha$, namely the inverse mass
 of the scalar particle whose propagator is precisely $K$. But that
 correlation  length $\xi_d$ was related to the distance the test-charge can
 propagate, while here $\xi_\ell=\ell-d$, where $\ell$ is the  length of the
 typical path the particle takes to propagate from $(0,0)$ to $(0,d)$:
\begin{equation}
am_\ell=\log{\left(1+\left(\frac{am_d}{2}\right)^2\right)}
\end{equation}
When  $am_d$ goes to zero, we have to leading order
\begin{equation}
am_\ell\simeq \left(\frac{am_d}{2}\right)^2  \label{mdmell}
\end{equation}
That is, in terms of correlation lengths in lattice units, 
\begin{equation}  \label{xidxiell}
\frac{\xi_d}{a}=\frac{1}{2}\sqrt{\frac{\xi_\ell}{a}},\quad \alpha\rightarrow 0
\end{equation}

This is precisely the behaviour of a \emph{Brownian motion}:
 the distance by which the particle is diplaced after a path of length $\ell$
 is asymptotically proportional to the square root of $\ell$\footnote{Another 
consequence of this result is that if one maintains
$d$ fixed in physical units as one takes the continuum limit, then the 
different scaling properties of $\xi_d$ and $\xi_\ell$ imply that $\xi_\ell$
diverges like $a^{-1}$}. The limit $am_d\rightarrow 0$ can also
be interpreted as  the continuum limit. 
Thus we have in fact established the connection
between the Brownian propagation of the test charge and the continuum limit. 
Since rotation invariance is only regained in  the continuum limit, 
we now know that \emph{it is only when our test-charge is 
following random walks that rotation invariance is restored}.

  When on the 
other hand $\alpha\rightarrow 0$, $\xi_d\rightarrow 0$ and therefore also
 $\xi_\ell\rightarrow 0$, so that the length of the typical path taken by the 
test-charge to move from $(0,0)$ to $(d,0)$ is simply $d$. Therefore, 
the exponent $\eta$ defined by
\begin{equation}
\eta(d)=\frac{\log{\bar{\ell}(d)}}{\log{d}}
\end{equation}
should always take values between 1 and 2; the first extreme value corresponds
to a completely directed motion, the second to a random walk. As $\alpha$ is
increased from 0 to $\frac{1}{4}$, we expect to cross over from the first to
the second regime.

One has to be careful about the range of validity of this analysis.
The conclusions are based on the idea that the dominant
contribution to the propagator comes from the long paths governed by an
exponential law with correlation length $\xi_\ell$.  But this only holds
for distances $d$ smaller than $\xi_d$. For propagation over larger
 distances, 
following a path of length $\ell=d+\Delta\ell$ is suppressed, not by 
$e^{-am_\ell \Delta\ell}$, but by $e^{-am_d \Delta\ell}$, which is a much 
stronger suppression in view of (\ref{mdmell}). Therefore, for $d\gg \xi_d$,
the scale $\xi_\ell$ becomes irrelevant and we have a directed type of 
motion, \emph{however close to the critical point we are}. Since, as we have
seen, $d$ can only be neglected in the propagation law $\bar{\ell}=d+4d^2$
if $d\gg 1$, \emph{the brownian motion regime only holds for distances}
\begin{equation}
1\ll d \ll \xi_d
\end{equation}
Thus, for $[1,\xi_d]$ to be a non-zero interval, we must work with 
$\alpha>\frac{1}{5}$, otherwise the Brownian motion regime does not exist
on any length scale at all.

\begin{figure}[ht]
\begin{center}
\mbox{\epsfig{file=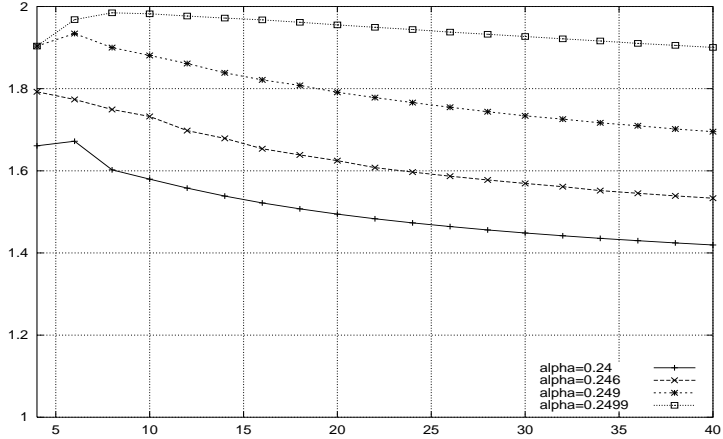, width=10cm, height=6cm}}
\caption{$\eta\equiv\frac{\log{\ell_{max}(d)}}{\log{d}}$ as a function of 
$d$}
\label{loc_exp}
\end{center}
\end{figure}

\paragraph{Illustration\\}
To show to which power of '$d$' $\ell$ is proportional to  asymptotically, 
we plot $\eta\equiv\frac{\log{\ell_{max}(d)}}{\log{d}}$ as a function of $d$ 
in Fig. (\ref{loc_exp}), for several values of $\alpha$. The variable $\eta$
represents the local power law:
\begin{equation}
\ell_{max}(d)\propto d^\eta.\label{loc_powlaw}
\end{equation}

We observe that all exponents lie between 1 and 2, and are closer to 2. 
 Inverting eqt. (\ref{loc_powlaw}),
we see that $\eta\simeq2$ corresponds to $d\sim \sqrt{\ell}$. This is precisely
the scaling law of the distance $d$ covered in a random walk
  in a time $\ell$. On the other hand, $\eta\simeq 1$ 
expresses that the paths are well orientated towards the final point $(d,0)$.
Because of this interpretation, we may consider $\eta$ as an ``order 
parameter''. The plot illustrates that \emph{the closer $\alpha$ is
 to its critical value, the more the propagation resembles a Brownian motion}.

\begin{figure}[ht]
\begin{center}
\mbox{\epsfig{file=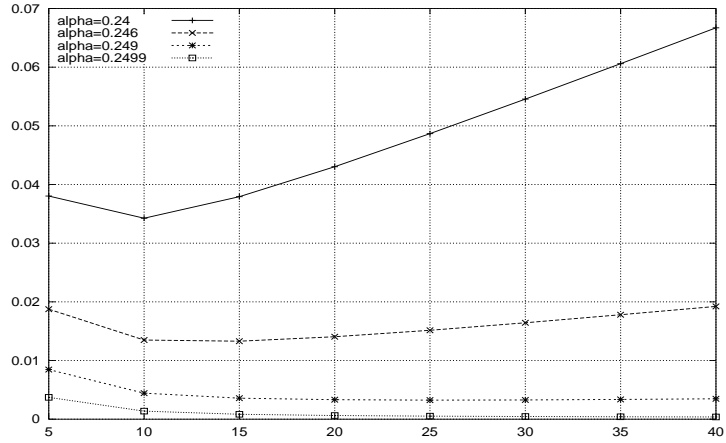, width=10cm, height=6cm}}
\caption{$\frac{2(K(\lambda,0)-K(\frac{3}{5}\lambda,\frac{4}{5}\lambda))}
{(K(\lambda,0)+K(\frac{3}{5}\lambda,\frac{4}{5}\lambda))}$ as function of $\lambda$, for several values of $\alpha$}
\label{rotinv}
\end{center}
\end{figure}

\paragraph{Rotation invariance test}
In the long-path limit, $r\equiv\sqrt{m^2+n^2}\rightarrow \infty$,
 we expect to obtain rotational invariance, in the sense that $K(x,y)$ 
is constant at fixed $r$. One easy way to check this is to choose 
Pythagorean numbers $(a,b,c)$, 
such as $(3,4,5),~(5,12,13),~(7,24,25)$, for which $a^2=b^2+c^2$, and to
compare $K(\lambda a,\lambda b)$ to $K(\lambda c,0)$. This is illustrated 
on Fig. (\ref{rotinv}), where the relative difference
\begin{equation}\Delta\equiv\frac{2~\left(K(\lambda,0)-K(\frac{3}{5}\lambda,\frac{4}{5}\lambda)\right)}
{K(\lambda,0)+K(\frac{3}{5}\lambda,\frac{4}{5}\lambda)}\label{Delta}
\end{equation}
is plotted as function of $\lambda$ for different values of $\alpha$.

At a fixed  size $\lambda$, the relative
difference decreases when $\alpha\rightarrow \frac{1}{4}$. 
This is as expected from
the preceding analysis, which showed that close to the critical behaviour, 
propagation can be thought of as a Brownian motion, which means that the
dominantly contributing paths are much longer than $\lambda$. In the 
``long-path dominance'' limit, 
it is expected that rotational invariance, since any
continuum path can be accurately approximated by a lattice path, as one 
would do to evaluate a path integral numerically.

At a fixed value of $\alpha$, and varying $\lambda$, the situation is more 
complicated: at first, for all the considered $\alpha$, the difference 
decreases with the length of the superlinks. But we observe that
 $\forall \alpha$, \emph{the difference ultimately increases at large
 $\lambda$}. The physical interpretation is that
for $\lambda$ large, the paths have a very accurately defined direction in
which to move, and however small  $|\alpha-\frac{1}{4}|$, $\lambda$ always
becomes larger than the finite size of the ``cloud'' of paths. Therefore 
the propagation becomes directional, and we end up in the regime where 
$\ell_{max}\propto d$, as we saw above in a more analytical way\footnote{ These conclusions are the consequence of the
 fact that (\ref{kfroz}) for $(m,n)=(d,0)$ is an asymptotic series in $d$.}

\paragraph{Finite volume effects}
We have seen that extremely long paths can contribute to the superlinks.
When working on a finite lattice with periodic boundary conditions, 
paths can easily go ``around  the world''. On a torus, the length of
such paths is greater if done along a diagonal than along a lattice axis.
Could that spoil our conclusions on the rotational invariance of $K$?
We  illustrate in  Fig. (\ref{finivol}) 
how the finite volume affects the quantity $\Delta$.

\begin{figure}[htb]

\centerline{~~\begin{minipage}[c]{6.5cm}
    \psfig{file=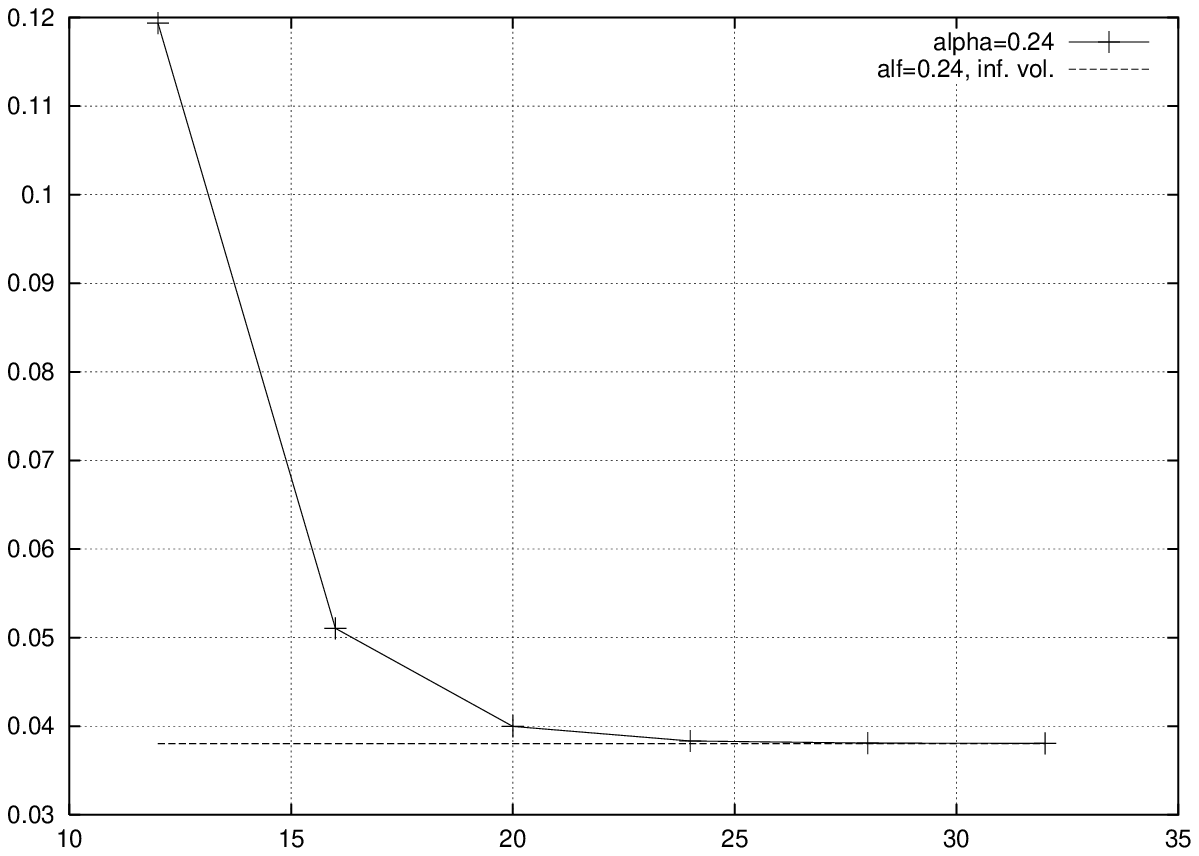,angle=0,width=6.5cm}
    \end{minipage}%
    ~~~~~\begin{minipage}[c]{6.5cm}
    \psfig{file=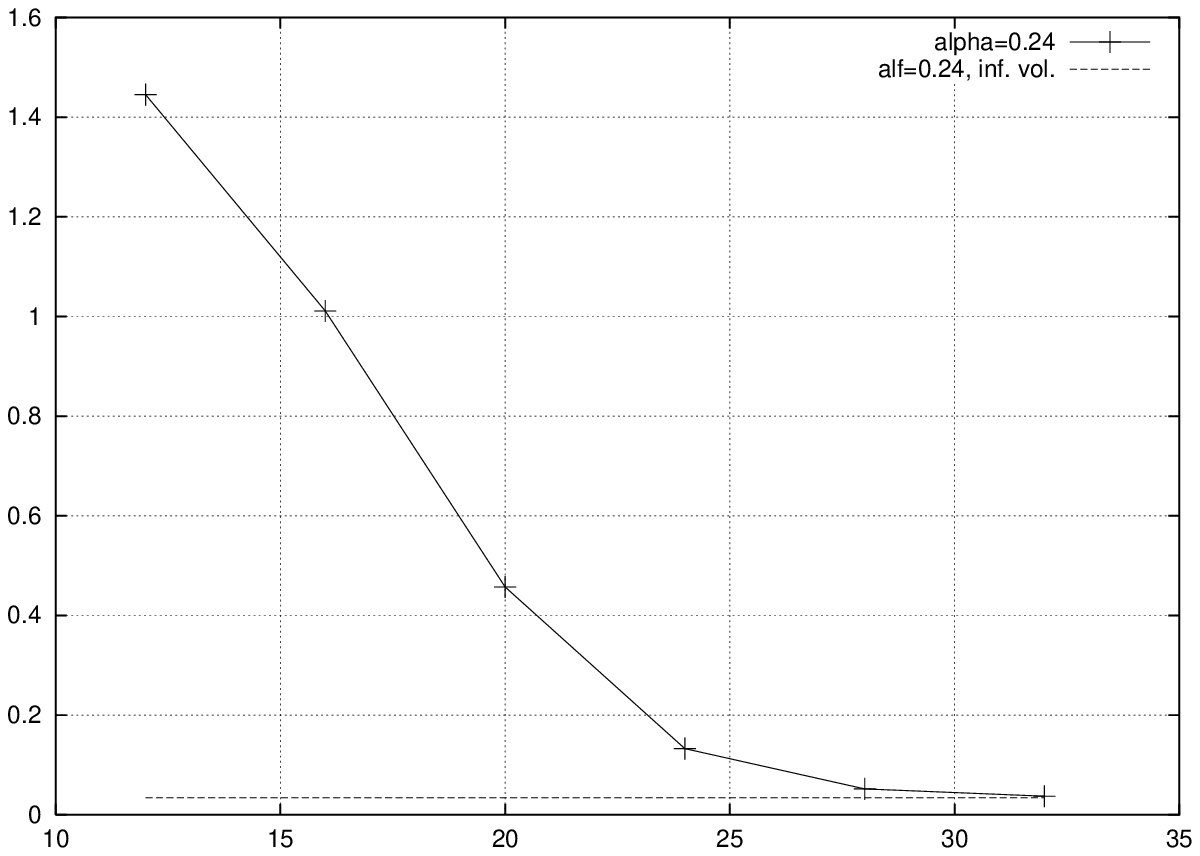,angle=0,width=6.5cm}
    \end{minipage}}

\vspace*{0.5cm}

\caption[a]{$\Delta$ as function of the lattice size $L$, 
for the couple of links a) $(3,4)$ and $(5,0)$, at $\alpha=0.24$ and b)
$(6,8)$ and $(10,0)$;
the asymptotic value for infinite volume is indicated as a constant}
\la{finivol}
\end{figure}

Due to ``paths around the world'', 
significant finite volume effects are seen when looking at the 
difference between superlinks of same Euclidean length, pointing in different 
directions.

\paragraph{Spectrum of the $M$ matrix}
It is useful to know how  the spectrum of 
the $M$ matrix in a finite $\beta$ configuration differs from the frozen 
configuration spectrum. 
Fig. (\ref{fig:eigv}) shows the spectrum of $M$ on a $16^2$ lattice
 both in the frozen configuration and at  $\beta=6$.
For individual Monte-Carlo generated gauge field configurations, the spectrum
of the frozen configuration gets smeared out by statistical fluctuations. 
Its qualitative features do not vary much with $\beta$ or $L$ at values that
are customary in simulations. The maximum of the spectrum is in $90\%$
of cases situated between 4 and 4.1. Therefore a  large enough, yet safe
 value of $\alpha$ to work with is 0.24.

\begin{figure}[htb]

\centerline{~~\begin{minipage}[c]{6.5cm}
    \psfig{file=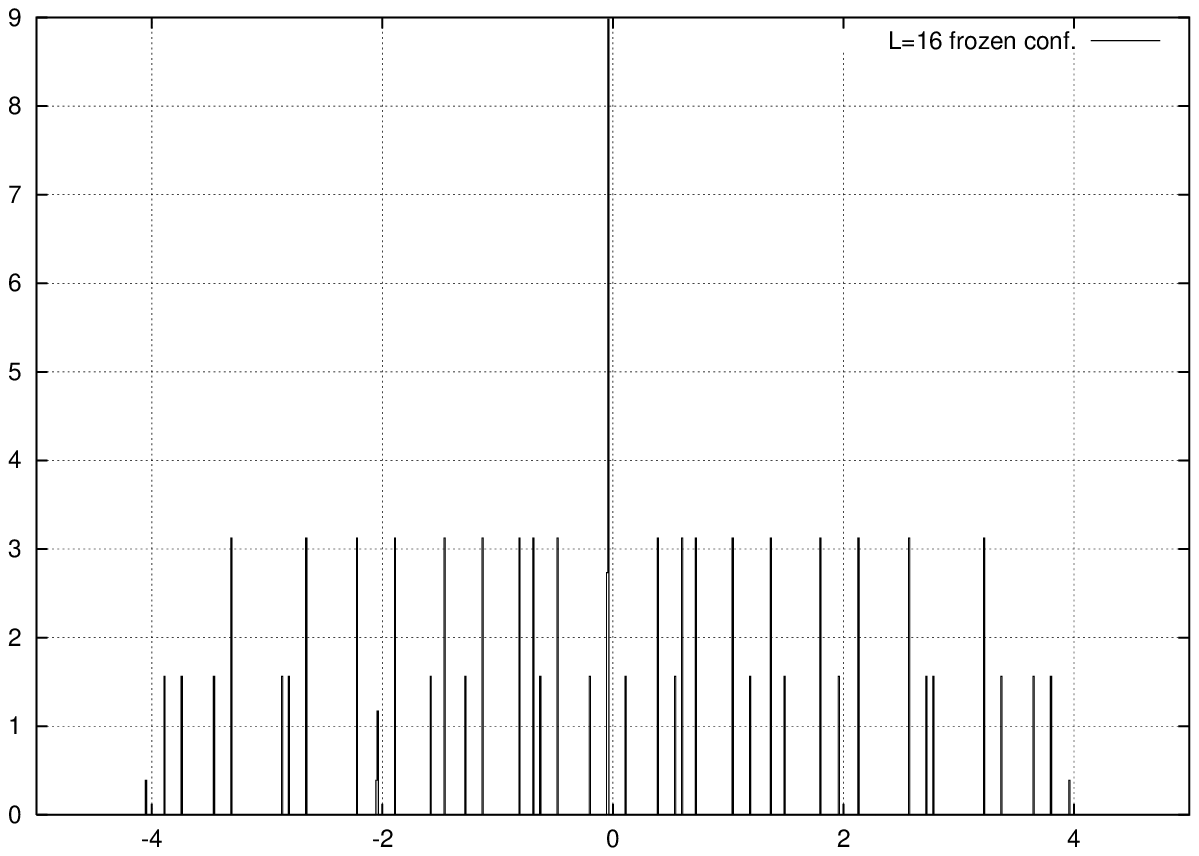,angle=0,width=6.5cm}
    \end{minipage}%
    ~~~~~\begin{minipage}[c]{6.5cm}
    \psfig{file=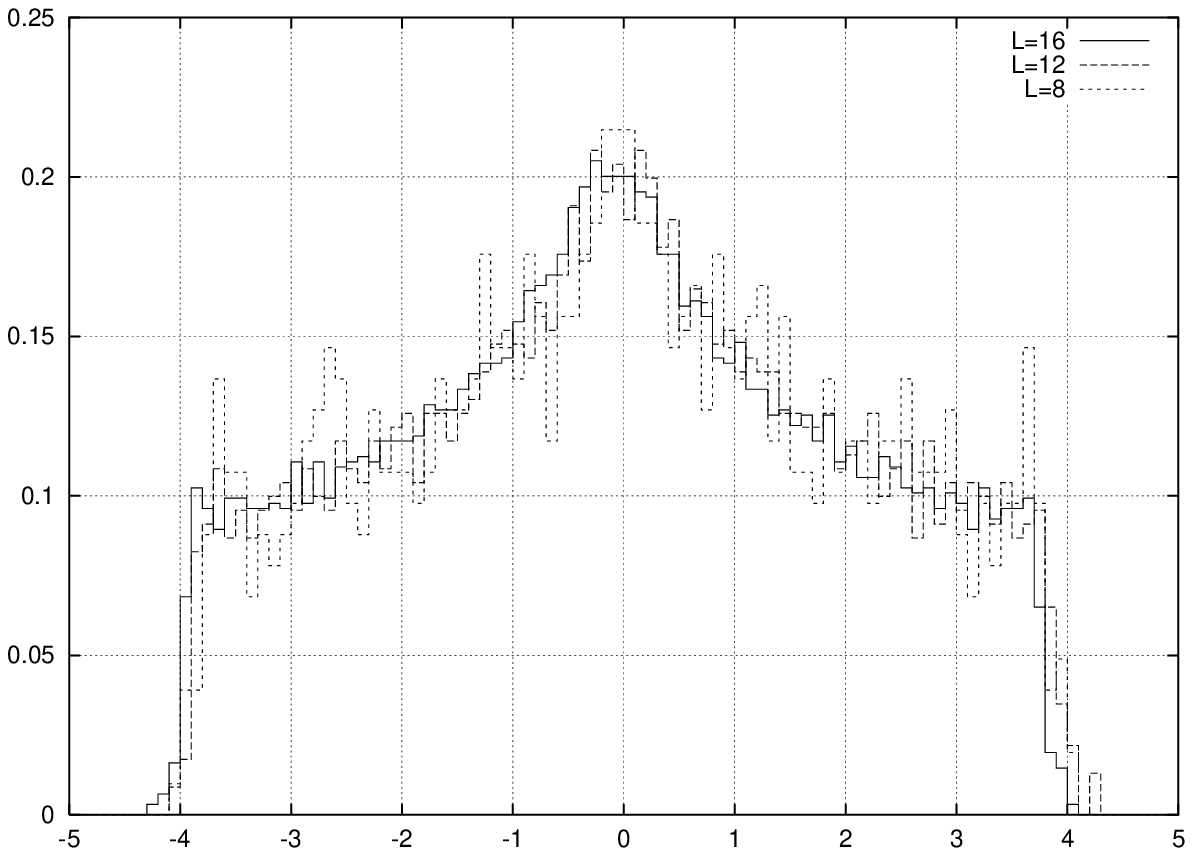,angle=0,width=6.5cm}
    \end{minipage}}

\vspace*{0.5cm}

\caption[a]{The eigenvalues of the $M$ matrix a) on a frozen $16^2$ lattice 
 b) at $\beta=6$ and different lattice sizes }

\la{fig:eigv}
\end{figure}

\section{Parity and rotations in continuous (2+1) dimensions}

The rotation group $SO(2)$ being abelian, its irreducible representations are 
one-dimensional: 
\[ \langle \phi | j \rangle = e^{ij\phi}\]
Here $j$ takes all positive an negative values. 

The parity transformation, i.e. a flip around an axis, 
takes a clockwise-winding state into an
anticlockwise-winding state, so that
\[ P |j \rangle = e^{i\theta} |-j\rangle, \]
which implies that $P$ and $J$ do not commute and therefore
cannot be diagonalised simultaneously. 
For a particular choice of axis, $\theta$ can be chosen to be zero.
The fact that the Hamiltonian
is parity-invariant implies that the $|j\>$ and  $|-j\>$ are degenerate:
\[ E_j=\< j | H | j\> = \< Pj | PHP | Pj\>= \< -j | H | -j\>= E_{-j}~.\]
This fact is called ``parity doubling''.

It is also straightforward to show that 
\[ \{J,P\}=0. \]
As a consequence, 'parity' as defined with an axis rotated
 by an angle $\phi$ with respect to the 
reference axis will be related to $P$ according to
\[P_\phi\equiv e^{iJ\phi} P  e^{-iJ\phi}= e^{2iJ\phi} P=P  e^{-2iJ\phi} \]
In particular, acting on a spin $j\neq 0$, this relation implies that
\[ P_\phi | j \rangle = - P| j \rangle,\quad \phi=\frac{\pi}{2j}\]
Thus an elegant way to understand parity doubling is that
the $P=\pm 1$ labelling can be reversed by the use of  another 
convention; except for the spin 0, where all choices of parity axis
 will label the states in the same way.

\section{Irreducible representations of the square group}
The character table of the symmetry group of a 2-dimensional time-slice is 
given below. $C_4$ are the rotations by $\frac{\pi}{2}$, $C_2$ is the rotation
by $\pi$, $\sigma$ is the reflexion around the $x$ axis, $\sigma'$ is the 
reflexion around the $y=x$ axis.
\begin{center}
\begin{tabular}{|c|c||c|c|c|c|c|}
\hline
fct. &     & $E$  & $2C_4$  & $C_2$  &  $2\sigma$ & $2\sigma'$ \\
\hline
1 &$(0^+,~4^+,\dots) \rightarrow A_1$ &  1   &   1    &   1     &   1    &   1 \\
$xy(x^2-y^2)$&$(0^-,~4^-,\dots)\rightarrow A_2$ &  1   &   1    &   1     &   -1   &   -1 \\
$x^2-y^2$& $(2^+,~6^+,\dots)\rightarrow A_3$ &1   &   -1    &   1     &   1    &   -1 \\
$xy$ &$(2^-,~6^-,\dots)\rightarrow A_4$ &1   &   -1    &   1     &   -1    &   1 \\
$(x,y)$&$(1^\pm,~3^\pm,\dots)\rightarrow E$   &2   &   0    &   -2   &   0   &   0 \\
\hline
\hline
$(x+iy)^j$& $D_j$      &   2 &$2\cos{\frac{j\pi}{2}}$ &  $2\cos{j\pi}$ &  0  & 0 \\
\hline
\end{tabular}
\end{center}
It is interesting that $1^+$ and $1^-$ are exactly degenerate on the lattice 
--  they belong to the same representation on the lattice --, 
while the $2^-$ and $2^+$ are not.
Applying the projection rules for characters, we can immediately find 
how  the spin $J$ representation $D_J$ decomposes onto the irreducible 
representations of the square group. For instance:
\[  D_4 = A_1 \oplus A_2 \]

\section{Irreducible representations of the cubic group}
The character table of the rotation symmetry group (432) 
of a 3-dimensional time-slice is given below. The insertion of parity in the 
group ((m3m): $O_h=O\times i$) does not introduce any complications as in 
two dimensions, because parity commutes with rotations and is realised 
exactly on the lattice.
$C_4$ are the rotations by $\frac{\pi}{2}$, $C_2$ 
by $\pi$ (3 along the axes and 6 along face diagonals) and $C_3$ are the 
ternary axes along the volume diagonal.
\begin{center}
\begin{tabular}{|c|c||c|c|c|c|c|}
\hline
fct.& repr.  & $E$  & 8$C_3$ & 3$C_2$  & $6C_2$   &  $6C_4$   \\
\hline
1 & $A_1$ &  1   &   1    &   1     &   1    &   1 \\
$xyz\propto Y_3^2-Y_3^{-2}$ &$A_2$ &  1   &   1    &   1     &   -1   &   -1 \\
$(Y_2^0,Y_2^2+Y_2^{-2})$& $ E$ &2   &   -1    &   2     &   0    &   0 \\
$(x,y,z)$,$(Y_1^1,Y_1^{-1},Y_1^0)$ & $T_1$ &3   &   0    &   -1     &   -1    &   1 \\
$(Y_2^1,Y_2^{-1},Y_2^2-Y_2^{-2})$&$ T_2$   &3   &   0    &   -1   &   1   &   -1 \\
\hline
\hline
$Y_j^m$& $D_j$ & $2j+1$ & $(1,0,-1) $ &  $(1,-1)$   &  (1,-1) & (1,1,-1,-1) \\
\hline
\end{tabular}
\end{center}
In the last line, the different values of $\chi_j(C_n)$ correspond to
$j\equiv 0,\dots,n-1~(\mathrm{mod}~n)$. 
These values are easily obtained from the general formula
\[\chi_j(\alpha)=\frac{\sin{(j+\frac{1}{2})\alpha}}{\sin{\frac{\alpha}{2}}}.\]
Thus the smallest spins coupling to the various lattice representations are
\ba
A_1 \quad&\rightarrow &\quad {\rm spin}~0 \nonumber\\
T_1 \quad&\rightarrow &\quad {\rm spin}~1 \nonumber\\
E \quad&\rightarrow &\quad {\rm spin}~2 \nonumber\\
T_2 \quad&\rightarrow &\quad {\rm spin}~2 \nonumber\\
A_2 \quad&\rightarrow &\quad{\rm spin}~3 \nonumber
\ea
Inversely, a few useful decompositions of the continuum representations read
\ba
D_0&=& A_1 \qquad ({\rm scalar})\nonumber\\
D_1&=& T_1 \qquad ({\rm vector})\nonumber\\
D_2&=& E\oplus T_2 \qquad ({\rm tensor})\nonumber\\
D_3&=& A_2\oplus T_1\oplus T_2 \nonumber\\
D_4&=&A_1\oplus E \oplus T_1\oplus T_2\nonumber\\
D_5&=&E \oplus2 T_1\oplus 2T_2\nonumber\\
D_6&=&A_1 \oplus A_2\oplus E \oplus T_1  \oplus2 T_2
\ea

\vfil\eject
\section*{Tables}
\begin{table}[ht]
\centerline{
\begin{tabular}{|c|c|c|c|}
\hline
10&     -3&     -10&      3\\
6&      5&      -6&     -5\\
 \hline
 14&     -4&     -14&      4\\
   8&      7&       -8&     -7\\
 \hline
 6&      5&     -6&     -5 \\
  0&      3&      0&     -3  \\
 \hline
 8&      7&     -8&     -7  \\
  0&      4&      0&     -4  \\
\hline
\end{tabular}
~~~~~~~\begin{tabular}{|c|c|c|c|}
\hline
          10&   -3&   -10&    3 \\
           6&    5&   -6&   -5\\
 \hline
           15&    -3&   -15&    3\\
           9&      5&   -9&    -5\\
 \hline
          14&   -4&  14&    4\\
          8&    7&   -8&   -7\\
 \hline
           6&    5&   -6&   -5\\
           0&    3&    0&   -3\\
 \hline
          12&   10&  -12&  -10\\
           0&    6&    0&   -6\\
 \hline
           8&    7&   -8&   -7\\
           0&    4&    0&   -4\\
\hline
\end{tabular}
~~~~~~~\begin{tabular}{|c|c|c|c|}
\hline
         13&   -8&   -5 &\\
           0&   12  &       -12  &\\
 \hline
            13&   -1&  -12&\\
           0&    5      &    -5&\\
 \hline
          18&   -1   & -17&\\
           0&    7&   -7& \\
 \hline
          13&   12&  -13&  -12\\
           0&    5&    0&   -5\\
 \hline
          21&   -6&  -21&    6\\
          12&   14&  -12 &   -14\\
 \hline
           24&   -5&  -24&    5\\
            10&   12&   -10&  -12\\
\hline
\end{tabular}
}
\caption{Vector coordinates of the operators used at $\beta=6$ 
(12-fold rotations; left), at $\beta=9$ (12-fold rotations; center) and 
$\beta=12/~ 14.5$ (16-fold rotations; right). 
They are given in line, the upper/lower corresponding to the $x$ and $y$
components of the vector joining one point of the polygon to the next one.}
\label{coords}
\end{table}

\begin{table}
\begin{center}
\begin{tabular}{|c|c|c|c|c|}
\hline
$\mathbf{\beta=6}$ & $am$ ($t=a$)& $am$ ($t=2a$)& quality$[\%]$ & overlap$[\%]$\\
\hline
$0^+$ &  1.2309(77)& $\mathbf{ 1.203(27)}$      &    97.2(36)  & 6.3 \\
$0^{+*}$ &1.995(23)  &$\mathbf{ 1.79(12)  }$    &  81.4(12) & 4.8 \\
$2^+$ & 1.998(12) & $\mathbf{1.777(80)}$       & 80.1(74) & 4.1 \\
$2^-$ & 1.947(11) & $\mathbf{1.70(12)}$       & 78(10) & 9.9 \\
$4^+$ &  2.509(24)& $\mathbf{  2.44(27)  }$     &   93(27) & 13.8 \\  
$4^-$ & 2.536(35) & $\mathbf{  2.42(37)}$       &   89(36) & /\\
 \hline
\end{tabular}
\end{center}
\begin{center}
\begin{tabular}{|c|c|c|c|c|c|}
\hline
$\mathbf{\beta=9}$ & $am$ ($t=a$)& $am$ ($t=2a$)& $am$ ($t=3a$)&
 quality$[\%]$ & overlap$[\%]$\\
\hline
$0^+$    &    0.8053(85) &$\mathbf{0.7681(93)}$& 0.739(22)  &96.3(21) & 6.9 \\
$0^{+*}$ &  1.1904(82)&$\mathbf{1.159(28)}$& 0.995(12)     & 96.9(36) & 5.9 \\
$2^+$ & 1.3311(71) &$\mathbf{ 1.287(47) }$&  1.156 (97)    &  95.6(54) & 5.5\\
$2^-$ & 1.410(13) &$\mathbf{ 1.301(49) }$&  1.16 (21)    &  89.7(58) & 9.0\\
$4^+$ &   1.721(14) &$\mathbf{ 1.623(66)}$ & 1.70(45)     & 90.7(76) & 2.0 \\
$4^-$ &   1.709(18) &$\mathbf{ 1.67(11)}$ & 1.58(52)     & 96(12) & / \\
 \hline
\end{tabular}
\end{center}
\begin{center}
\begin{tabular}{|c|c|c|c|c|c|c|}
\hline
$\mathbf{\beta=12}$ & $am~(t=a)$  & $am~(t=2a)$ & $am~(t=3a)$ & $am~(t=4a)$ &
 quality$[\%]$ & overlap$[\%]$\\
\hline
$0^+$ & 0.6337(54) & 0.5845(66)&$\mathbf{0.567(14)}$&0.558(27)&91.8(41)&1.8 \\
$0^{+*}$ &1.054(11)&0.946(22)&$\mathbf{ 0.899(47)}$ & 0.95(14)& 81.6(96)&3.3\\
$2^+$   & 1.0991(55) & 1.030(20) &$\mathbf{0.991(54)}$& 0.86(11)&86(11)&4.0 \\
$2^-$   & 1.0928(71) & 1.009(29) &$\mathbf{0.946(98)}$& 0.70(14)&81(19)&2.6 \\
$4^+$  &1.4105(86) &1.364(45)&$\mathbf{1.24(14)}$&0.70(24) & 95.5(53)&5.7\\  
$4^-$  &1.412(10) &1.328(50)&$\mathbf{1.20(15)}$&0.95(28) &91.9(58)&/\\ 
\hline
$3^+$  &1.611(12) &$\mathbf{ 1.525(69)}$&2.09(63)& / &91.8(76)&7.5\\ 
$1^+$  &1.863(25) &$\mathbf{ 1.72(13)}$& /& / &87(14)&1.7\\ 
$3^-$  &1.644(12) &$\mathbf{ 1.680(83)}$&1.64(47)&/ &100(10)&7.0\\ 
$1^-$  &1.889(21) &$\mathbf{ 1.79(13)}$& /& / &91(14)&1.8\\ 
 \hline
\end{tabular}
\end{center}
\begin{center}
\begin{tabular}{|c|c|c|c|c|c|c|}
\hline
$\mathbf{\beta=14.5}$ & $am~(t=a)$  & $am~(t=2a)$ & $am~(t=3a)$ & $am~(t=4a)$ &
 quality$[\%]$ & overlap$[\%]$\\
\hline
$0^+$ & 0.5823(42)& 0.5107(57)&0.4921(79)&0.486(15)& 79(11)$$ & 3.3 \\
$0^{+*}$&0.7999(51)&0.732(11)&$\mathbf{0.666(20)}$&0.601(40) & 82.0(46)&8.2 \\
$2^+$ & 0.9851(52)& 0.883(12)&$\mathbf{ 0.800(22)}$&0.786(69) &76.4(47)&1.1\\
$2^-$ & 0.9413(70)& 0.867(12)&$\mathbf{ 0.826(30)}$&0.822(84) &85.4(68)&3.7\\
$4^+$    &1.413(13)&1.184(41)&$\mathbf{0.98(11)}$&1.14(38) &91.0(26)& 9.2\\  
$4^-$  &1.2698(91)&1.196(43)&$\mathbf{1.028(97)}$&/ &92.9(51)& /\\  
\hline
$3^+$  &1.527(26)& 1.64(11) & 1.66(77)&/ &100(15)&2.5\\ 
$3^-$  &1.563(16)& $\mathbf{ 1.569(54)}$ & 1.73(40)&/ &100(72)&2.7\\
$1^+$  &1.732(20)& $\mathbf{1.616(86)}$ & 1.65(63)&/ &89.0(96)&4.4\\ 
$1^-$  &1.68(17)& 1.7(1.0) & /&/ & /&3.9\\
 \hline
\end{tabular}
\end{center}
\caption{Strategy I: The local effective masses, quality factors and overlaps 
(as defined in the text) between states of different wave functions
 at $\beta=6,~9,~12$ and 14.5. For the $0^+$ at $\beta=14.5$,
 we used the effective mass at five lattice spacings $\mathbf{0.460(30)}$.}
\la{sIdata}
\end{table}
\clearpage


\begin{table}
\begin{center}
\begin{tabular}{|c|c|c|c|}
\hline
$\mathbf{\beta=6}$ & $am~[\bar t]$ & $c$ & $c'$ \\
\hline
$A_1$     & 1.190(24)[1.5] & 1.000(10)  &  0.017(14)  \\
$A_1^*$  & 1.804(96)[1.5] & 0.990(21)   &  0.142(28)  \\
$A_3$     &1.666(81)[1.5] &   1         &   0          \\
 \hline
\end{tabular}
\end{center}
\begin{center}
\begin{tabular}{|c|c|c|c|}
\hline
$\mathbf{\beta=9}$ & $am~[\bar t]$ & $c$ & $c'$ \\
\hline
$A_1$     & 0.7731(79)[1.5]  &   1.000(14)  &  0.028(21)  \\
$A_1^{*}$  & 1.179(37) [1.5] &   0.998(10)  & 0.063(16)  \\
$A_1^{+**}$ & 1.433(64)[1.5]  &   0.973(25)   &  0.231(32))   \\
$A_1^{+***}$     & 1.70(13) [1.5]  &   0.786(98) &    0.618(12)\\
$A_3^+$     &  1.303(47)[1.5] &   1     &        0     \\
 \hline                    
\end{tabular}
\end{center}
\begin{center}
\begin{tabular}{|c|c|c|c|c|}
\hline
$\mathbf{\beta=12}$ & $am~[\bar t]$ & $c$ & $c'$ \\
\hline
$A_1$    & 0.572(15)~[2.5] &   1.000(18) & 0.000(25)\\
$A_1^{*}$ & 0.856(53)~[2.5] &   0.992(26) & 0.124(37)\\ 
$A_1^{**}$& 0.943(39)~[2.5] &   0.988(38) & 0.152(53)\\ 
$A_1^{***}$    &1.294(59) ~[1.5] &   0.680(68) & 0.734(97)\\ 

$A_2$    & 1.365(57)~[1.5] &  0         & 1  \\
$A_3$    & 0.990(60)~[2.5] &   1 &  0  \\
$A_4$    & 1.03(10)~~[2.5] &   0.999(32) & 0.035(21) \\
\hline
$E^-$    & 1.547(72) [1.5]&  0         & 1  \\
$E^+$    & 1.650(65) [1.5]&   0         & 1  \\
 \hline
\end{tabular}
\end{center}
\begin{center}
\begin{tabular}{|c|c|c|c|}
\hline
$\mathbf{\beta=14.5}$ & $am~[\bar t]$  & $c$ & $c'$ \\
\hline
$A_1$    & 0.489(13)~[2.5] &  1.000(41) & 0.016(58)\\ 
$A_1^{*}$ & 0.669(24)~[2.5]  &  0.998(16) & 0.0619(22) \\
$A_1^{**}$& 0.816(56)~[2.5]  &  0.985(40) & 0.172(56)  \\  
$A_1^{***}$    &1.11(12) ~[2.5]  &  0.577(48) & 0.816(68)\\
$A_2$    & 1.04(12)~[2.5]   &  0           & 1 \\  
$A_3$    & 0.776(34)~[2.5]  &  1  &  0 \\ 
$A_4$    & 0.71(13)~~[3.5]  &  0.995(30) & 0.097(27) \\ 
\hline
$E^+$    & 1.40(38)[1.5]  & 0            & 1 \\  
$E^-$    & 1.52(53)[1.5]  & 0            & 1 \\  
 \hline 
\end{tabular}
\end{center}
\caption{Strategy II: The local effective masses, and Fourier coefficients
obtained  at $\beta=6,~9,~12$ and 14.5. 
The wave function coefficients $c$ and $c'$, 
all obtained at two lattice spacings,  correspond  respectively to the 
smallest and second-smallest spin wave function compatible with the lattice
representation (e.g., $c=c_0$ and $c'=c_4$ for the $A_1$ representation).
The number in brackets indicates at what time separation the local effective
mass was evaluated.}
\la{sIIdata}
\end{table}
\clearpage

\begin{table}
\begin{center}
\begin{tabular}{|c|c|c|c|}
\hline
State & $L=16$ &  $L=24$ &  $L=32$ \\
\hline
$0^+$ & 0.764(11) & 0.7681(93) & 0.766(27)\\
$0^{+*}$ &1.065(21) & 1.159(28)  & 1.113(48) \\
$2^+$  &  1.194(22) & 1.287(47) & 1.295(62)\\
$4^+$ & 1.620(37) & 1.623(66) & 1.57(13) \\
\hline
\end{tabular}
\end{center}
\caption{Strategy I: volume dependence of various glueball masses in lattice 
units at $\beta=9$}
\label{vdep_tab}
\end{table}

\begin{table}
\begin{center}
\begin{tabular}{|c|c|c|}
\hline
$\beta$ & $L$ &  $a\sqrt{\sigma}$ \\
\hline
6.0  & 16 & 0.2538(10) \\
9.0 & 24& 0.1616(6)\\
12.0 & 32 & 0.1179(5)\\
14.5 & 40 & 0.09713(20)\\
\hline
\end{tabular}
\end{center}
\caption{The string tension values obtained in [\ref{mtd3}]}
\la{tensions}
\end{table}

\begin{table}
\begin{center}
\begin{tabular}{|c|c|c|}
\hline
State & $m/\sqrt{\sigma}$ & C.L.$[\%]$ \\
\hline
$0^+$ & 4.76(11) &  99 \\
\hline
$0^{+*}$ & 6.88(23) & 14 \\
\hline
$2^+$    & 8.44(24)& 99  \\
$2^-$ &    8.81(34) & 90\\
\hline
$4^+$ &10.28(81) & 94 \\  
$4^-$ &10.70(93) & 97 \\   
\hline
\end{tabular}
\end{center}
\caption{Strategy I: Continuum extrapolation of the spectrum in units 
of the string tension obtained with strategy I.
}
\la{contI}
\end{table}

\begin{table}[b]
\begin{center}
\begin{tabular}{|c|c|c|}
\hline
State & $m/\sqrt{\sigma}$ & C.L.$[\%]$ \\
\hline
$0^+$ & 4.934(98) &  50 \\
\hline
$0^{+*}$ & 7.03(26) & 55 \\
\hline
$0^{+**}$ & 7.54(70)  & 36 \\
\hline
$2^+$    & 8.65(33)& 38  \\
\hline
$4^+$ & 11.6(1.3) & 84 \\
\hline
\end{tabular}
\end{center}
\la{contII}
\caption{Strategy II: we obtain here the following continuum-extrapolated
 glueball masses, in units of the string tension. The continuum quantum 
numbers have been attributed after extrapolation of the Fourier coefficients.}
\end{table}
\end{document}